\documentclass[12pt]{article}

\makeatletter
\@ifundefined{pdfoutput}
{}
{\pdfoutput=1 \pdfminorversion=6}
\makeatother

\usepackage{ucs} 
\usepackage[T1]{fontenc}

\usepackage[a4paper,text={16.8cm,22.4cm}]{geometry}
\usepackage{amsfonts,amsmath,amssymb,bm,graphicx,color}
\usepackage{bbm}
\usepackage{cite}
\usepackage{multirow}
\usepackage{float}
\usepackage{slashed}
\usepackage{arydshln}

\allowdisplaybreaks 
\let\origtitlepage\titlepage
\def\titlepage{\origtitlepage\count0-1}

\newcommand\slashnext[1]{{\setbox0\hbox{\(#1\)}\setbox1\hbox to \wd0{\hss/\hss}\rlap{\box1}{\box0}}}

\newcommand\Aslash{\mathord{\slashnext A}}

\newcommand\qslash{\mathord{\slashnext q}}

\newcommand{\re}{\mathop{\mathrm{Re}}}
\newcommand{\im }{\mathop{\mathrm{Im}}}

\newcommand{\epol}{\epsilon^*}

\newcommand{\MSbar}{\ensuremath{\overline{\rm MS}}}

\newcommand{\be}{\begin{equation}}
\newcommand{\ee}{\end{equation}}
\newcommand{\bea}{\begin{eqnarray}}
\newcommand{\eea}{\end{eqnarray}}
\newcommand{\nn}{\nonumber}

\newcommand\dodot{{.\spacefactor=1000} \afterassignment\myowntemp\let\myowntemp=}
{\catcode`\.=\active\relax\global\let.=\dodot}
\newcommand\dojournal[1]{#1}
\newcommand\journal{{\catcode`\.=\active\relax\expandafter}\dojournal}
{\catcode`\/=12\relax
\gdef\makesplittable#1/#2\endsplittable{%
  \dodotsplittable{#1}\ifx\relax#2\relax\else/\discretionary{}{}{}%
  \makesplittable#2\endsplittable\fi}
\gdef\dosplittable#1{{\makesplittable#1/\endsplittable}}}
{\catcode`\.=12\relax
\gdef\makedotsplittable#1.#2\endsplittable{%
  \dolparensplittable{#1}\ifx\relax#2\relax\else.\discretionary{}{}{}%
  \makedotsplittable#2\endsplittable\fi}
\gdef\dodotsplittable#1{{\makedotsplittable#1.\endsplittable}}}
{\catcode`\(=12\relax
\gdef\makelparensplittable#1(#2\endsplittable{%
  \dorparensplittable{#1}\ifx\relax#2\relax\else\discretionary{}{}{}(%
  \makelparensplittable#2\endsplittable\fi}
\gdef\dolparensplittable#1{{\makelparensplittable#1(\endsplittable}}}
{\catcode`\)=12\relax
\gdef\makerparensplittable#1)#2\endsplittable{%
  #1\ifx\relax#2\relax\else)\discretionary{}{}{}%
  \makerparensplittable#2\endsplittable\fi}
\gdef\dorparensplittable#1{{\makerparensplittable#1)\endsplittable}}}
\newcommand\dourl[1]{{\edef\temp{\noexpand\dosplittable{#1}}\expandafter}\temp}
\newcommand\dodoi[1]{{\edef\temp{\noexpand\dosplittable{#1}}\expandafter}\temp}

\usepackage[unicode]{hyperref}

\makeatletter
\@ifundefined{unichar}
{\newcommand\myunichardef[3]{\expandafter\providecommand\csname text#1\endcsname
                            {#2}}}
{\newcommand\myunichardef[3]{\expandafter\providecommand\csname text#1\endcsname
                            {\unichar{"#3}}}}
\makeatother
\myunichardef{leq}{<=}{2264}
\myunichardef{partial}{del}{2202}
\myunichardef{cdot}{.}{00B7}
\myunichardef{composetilde}{\string~}{0303}
\myunichardef{composemacron}{bar}{0305}
\myunichardef{subscriptone}{\string_1}{2081}
\myunichardef{subscripti}{\string_i}{1D62}
\myunichardef{subscriptj}{\string_j}{2C7C}

\begin{document}

\begin{titlepage}
\begin{flushright}
LAUR-15-21109
\end{flushright}

\vspace{0.2cm}
\begin{center}
\LARGE\bf
Dimension-5 CP-odd  operators: \\ QCD  mixing  and renormalization
\end{center}

\vspace{0.2cm}

\begin{center}
Tanmoy Bhattacharya, 
Vincenzo Cirigliano, 
Rajan Gupta, \\
Emanuele Mereghetti, 
Boram Yoon

\vspace{0.6cm}
{\sl 
Theoretical Division, Los Alamos National Laboratory \\
Los Alamos, NM 87545, U.S.A.\\[2mm]
}
\end{center}

\vspace{0.2cm}

%\date{\today}

\begin{abstract}
\vspace{0.2cm}
\noindent 
We study the  off-shell mixing and renormalization of flavor-diagonal dimension-5 T- and P-odd operators 
involving  quarks, gluons, and photons,
including quark electric dipole 
and chromo-electric dipole operators. 
We  present the renormalization matrix to one-loop in the  $\overline{\rm MS}$ scheme.
We also provide a definition of the  quark chromo-electric dipole operator in 
a regularization-independent momentum-subtraction scheme suitable for non-perturbative lattice calculations  
and present  the matching coefficients with the $\overline{\rm MS}$ scheme to one-loop in perturbation theory, 
using both the na\"{\i}ve dimensional regularization and 't Hooft-Veltman  prescriptions for $\gamma_5$. \\

\end{abstract}
\vfil

\end{titlepage}

\section{Introduction}

Permanent electric dipole moments (EDMs) of non-degenerate systems
violate invariance under parity (P) and time reversal (T), or,
equivalently~\cite{Luders:1954zz}, CP, the combination of charge
conjugation and parity.  Given the smallness of Standard Model (SM)
CP-violating (CPV) contributions induced by quark
mixing~\cite{Kobayashi:1973fv} (for a review see
\cite{Pospelov:2005pr}), nucleon, nuclear, and atomic/molecular
EDMs~\cite{Baron:2013eja,Hudson:2011zz,Regan:2002ta,Baker:2006ts,Griffith:2009zz}
are very deep probes of the SM $\theta$-term (already constrained at
the level of $\theta \sim 10^{-10}$) and of possible new sources of CP
violation beyond the Standard Model (BSM).

In fact, EDMs at the sensitivity level of ongoing and planned
experiments probe BSM CPV interactions originating at the TeV scale or
above (up to hundreds of TeV depending on assumptions about the BSM
scenario).  These new CPV interactions may be a key
ingredient of relatively low-scale baryogenesis mechanisms such as
electroweak baryogenesis (see \cite{Morrissey:2012db} and references
therein), making the study of EDMs all the more interesting.  EDMs of
the nucleon, nuclei, and atoms are sensitive to a number of new
sources of CP violation, in a complementary way~\cite{Chupp:2014gka},
so that a broad experimental program to search for EDMs in various
systems is called for (a summary of current status and prospects can
be found in Refs.~\cite{Kumar:2013qya,Engel:2013lsa}).
 
Extracting robust information on the new CPV sources from the
(non)observation of EDMs is a challenging theoretical problem, that
involves physics at scales ranging from the TeV (or higher) down to
the hadronic, nuclear, and atomic scales, depending on the system
under consideration.
The relevant physics at the hadronic and nuclear scale involves strong
interactions, and requires the calculation of non-perturbative matrix
elements.
While interesting model-independent statements can be made within a
nucleon-level chiral effective theory
approach~\cite{Mereghetti:2010tp,deVries:2012ab,deVries:2011an,Bsaisou:2014oka,Dekens:2014jka,Bsaisou:2014zwa},
ultimately the computation of a number of hadronic matrix elements is
necessary.
Existing calculations of the impact of BSM operators on hadronic EDMs  
typically rely on modeling the strong dynamics in ways consistent with 
the Quantum Chromodynamics (QCD) symmetries, using methods such as 
QCD sum rules~\cite{Pospelov:1999ha,Pospelov:1999rg,Pospelov:2000bw,Hisano:2012sc}
and the Dyson-Schwinger equations~\cite{Pitschmann:2012by,Pitschmann:2014jxa} (see Refs.~\cite{Pospelov:2005pr,Engel:2013lsa} for reviews).
Since models do not rely on systematic approximations to the strong dynamics of quarks and gluons in the nucleons, 
current results represent in some cases only  crude estimates, with different model-calculations differing by up to an order of magnitude, 
depending on the operator under study~\cite{Engel:2013lsa}.  
Needless to say,  this state of affairs greatly dilutes the impact of EDM experimental searches 
in probing short-distance physics.  Moreover, the uncertainties 
affect the robustness  of the phenomenological studies relating new sources of 
CP violation to baryogenesis mechanisms (depending on what is the dominant mechanism and operator  generating the EDM). 

In this context, lattice QCD calculations offer the opportunity to perform  
{\em systematically improvable} calculations of the CPV  hadronic dynamics.
Historically, lattice QCD  efforts have mostly focused on the determination of  the nucleon EDM 
induced by the SM $\theta$ term~\cite{Aoki:1989rx,Aoki:1990ix,Guadagnoli:2002nm,Shintani:2005xg,Berruto:2005hg,Shintani:2008nt,Aoki:2008gv,Shintani:2014zra,Guo:2015tla}. 
Only recently there has been interest in studying the impact of 
the leading CP-odd operators on the nucleon EDM~\cite{Bhattacharya:2012bf,Bhattacharya:2014cla,Bhattacharya:2013ehc}  and the T-odd pion nucleon couplings~\cite{andreWL}.

This program, however, comes with several challenges, ranging from controlling the signal-to-noise ratio on the lattice 
to studying operator mixing,  and matching  suitably renormalized lattice operators 
to the minimally-subtracted operators typically used in phenomenological applications. 
In this paper we focus on defining UV finite CP-odd operators of dimension five and lower, 
using a renormalization scheme suitable for implementation on the lattice, 
and matching   this scheme to the perturbative $\MSbar$ scheme to one-loop.

The paper is organized as follows. 
In Section~\ref{Sec2}, we describe the effective theory framework 
parameterizing BSM effects at low-energy and identify the leading dimension-5 
CPV  operators. 
In Section~\ref{sect:3}, we construct the basis of operators needed to study 
the renormalization of the quark chromo-electric dipole moment (CEDM) operator 
in an off-shell momentum subtraction scheme with non-exceptional momenta. 
In Section~\ref{sect:4}, we present the one-loop calculations needed
to determine the full mixing matrix to $O(\alpha_s)$ 
for the operator basis 
discussed in Section~\ref{sect:3}.
In Section~\ref{sect:5}, we give our results for the matrix of
renormalization constants in the $\MSbar$ scheme, while in
Section~\ref{sect:MStoRISMOM} we specify the renormalization
conditions that define a regularization-independent (RI) momentum
subtraction scheme and provide the $O(\alpha_s)$ matching coefficients
to the $\MSbar$ scheme.
In Section~\ref{sect:WI}, we discuss the consistency of our
renormalization conditions with the singlet axial Ward identities.
In Section~\ref{sect:comparison} we compare our results to 
recent related work~\cite{Constantinou:2015ela} that studies  the renormalization 
of the strangeness changing chromo-magnetic quark operator. 
We end with our  conclusions and outlook in
Section~\ref{sect:conclusion}.

A number of technical issues are discussed in the Appendices. In
Appendix~\ref{app:CPtransformation} we summarize our choice of phase
convention used to define the CP transformations. The
regularization-independent calculation is done using off-shell matrix
elements with quarks and gluons as external states in a fixed
gauge. In Appendix~\ref{app:BRST} we derive the constraints on the
mixing with gauge dependent and off-shell operators imposed by BRST
symmetry. The Peccei-Quinn mechanism and its implications for CPV 
operators are discussed in Appendix~\ref{app:axion}.  In
Appendix~\ref{app:projections} we discuss the subtleties that arise in
the isospin symmetry limit. Finally, in Appendix~\ref{app:Matching}, we
summarize the matching  coefficients between the $\MSbar$ and the RI
scheme.

\section{Framework} \label{Sec2}

In this section we describe in some detail   the hadronic-scale  CPV 
effective Lagrangian induced by  BSM physics at the high scale.
The identification of the CPV combinations of short-distance parameters 
involves several steps. 
We start our discussion in Section \ref{sect:2.1} 
by classifying  the leading  BSM-induced operators 
that can lead to CPV effects  at the quark  and gluon level. 
We then discuss in Section~\ref{sect:vacalign} the relation between CP
and chiral symmetry in presence of operators that explicitly break
chiral symmetry: the CP symmetry that remains unbroken by the vacuum
takes the standard form given in Appendix~\ref{app:CPtransformation}
only  after performing an appropriate chiral rotation of the fields (``vacuum alignment'') 
that eliminates pion tadpoles~\cite{Dashen:1970et,Baluni:1978rf,Crewther:1979pi}.
In Section \ref{sect:invariantCPV} we implement the vacuum alignment
in presence of higher-dimensional operators induced by BSM physics and
in Section~\ref{sect:2.4} we summarize the vacuum-aligned effective
Lagrangian including operators up to dimension five.

\subsection{CP violation in the  Standard Model and beyond} 
\label{sect:2.1}

Assuming the existence of new physics beyond the Standard Model (BSM)
at a scale $\Lambda_{\rm BSM} \gg v_{ew}$, we can parameterize the BSM
effects in terms of local operators of dimension five and higher,
suppressed by powers of the scale $\Lambda_{\rm BSM}$.  The new
operators are built out of SM fields and respect the $SU(3)_C \times
SU(2)_W \times U(1)_Y$ gauge symmetries of the SM.  The leading
CP-violating operators appear only at dimension
six~\cite{Buchmuller:1985jz,Grzadkowski:2010es}.  Their
renormalization group evolution from the new physics scale down to the
hadronic scale has been studied in several papers, most recently in
Refs.~\cite{Dekens:2013zca,Hisano:2012cc}, and the resulting effective
chiral Lagrangian at the hadronic level has been discussed in
Refs.~\cite{deVries:2012ab,Bsaisou:2014oka} (for a review see
Ref.~\cite{Engel:2013lsa}).

In this work, we are primarily interested in the structure of the effective Lagrangian including  
new sources of CP violation  below the weak scale. 
After integrating out the top quark, the  Higgs boson, and the $W^\pm$ and $Z$  gauge bosons, 
the needed  operators are invariant under  the $SU(3)_C \times U(1)_{\rm EM}$ gauge group. 
At  a scale $\mu  <  M_{W,Z}$, the effective Lagrangian including the leading (i.e., originating at dimension six) flavor-conserving CP-violating 
effects at the  quark- and  gluon-level can be written as follows:\footnote{%
Without loss of generality, we have performed a $SU(n_F)_L \times SU(n_F)_R$ transformation to put the 
quark mass matrix in diagonal form, with complex masses sharing a common phase $\rho$, namely  $m_i = |m_i| e^{i \rho}$. 
Moreover, note that the masses $m_i$ and  $\theta$ include (i) possible threshold corrections, 
i.e., effects that originate from higher dimensional operators, such as $H^\dagger H G \tilde G$ and  $H^\dagger H  \bar{q}_L  q_R'  H$, 
and (ii) corrections induced by mixing with the chromo-electric dipole moment at finite quark mass.}
\bea
{\cal L}_{\rm eff}  &=& {\cal L}_{\rm SM} \big\vert_{m_i=0} 
-   m_i   \bar{\psi}_{Li}  \, \psi_{Ri}  
-   m_i^*    \bar{\psi}_{Ri}  \, \psi_{Li}  
 - \frac{g^2}{32 \pi^2}  {\theta}  G \tilde G
\nn \\
&-&  \frac{v_{ew}}{2 \Lambda_{\rm BSM}^2} e  \left(  d_i^{(\gamma)}  \,   \bar{\psi}_{Li}  \sigma_{\mu \nu}  F^{\mu \nu} \, \psi _{Ri}  \ + \    d_i^{(\gamma)*}  \,   \bar{\psi}_{Ri}  \sigma_{\mu \nu}  F^{\mu \nu} \, \psi _{Li}  \right) 
\nn \\ 
&-& \frac{v_{ew}}{2 \Lambda_{\rm BSM}^2}    g  \left( d_i^{(g)}  \,   \bar{\psi}_{Li}  \sigma_{\mu \nu}  G^{\mu \nu} \, \psi _{Ri}  \ + \   d_i^{(g)*}  \,   \bar{\psi}_{Ri}  \sigma_{\mu \nu}  G^{\mu \nu} \, \psi _{Li}  \right) 
\nn \\
&+&     \frac{d_G}{\Lambda_{\rm BSM}^2}  \  f^{abc}  G_{\mu \nu}^a  \tilde{G}^{\nu \beta,b} G_\beta^{\mu,c}
\ + \   \hbox{4-quark} \ \ {\rm operators}~, 
\label{eq:Leff0}
\eea
where $e$ and $g$ are the electric and color charges, $v_{ew}$ is
the Higgs VEV (vacuum expectation value), the index $i$ runs over the active quark flavors (at
$\mu \sim 1$~GeV one has $i \in \{ u,d,s \}$), and $ \tilde G^{\mu
  \nu, b} = \varepsilon^{\mu \nu \alpha \beta} G^b_{\alpha \beta}/2$.

The first line in Eq.~\eqref{eq:Leff0}  contains the Standard Model dimension-4 operators, 
including the mass matrix put in the standard diagonal form and a common phase,  and the QCD  $\theta$-term. 
Because of the anomalous Ward identity, a choice of fermion phases can
be used to  rotate the $\theta$-term into  a CP-odd pseudoscalar quark mass term,  instead. 

The second and third lines in Eq.~\eqref{eq:Leff0} contain the BSM
contribution due to the quark magnetic (MDM) and electric dipole
moment (EDM), and chromo-magnetic (CMDM) and chromo-electric dipole
moment (CEDM) operators, respectively.  Below the weak scale these
operators are of mass-dimension five: their origin as dimension-6
operators at the high scale is hidden in the overall dimensionless
factor of $v_{\rm ew}/\Lambda_{\rm BSM}$. It is important to note that
the physical meaning of these operators as CP-violating electric or
CP-conserving magnetic moments relies on an implicit chiral phase
convention.  Similar to the quark mass terms, however, these operators
explicitly break chiral symmetry, thus contributing to vacuum
alignment~\cite{Dashen:1970et}.  As discussed in
Section~\ref{sect:vacalign}, the vacuum alignment, in turn, determines
the unbroken CP symmetry, and a chiral rotation---which mixes the EDM
and MDM, as well as CEDM and CMDM operators---may be needed to put the
symmetry transformation in the standard form. If no complex phases
appear in the Lagrangian after such a rotation, no physical CP
violation can arise.  In Section~\ref{sect:invariantCPV}, we,
therefore, discuss the combinations of $m_i, \theta, d_i^{(\gamma)}$,
and $d_i^{(g)}$ that are independent of such phase choices, and give
CP-violating contributions to observables.

Finally, the fourth  line in Eq.~\eqref{eq:Leff0}  contains the  CP-odd BSM operators 
that are genuinely of mass-dimension six at low energy, 
such as the Weinberg three-gluon operator and four-quark operators.  

In order to convert experimental results on nucleon and nuclear EDMs 
into bounds or ranges for the short-distance CP-odd couplings,  one 
needs to compute the effect of the CP-odd operators in Eq.~\eqref{eq:Leff0}  
on hadronic observables, such as the nucleon EDM and the T-odd $\pi N N$ couplings. 
One essential step in connecting the short-distance physics to
hadronic observables involves defining UV finite operators in a
suitable scheme, whose matrix elements can then be computed
non-perturbatively using lattice QCD.
In this work, we focus on the ultraviolet divergences and mixing
structure of the leading gauge-invariant CP-odd dimension-5 operators,
namely the quark CEDM and EDM.  
%The reason for this is twofold:  first,  
These operators are of great
phenomenological interest, being the leading sources of
flavor-diagonal CP violation in several extensions of the
SM~\cite{Pospelov:2005pr,Engel:2013lsa}.
Moreover,   since dimension-5 operators can mix only with operators of dimension up to five 
(mixing with lower dimensional operators occurs in mass-dependent renormalization schemes),  
we can consistently ignore operators of dimension six and higher, which we leave for future work.

%We consider the mixing of these dimension-5 operators with other
%operators of dimension five and lower. Mixing with operators of
%different dimensions occurs in mass-dependent renormalization schemes.
%We consistently ignore operators of dimension six and higher, which we
%leave for future work.  

\subsection{CP symmetry and chiral symmetry breaking} 
\label{sect:vacalign}

In this subsection we discuss the connection between CP and chiral
symmetries.  The main point is that explicit chiral symmetry breaking
selects the vacuum of the theory~\cite{Dashen:1970et}, as well as the
unbroken CP symmetry.  The unbroken CP symmetry takes the standard
form given in Appendix \ref{app:CPtransformation} only after a chiral
rotation that eliminates pion tadpoles, {\it i.e.}, after implementing
vacuum alignment~\cite{Dashen:1970et} discussed in Section~\ref{sect:invariantCPV}.

The CP transformation interchanges left-chiral particles with right-chiral  anti-particles. 
It is implemented on chiral fermion fields by
\begin{eqnarray}
  {\cal CP}^{-1}\psi_L \, {\cal CP} &=& i  \gamma_2    \bar{\psi_L}^T\,\nonumber\\
  {\cal CP}^{-1}\psi_R \, {\cal CP} &=& i \gamma_2  \bar{\psi_R}^T\,
\end{eqnarray}
where ${\cal CP}$ is the CP operator  (see Appendix~\ref{app:CPtransformation} for details). 
The CP operation does not commute with chiral rotations, so we can consider its outer automorphisms.  
In  fact, defining the chiral rotation operator  \(\hat\chi\)  via ($i$ labels quark flavors)  
\begin{eqnarray}
    {\hat\chi}^{-1}\psi_{L,i}{\hat\chi} &=& e^{-i\chi_i/2}\psi_{L,i}\,\nonumber\\
  {\hat\chi}^{-1}\psi_{R,i} {\hat\chi} &=& e^{i\chi_i/2}\psi_{R,i}\, ~, 
\end{eqnarray}
one finds 
\begin{eqnarray}
      {\cal CP}_\chi^{-1}\psi_{L,i} \, {\cal CP}_\chi &=& i e^{i\chi_i} \gamma_2 \bar{\psi_{L\rlap{$\scriptstyle ,i$}}}^T\,\nonumber\\
  {\cal CP}_\chi^{-1}\psi_{R,i} \, {\cal CP}_\chi &=&  i e^{-i\chi_i}  \gamma_2    \bar{\psi_{R\rlap{$\scriptstyle ,i$}}}^T\,
\end{eqnarray}
where \({\cal CP}_\chi \equiv {\hat\chi}^{-1}{\cal CP}\hat \chi\). 
If chiral symmetry is a good symmetry of the Lagrangian ${\cal L}_0$, then each of these is an equivalent CP symmetry.

Because of the spontaneous breaking of chiral symmetry, almost all the
\({\cal CP}_\chi\)
are spontaneously broken by the vacuum of the theory. In this case, it
is convenient to make a chiral phase choice such that the vacuum has a
zero expectation value for all the flavor  bilinears of the
form \(\langle\bar\psi_i\gamma_5\psi_j\rangle\). 
In fact, it is only with  this phase choice that the pions, the Goldstone
modes of the broken chiral symmetry, correspond to the operator
\(\bar\psi_i\gamma_5\psi_j\).
With this choice of phases, in the ``reference vacuum'', 
the CP symmetry \({\cal CP}_0\) stays unbroken by the vacuum;
we implicitly make this choice throughout this paper.

We next consider the effect of explicit chiral symmetry breaking. 
For a small explicit breaking of chiral symmetry, 
encoded in a new term  $\delta {\cal L}$  in the Lagrangian
${\cal L} = {\cal L}_0+ \varepsilon  \, \delta \cal L$ (with $\varepsilon \ll 1$), 
chiral perturbation theory is expected to be a good guide to understanding the structure of
the theory. But, because of the explicit breaking of the chiral
symmetry, the vacuum is no longer degenerate: the explicit breaking
chooses a direction in chiral space with which the vacuum aligns~\cite{Dashen:1970et}. 
If this does not match the  ``reference vacuum'', large
corrections appear due to degenerate perturbation theory.

To avoid this problem, it is convenient to perform a chiral transformation $\hat \chi$
so that the explicit chiral symmetry breaking  $\delta \cal L$  
selects the reference vacuum,  
in which the unbroken CP symmetry takes the standard form, namely  \({\cal CP}_0\).
The way to do this is to impose  
the condition that the vacuum state does not mix with the Goldstone state~\cite{Dashen:1970et,Baluni:1978rf,Crewther:1979pi}, {\it i.e.,}
\begin{equation}
\langle \pi | \delta {\cal L} | \Omega\rangle = 0\,,
\label{eq:nopion}
\end{equation}
where \(\delta {\cal L}\)
are the chiral breaking terms after such a rotation and
\(|\Omega\rangle\) and \(|\pi\rangle\)
are the reference vacuum and 
Goldstone pion states respectively. If the
only chiral breaking comes from the mass terms, this can be accomplished by 
rotating away the flavor non-singlet CP-violating mass terms  in Eq.~\eqref{eq:Leff0}  by the appropriate
chiral transformation $\hat \chi$.

\subsection{Vacuum alignment in presence of higher-dimensional operators}
\label{sect:invariantCPV}

We now discuss the vacuum alignment in presence of higher-dimensional
operators induced by BSM physics.  After a general discussion of the
chiral transformation needed to enforce Eq.~\eqref{eq:nopion}, we
specialize to the case in which the dominant source of chiral symmetry
breaking is provided by the quark masses, and the dominant BSM
operators are the quark (C)EDM and (C)MDM. In this case we present the
vacuum-aligned effective Lagrangian in both scenarios with and without
the Peccei-Quinn (PQ) mechanism~\cite{Peccei:1977hh}.

Except for \(G\tilde G\), all terms of dimension five and lower in the
Lagrangian defined in Eq.~\eqref{eq:Leff0} that violate CP are fermion
bilinears that also violate chiral symmetry.  Each of these terms
mixes with a CP-conserving one under chiral rotation, and it is
conventional to treat the two as real and imaginary parts of a single
operator.  Generalizing Eq.~\eqref{eq:Leff0} let us write the chiral
and CP-violating part of the Lagrangian involving quark bilinears as
\bea \delta {\cal L} &=& - \sum_{i,\alpha} \big[ d_i^\alpha O_i^\alpha
  + {\rm h.c.} \big] = - \sum_{i,\alpha} \big[ \re d_i^\alpha\ \re
  O_i^\alpha + \im d_i^\alpha \ {\rm Im} \, O_i^\alpha \big]
\label{eq:dL0}  
\eea
where 
\bea
O_i^\alpha &=& \bar{\psi}_{L,i}  \Gamma^\alpha \psi_{R,i}~,  \
\qquad  \re O_i^\alpha  = O_i^\alpha + O_i^{\alpha \dagger} ~, 
\qquad   \im O_i^\alpha = i \left[  O_i^\alpha - O_i^{\alpha \dagger} \right] ~,
\eea
\(i\) is a flavor index and \(\alpha\) parameterizes the
different operators, characterized by the  structure $\Gamma^\alpha$.  
The first few operators are the mass term (\(\alpha=0\)), 
the  quark CEDM ($\alpha=1$) and the  quark EDM ($\alpha=2$)
\bea
d_i^0 O_i^0  + {\rm h.c.}  &=& \bar \psi_i [ (\re m_i) + i (\im m_i) \gamma_5] \psi_i
\\ 
d_i^1 O_i^1  + {\rm h.c.}  &=& 
\frac{v_{ew}}{2 \Lambda_{\rm BSM}^2} \ g \ 
\bar \psi_i \Big[ (\re d_i^{(g)}) \sigma_{\mu \nu} G^{\mu \nu} + i (\im d_i^{(g)}) \sigma_{\mu \nu} G^{\mu \nu} \gamma_5 \Big] \psi_i \\ 
d_i^2 O_i^2  + {\rm h.c.}  &=& 
\frac{v_{ew}}{2 \Lambda_{\rm BSM}^2} \ e  \ 
\bar \psi_i \Big[ (\re d_i^{(\gamma)}) \sigma_{\mu \nu} F^{\mu \nu} + i (\im d_i^{(\gamma)}) \sigma_{\mu \nu} F^{\mu \nu} \gamma_5 \Big] \psi_i  ~.
\eea
In this notation, under a chiral rotation $\hat \chi$  (parameterized by  $\chi_i$) 
\begin{eqnarray}
d_i^\alpha &\to& d_i^\alpha e^{i\chi_i}\nonumber\\
\theta &\to& \theta + \chi_1 + \cdots + \chi_{n_F}   \ ~,
\label{eq:nopioncond}
\end{eqnarray}
and we seek a  chiral rotation   such that 
Eq.~\eqref{eq:nopion} holds and at the same time   $\theta \to  0$.

To implement  Eq.~\eqref{eq:nopion},  we need to introduce the 
non-perturbative matrix elements 
\be
\Delta^\alpha_{ij}  \equiv   \langle \pi_j | \im \ O^\alpha_i | \ \Omega \rangle
\ee
where the state $|\pi_j \rangle$ is interpolated by the field $\bar{\psi}_j i \gamma_5 \psi_j$. 
Then the mixing of the vacuum with the neutral Goldstone modes $(|\pi_j\rangle - |\pi_k\rangle)/\sqrt{2}$ is proportional to 
$\sum_{i \alpha}  \ \im d_i^\alpha  \, (\Delta^\alpha_{ij} - \Delta^\alpha_{ik})$.
The condition in Eq.~\eqref{eq:nopion} for each 
neutral Goldstone mode $(|\pi_j\rangle - |\pi_k\rangle)/\sqrt{2}$ becomes 
\be
\sum_{i,\alpha}  \im \left( d_i^\alpha   e^{i \chi_i} \right)  \ 
 \left[  \Delta^\alpha_{ij} - \Delta^\alpha_{ik} \right]
  = 0~, \qquad \qquad 
 k=1,  j = 2, ..., n_F~.
\label{eq:va1}
\ee
Since the unperturbed Lagrangian ${\cal L}_0$ is $SU(n_F)_V$ symmetric,  
the matrix elements can be written in terms of two constants, 
the diagonal $\Delta^\alpha_{S}$  and the off-diagonal $\Delta^\alpha_{V}$, 
defined by 
$\Delta^\alpha_{ij} =  \Delta_S^\alpha  \delta_{ij}  + \Delta_V^\alpha (1 - \delta_{ij})$. 
% Defining $r^{(\alpha)}  \equiv   (\Delta^\alpha_S - \Delta^\alpha_V)/(\Delta^0_S - \Delta^0_V)$, 
Eq.~\eqref{eq:va1} implies, for each flavor $i=1, ..., n_F$, 
\be
\sum_\alpha  \ \im \left( d_i^\alpha   e^{i \chi_i} \right)  \ r^{(\alpha)}= \kappa  
\ee
where $r^{(\alpha)}  \equiv   (\Delta^\alpha_S - \Delta^\alpha_V)/(\Delta^0_S - \Delta^0_V)$ 
(we  divided out the matrix elements of the 
dimension-3 operator  $\im O_i^0 = \bar{\psi}_i i \gamma_5 \psi_i$) 
and  $\kappa$ is a flavor-independent constant. Defining
\begin{equation}
d_i \equiv |d_i| e^{i\phi_i} \equiv \sum_\alpha  d_i^\alpha \, r^{(\alpha)}, 
\end{equation}
the chiral rotation we want needs to satisfy, for each $i$, 
\begin{equation}
|d_i| \sin (\chi_i+\phi_i)  =  \kappa  ~.
\label{eq:s1}
\end{equation}
Moreover, to implement $\theta \to 0$,  one needs $\theta + \sum_i \chi_i = 0$, or, equivalently, 
the constant $\kappa$  needs to satisfy
\begin{equation}
\theta - \sum_i \phi_i + \sum_i \sin^{-1} (\kappa |d_i|^{-1}) = 0\,.
\label{eq:s2}
\end{equation}
Eqs.~\eqref{eq:s1} and \eqref{eq:s2} provide a system of equations for $\chi_i$ and  $\kappa$, 
which does not have a closed form solution for $n_F>2$. 
On making the  chiral transformation dictated by Eqs.~\eqref{eq:s1} and \eqref{eq:s2}  we find that CP violation is  proportional to
\begin{eqnarray}
\delta {\cal L}_{CPV}  &= & \sum_{i, \alpha}
   \left[  \kappa  \re \frac{d_i^\alpha}{d_i} + \sqrt{|d_i|^2 -
  \kappa^2} \im \frac{d_i^\alpha}{d_i} \right] \im
   O_i^\alpha
   \nonumber\\
&\approx&
 \sum_{i, \alpha}
\left[-  \bar d (\theta - \phi_{\rm tot}) \re
          \frac{d_i^\alpha}{d_i}
        + |d_i| \im \frac{d_i^\alpha}{d_i}\right]
          \im O_i^\alpha\,,
\label{eq:vac-al1}
\end{eqnarray}
where \(\bar d^{-1} \equiv \sum_i |d_i|^{-1}\) and \(\phi_{\rm  tot} \equiv \sum_i \phi_i\), 
and the second line is obtained by solving Eq.~\eqref{eq:s2} for small $\kappa/|d_i|$, 
which is appropriate when $\theta$ is small and the dominant chiral
violation comes from a real mass term (the latter condition implies   $\phi_i \ll 1$).

Notice that if there is a single operator \(O_i^\alpha\)
that is the  only source of CP violation, then
\(d_i \propto d_i^\alpha\),
and the second term is zero.  This is because in this case this term
is also the only term that explicitly breaks the chiral symmetry and
the vacuum aligns itself with this direction.  As a result, performing a
chiral rotation to make the vacuum have the conventional chiral phase
removes any imaginary part from the operator, and CP
violation can only come from the anomalous chiral rotations.  In this case,
however, the CP violation is proportional to the harmonic sum of the
chiral violations from each flavor, and therefore vanishes if any flavor
remains chirally symmetric.

In what follows, we will instead consider the situation where the
dominant chiral breaking is always due to the \(\alpha=0\)
mass term, i.e., \(d_i \propto d_i^0\)  
approximately, and consider the case where all flavors are massive.
Only in this case, the dominant source of CP violation is proportional to \(\im d_i^\alpha\).
Consistent with this assumption, when studying mixing and renormalization we will keep in the operator basis 
terms proportional to the quark mass matrix. 

With these assumptions and after vacuum alignment, the
explicit form of Eq.~\eqref{eq:vac-al1}, specialized to the case of a
Lagrangian containing a mass term, quark EDM, and quark CEDM, is 
\begin{eqnarray}\label{eq:aligned}
\delta \mathcal L_{CPV}  &=& \bar \psi i \gamma_5 \psi 
\, m_{*}   \left( \bar\theta  
-   \frac{r}{2}  \textrm{Tr}\left[ \mathcal M^{-1} \left(\left[d_{CE}\right]  - m_*\bar\theta \mathcal M^{-1} \left[d_{CM} \right]   \right) 
\right] \right)
\nonumber \\ 
&+ &   \frac{r}{2} \bar \psi i \gamma_5   \left( \left[d_{CE} \right] - m_* \bar\theta \mathcal M^{-1} \left[d_{CM} \right] \right)  \psi   
\nn \\
&-& \  \frac{i g}{2}  \  \bar{\psi}   \sigma_{\mu \nu} \gamma_5  \, G^{\mu \nu}    \,\left(  [d_{CE}] -  m_* \bar\theta \mathcal M^{-1} \left[d_{CM} \right] \right)  \,   \psi
\nn \\
&-& \  \frac{i e}{2}  \  \bar{\psi}   \sigma_{\mu \nu} \gamma_5  \, F^{\mu \nu}    \,\left(  [d_{E}] -  m_* \bar\theta \mathcal M^{-1} \left[d_{M} \right] \right)  \,   \psi~, 
\end{eqnarray}
where we defined  $r  \equiv  r^{(1)} =  (\Delta^1_S - \Delta^1_V)/(\Delta^0_S - \Delta^0_V)$  and neglected $r^{(2)} = O(\alpha_{\rm EM}  \,  r^{(1)} )$. 
We further defined 
\be
\psi = 
\left(
\begin{array}{c}
u \\
d \\
s
\end{array}
\right)~, 
\qquad  \qquad 
{\cal M} = 
\left(
\begin{array}{ccc}
m_u  & 0 & 0 \\
0 & m_d & 0 \\
0 & 0 & m_s
\end{array}
\right)~, 
\ee
and the matrix-valued CEDM and CMDM couplings as
\begin{equation}
\left[d_{CE} \right] = \frac{v_{\textrm{ew}}}{\Lambda_{\rm BSM}^2}\left( 
\begin{array}{ccc}
\textrm{Im} \,d_u^{(g)}  & 0 & 0 \\
0 & \textrm{Im} \,d_d^{(g)} & 0 \\
0 & 0 & \textrm{Im} \,d_s^{(g)}
\end{array} \right), \qquad 
\left[d_{CM} \right] = \frac{v_{\textrm{ew}}}{\Lambda_{\rm BSM}^2}\left( 
\begin{array}{ccc}
\textrm{Re} \,d_u^{(g)}  & 0 & 0 \\
0 & \textrm{Re} \,d_d^{(g)} & 0 \\
0 & 0 & \textrm{Re} \,d_s^{(g)}
\end{array} \right) \quad 
, \label{eq:dCdE}
\end{equation}
with analogous definitions for the electric $[d_E]$ and magnetic $[d_M]$ couplings. 
Finally, $\bar\theta = \theta - n_F \rho$ with $n_F \rho$ the phase of
the determinant of the mass matrix before the anomalous chiral
rotation renders it real, and $m_{*}$ is the reduced quark mass
\begin{equation}
m_{*} = \frac{m_s m_d m_u}{m_s (m_u + m_d) + m_u m_d}~.
\end{equation}
The first term in Eq.~\eqref{eq:aligned} is the familiar $\bar\theta$
term, shifted by a correction proportional to the quark CEDM and a
second correction, proportional to the coefficients of the CMDM
multiplied by $\bar\theta$.  The third and fourth lines of
Eq.~\eqref{eq:aligned} contains the quark (C)EDM operators, which
after vacuum alignment receive a correction proportional to the (C)MDM
coefficient multiplied by $\bar\theta$.  Moreover, vacuum alignment
causes the appearance of a complex mass term, proportional to the same
combination of the CEDM and CMDM coefficients (second line of
Eq.~\eqref{eq:aligned}).

The above discussion is valid in absence of PQ
mechanism~\cite{Peccei:1977hh}.  As we review in
Appendix~\ref{app:axion}, if CP violation arises only from the mass
term, the PQ mechanism dynamically relaxes $\bar{\theta}$ to zero.  In
the presence of other CP-violating sources, like the quark CEDM, the
Peccei-Quinn (PQ) mechanism causes $\bar\theta$ to relax to a non-zero
value $\bar\theta_{\textrm{ind}}$, proportional to the new source of
CP violation. In particular, as we discuss in further detail in
Appendix \ref{app:axion}, in the presence of the quark CEDM
\begin{equation}
\bar\theta_{\textrm{ind}} = \frac{r}{2} \textrm{Tr} \left[\mathcal M^{-1} \left[d_{CE} \right] \right],
\end{equation}
thus enforcing a  cancellation between 
the first two terms  in Eq.~\eqref{eq:aligned}.  Since $\bar\theta_{\textrm{ind}}$ is suppressed by two powers of $\Lambda_{\textrm{BSM}}$,
terms proportional to $\bar\theta  [d_{CM}]$ in Eq.~\eqref{eq:aligned} become effectively dimension eight, and can be neglected. 
Thus, if the PQ mechanism is at work, the first line of Eq.~\eqref{eq:aligned} vanishes 
and the terms proportional to $\bar{\theta}$ in the second and third line of Eq.~\eqref{eq:aligned} can be neglected, 
leading to 
\begin{eqnarray}\label{eq:aligned2}
\delta \mathcal L_{CPV}^{PQ}  &=& 
  \frac{r}{2} \  \bar \psi i \gamma_5  \left[ d_{CE} \right]   \psi   
  -\  \frac{i g}{2}  \  \bar{\psi}   \sigma_{\mu \nu} \gamma_5  \, G^{\mu \nu}   \,  [d_{CE}] \,   \psi  
    -\  \frac{ie}{2}  \  \bar{\psi}   \sigma_{\mu \nu} \gamma_5  \, F^{\mu \nu}   \,  [d_{E}] \,   \psi~,  
\end{eqnarray}
with both CEDM and pseudoscalar quark density with flavor structure dictated by $[d_{CE}]$.

Eq.~\eqref{eq:aligned} and Eq.~\eqref{eq:aligned2} provide the vacuum-aligned low-energy Lagrangians, 
in  presence of BSM sources of CP and chiral symmetry violation. 
They are particularly useful within the chiral perturbation theory framework, as they guarantee  the cancellation of 
tadpole diagrams in which Goldstone modes are absorbed by the vacuum. 
This form of the CP-violating perturbation allows one to identify   what non-perturbative matrix elements are needed 
in order to address the impact of a BSM-induced CEDM operator on the nucleon EDM, i.e.\hbox{},  the dependence of $d_n$  on  $[d_{CE}]$. 
Both with and without PQ mechanism  the effective Lagrangian involves  the CEDM operator as well as 
flavor singlet and non-singlet pseudoscalar quark operators.  Moreover, 
at the lowest order, the effect of flavor non-singlet $\bar \psi  i \gamma_5 t^{3,8} \psi$ operators  is  proportional  to 
insertions of  the flavor-singlet   density  $\bar \psi  i \gamma_5  \psi$. 
This is very simple to see within the functional integral approach, in which $\bar \psi  i \gamma_5 t^{3,8} \psi$ can be eliminated 
through a non-anomalous axial rotation.  
The same result can be obtained within an operator approach.  In this framework, using soft-pion 
techniques, one can show that  a cancellation 
occurs between non-tadpole and tadpole diagrams with insertion of $\bar \psi  i \gamma_5 t^{3,8} \psi$, 
leaving a term proportional to the insertion of $\bar \psi  i \gamma_5  \psi$. 
In absence of PQ mechanism,  the  resulting  flavor-singlet pseudoscalar insertion proportional to $[d_{CE}]$ cancels exactly the existing 
singlet term in Eq.~\eqref{eq:aligned}.   If the PQ mechanism is operative, the resulting flavor-singlet pseudoscalar  insertion is proportional to 
$m_* \bar \theta_{\rm ind}$. The net effect is equivalent to  replacing Eqs.~\eqref{eq:aligned} and \eqref{eq:aligned2}  with 
\begin{subequations}
\label{eq:aligned3}
\begin{eqnarray}
\delta \mathcal L_{CPV}  &=&  m_* \ \bar \theta \   \bar \psi i \gamma_5 \psi \   
- \  \frac{i g}{2}  \  \bar{\psi}   \sigma_{\mu \nu} \gamma_5  \, G^{\mu \nu}    \,\left(  [d_{CE}] -  m_* \bar\theta \mathcal M^{-1} \left[d_{CM} \right] \right)  \,   \psi 
\nn \\
&  & \qquad \qquad  \quad  \ \ - \  \frac{ie}{2}  \  \bar{\psi}   \sigma_{\mu \nu} \gamma_5  \, F^{\mu \nu}    \,\left(  [d_{E}] -  m_* \bar\theta \mathcal M^{-1} \left[d_{M} \right] \right)  \,   \psi 
\\
\delta \mathcal L_{CPV}^{PQ}  &=&  m_* \ \bar \theta_{\rm ind}   \   \bar \psi i \gamma_5 \psi  \    -\  \frac{i g}{2}  \  \bar{\psi}   \sigma_{\mu \nu} \gamma_5  \, G^{\mu \nu}   \,  [d_{CE}] \,   \psi  
 -\  \frac{ie}{2}  \  \bar{\psi}   \sigma_{\mu \nu} \gamma_5  \, F^{\mu \nu}   \,  [d_{E}] \,   \psi~.  
\end{eqnarray}
\end{subequations}
These can be regarded as partially aligned effective Lagrangians, in which only the dominant mass term has been 
aligned to eliminate pion tadpoles, while  the BSM perturbation is not  aligned.  
While the physics cannot depend on the choice of equivalent parameterization   Eq.~\eqref{eq:aligned},  Eq.~\eqref{eq:aligned2}  and  Eq.~\eqref{eq:aligned3}, 
use of different effective Lagrangians is a matter of convenience, depending on the non-perturbative approach employed to 
study  hadronic physics. 
Starting from the Lagrangian in Eq.~\eqref{eq:aligned3}, in the chiral effective theory approach   tadpole diagrams arise, that can be dealt within perturbation theory~\cite{deVries:2012ab}.
On the other hand,  in a non-perturbative approach based on the functional integral, such as lattice QCD,  
the partially aligned Lagrangian can be more convenient:  it shows that the only needed 
non-perturbative matrix elements involve the (C)EDM operator and the  singlet 
pseudoscalar density (or equivalently $G \tilde G$).

\subsection{CP-violating effective Lagrangian at the hadronic scale}
\label{sect:2.4}

To summarize the above discussion,  at  the hadronic scale ($\mu \sim 1$~GeV)  the  vacuum-aligned flavor-conserving  effective Lagrangian 
including the leading BSM sources of CP violation  (up to dimension five)
% up to dimension five 
can be written as follows, 
\bea
{\cal L} &=& {\cal L}_{QCD+QED}  -   \bar{\psi}  {\cal M} \psi     
-   \bar{\psi}   \left[ \delta {\cal M}  \right]  i \gamma_5  \psi    
\nn \\
&-& \frac{i e}{2}  \,   \bar{\psi} \sigma_{\mu \nu} \gamma_5  F^{\mu \nu} \,   [D_E]  \, Q\, \psi 
\  - \  \frac{i g}{2}  \,   \bar{\psi}   \sigma_{\mu \nu} \gamma_5  \, G^{\mu \nu}    \, [D_{CE}]  \,  \psi   ~, 
\label{eq:Leff1}
\eea
where 
\be
Q = 
\left(
\begin{array}{ccc}
q_u  & 0 & 0 \\
0 & q_d & 0 \\
0 & 0 & q_s
\end{array}
\right)~.  
\label{eq:mQ}
\ee
Here we are neglecting 
operators that are total derivatives and/or vanish by 
using the equations of motion (EOM), needed later on  when we impose off-shell renormalization conditions at finite momentum insertion.   
The  matrix-valued CP-violating couplings  $[\delta {\cal M}]$,  $[D_{CE}]$, $[D_E]$  
are related to the short-distance couplings of Eq.~\eqref{eq:Leff0}  via Eq.~\eqref{eq:dCdE} and  Eq.~\eqref{eq:aligned} or  Eq.~\eqref{eq:aligned2}, 
depending on whether or not  the PQ mechanism is assumed. 
The pseudoscalar mass term $[\delta {\cal M}]$ in general  has a  non-singlet structure in flavor space, 
though at leading order, its physical effects can be related to a flavor-singlet mass term as discussed in Section~\ref{sect:invariantCPV} 
(see Eqs.~(\ref{eq:aligned3})).

\section{CP-odd operators of dimension \texorpdfstring{$\leq 5$}{\textleq 5}}
\label{sect:3}

The only T-odd and P-odd operators of dimension five appearing in the low-energy 
effective Lagrangian Eq.~\eqref{eq:Leff1} are  the quark EDM and CEDM, 
whose mixing and renormalization we wish to discuss.

The analysis of the quark EDM is  relatively simple: 
this operator is a quark bilinear from the point of view of strong interactions, 
and it is simply related to the tensor density.  Knowledge of the nucleon  tensor charges immediately 
allows one to extract the contribution of the quark EDM to the 
nucleon EDM~\cite{Ellis:1996dg,Bhattacharya:2015esa}.  
To lowest (zeroth) order in  electroweak interactions, 
this operator renormalizes diagonally, precisely as the tensor density. 
Since we are not interested in  the hadronic matrix elements to a precision of order $\alpha_{EM}/\pi < 1\%$, 
we neglect the quark EDM mixing with any other operator.  

On the other hand, the quark CEDM operator does not renormalize diagonally: 
it mixes with the quark EDM and other operators of dimension five or lower. 
The mixing structure is particularly rich if one considers  
renormalization within a so-called  regularization-independent (RI), momentum subtraction  (MOM) scheme, 
amenable to non-perturbative calculations in lattice QCD~\cite{Martinelli:1994ty}.
In this family of schemes the renormalization conditions are imposed on off-shell quark matrix elements in a fixed gauge, 
thus requiring the inclusion of operators that do not contribute to 
physical matrix elements, such as total derivatives and operators that vanish on-shell by 
using the equations of motion (EOM). 
We next discuss the relevant operator basis,  the mixing structure,  
and the strategy to determine the renormalization matrix. 

\subsection{Operator basis}
\label{sect:basis}

The implementation of  RI momentum subtraction schemes requires working in a fixed gauge.   
With gauge fixing, full gauge invariance is lost and  
the action is  only invariant under 
BRST  transformations~\cite{Becchi:1975nq,Tyutin:1975qk}.
A given gauge invariant operator $O$ (we have in mind the quark CEDM)
mixes under renormalization with two classes of operators of the same (or
lower) dimension~\cite{Deans:1978wn,Collins}: (i) gauge-invariant and
ghost-free operators with the same symmetry properties as $O$
(Lorentz, CP, P) that do not vanish by the EOM; (ii) ``nuisance''
operators allowed by the solution to the Ward Identities associated
with the BRST symmetry: these vanish by the EOM and need not be gauge
invariant.  The ``nuisance'' operators can be constructed as off-shell
BRST variation of operators that have ghost number $-1$, but otherwise
with same symmetry properties as $O$, as discussed in
Ref.~\cite{Deans:1978wn} and detailed in
Appendix~\ref{app:BRST}\rlap.\footnote{See Ref.~\cite{Hill:1979gj} for
  an application of this formalism to the CP-even sector of QCD.
  There is a one-to-one correspondence between our operator basis and
  the one of Ref.~\cite{Hill:1979gj}, provided we drop the
  total-derivative operators from our basis and set $m=0$, as done in
  Ref.~\cite{Hill:1979gj}.}

Following the above general prescription, we have constructed the basis 
of  CP-odd  (T-  and P-odd) operators that mix with the quark CEDM operator 
(the CP  transformation properties of fields are reviewed in Appendix~\ref{app:CPtransformation}). 
We present our results for $n_F = 3$. To restrict the possible structures in flavor space 
we use the  spurion method. 
While the effective Lagrangian in Eq.~\eqref{eq:Leff1} is not invariant under 
chiral transformations on the quark fields $\psi_{L,R}  \to  U_{L,R}  \psi_{L,R}$ 
with $U_{L,R} \in SU(3)_{L,R}$,  
one can formally recover chiral invariance by assigning spurion transformation properties 
to the CEDM coupling matrix  ($[D_{CE}] \to    U_L [D_{CE}] U_R^\dagger$), 
the mass matrix (${\cal M} \to    U_L{\cal M} U_R^\dagger$), and the charge matrix 
($Q \to    U_{L,R} QU_{L,R}^\dagger$).
One then includes in the basis operators that are chirally invariant in the 
spurion sense,  and are linear in the CEDM spurion  $[D_{CE}]$. 
Eventually, we set  $[D_{CE}] \to t^a$ ($a=0,3,8$), where  $t^0 = 1/\sqrt{6} I_{3 \times 3}$ 
is proportional to  the identity matrix in flavor space, 
while for $a = 3,8$,   $t^{a} = \lambda^a/2$, with $\lambda^a$ the SU(3) Gell-Mann matrices 
(normalizations are such that ${\rm Tr}_F  (t^a t^a) = 1/2$ for $a=0,3,8$).  

In our basis we include operators proportional to the quark mass matrix for two reasons: 
(i)  the identification of the CPV terms in  Eq.~\eqref{eq:Leff1}  assumed the quark mass 
to be the dominant source of explicit chiral symmetry breaking;  
(ii) we wish to include the effect of the strange quark, for which $m_s/\Lambda_{QCD}$ 
is not a big  suppression parameter. 

Finally, in order to present the operators that vanish by the EOM in a compact form,  
we introduce the combinations:
\bea
\psi_E & \equiv  & (i D^\mu \gamma_\mu -  {\cal M}) \psi~, \qquad \qquad \quad  \
 D_\mu = \partial_\mu - i g A_\mu^a T^a  - i e Q  A_\mu^{(\gamma)} \label{eq:defs1}
\\
\bar{\psi}_E  &\equiv & -  \bar{\psi}   \, (  i   \overleftarrow {D}^\mu \gamma_\mu  + {\cal M}) ~, \qquad \qquad 
\overleftarrow {D}_\mu =   \overleftarrow{\partial}_\mu + i g A_\mu^a T^a +   i e Q  A_\mu^{(\gamma)}   ~.
\label{eq:defs2}
\eea
Note that $\psi_E$ transforms under CP in the same way as $\psi$ (see Appendix~\ref{app:CPtransformation}).   

Next, we enumerate the operators of dimension five and lower 
that can mix with the quark CEDM: 
\be
C  = ig\,  \bar\psi t^a  \sigma^{\mu\nu} \gamma_5 G_{\mu\nu} \psi~,  
\label{eq:Cdef}
\ee
labeled by the flavor-diagonal structure $t^a$ ($a=0,3,8$). 
We will use the notation $O_i^{(d)}$ to indicate the $i^{\rm th}$ operator of dimension $d$. 
%%%%%%%%%%%%%%%%%%%%%%%%%%%%%%%%%%%%%%%%%%%%%%%%
If the regularization breaks chiral symmetry, {\it i.e.} an additional left-right spurion 
(proportional to the identity in the case of Wilson fermions) is present in the  effective Lagrangian,  
the CEDM operator can mix with additional operators. 
While we will restrict our analysis  to the case of good chiral symmetry 
(which can be attained on the lattice by using domain-wall~\cite{Kaplan:1992bt} or overlap~\cite{Narayanan:1993sk} fermions), 
we will  nonetheless identify the additional operators appearing at a given dimensionality. 
Finally, note that there are no CP-odd operators  containing ghost-antighost fields up to and including dimension five. 
%%%%%%%%%%%%%%%%%%%%%%%%%%%%%%%%%%%%%%%%%%%%%%%%

\subsubsection{Dimension 3}
\label{sect:dim3}
At dimension three there is only one operator allowed by the symmetries: 
\bea 
O^{(3)} &\equiv&    P =   \bar\psi i\gamma_5  t^a \psi  ~.
\eea 
This operator mixes with the quark CEDM even in the absence of 
other sources of chiral symmetry breaking, such as mass terms or regularization artifacts. 
Therefore,  the lattice operator $C_L$ requires subtraction of power divergences  due to mixing with the lower dimensional 
operator $P_L$\rlap.\footnote{Since  dimensionally regularized operators do not mix with lower dimensional operators at any finite order in perturbation theory, we will, when necessary, use a subscript $L$ for operators regularized in a scheme, like the lattice, that includes
a hard cutoff.}
Defining the subtracted operator  $C \equiv  C _L- \tilde Z P_L$, 
one can determine $\tilde{Z}$  by requiring   that the quark two-point function 
vanishes at a given symmetric kinematic point $p^2 = p'^2 = q^2 = - \Lambda_0^2$ for $m_q \to 0$, 
namely  ${\rm Tr}   \left(    \Gamma^{(2)}_{C}   \gamma_5 t^a   \right)_{\Lambda_0}= 0$.

\subsubsection{Dimension 4} 

Assuming good chiral symmetry,  there are no dimension-4 operators that 
mix with the quark CEDM operator.  
If the regularization breaks chiral symmetry in a flavor blind fashion,  
the CEDM can mix with the following operators: 
\be
G \tilde{G}~, \qquad 
\partial_\mu  ( \bar\psi\gamma^\mu\gamma_5 t^a \psi) , \qquad 
 \bar\psi  i\gamma_5    \left\{ \mathcal M,  t^a\right\} \psi  , \qquad 
  \textrm{Tr}\left[\mathcal M t^a\right]  \bar\psi  i\gamma_5  \psi ,  \qquad
  \textrm{Tr}  \left[{\cal M}\right]       \,      \bar\psi i\gamma_5  t^a \psi . 
\ee

\subsubsection{Dimension 5}  
\label{sect:dim5basis}

At dimension five, fourteen Hermitean  operators are present. 
The first ten operators are   gauge-invariant and do not vanish by the EOM. 
The latter four are ``nuisance'' operators. 
To all operators we assign  a number and also a more suggestive name: 
\bea
O^{(5)}_1 &\equiv&  C = ig\,  \bar\psi\tilde\sigma^{\mu\nu}G_{\mu\nu}  t^a\psi 
 \qquad \qquad \qquad  \tilde{\sigma}^{\mu \nu} \equiv \frac{1}{2} \left( \sigma^{\mu \nu} \gamma_5
 + \gamma_5 \sigma^{\mu \nu} \right)  
 \label{eq:CEDMdef} \\
O^{(5)}_2 &\equiv&   \partial^2 P =   \partial^2  \left(   \bar\psi i\gamma_5  t^a \psi \right) 
\\
 O^{(5)}_3 &\equiv&  E  = \frac{i e}{2} \,  \bar\psi\tilde\sigma^{\mu\nu}F_{\mu\nu} \{Q, t^a\} \psi 
\\
O^{(5)}_4  &\equiv&    (m \,  F \tilde{F})  =   
 \textrm{Tr} \left[ \mathcal M Q^2  t^a \right] \,
  \frac{1}{2} \epsilon^{\mu \nu \alpha \beta}  F_{\mu \nu}  F_{\alpha \beta}
\\
O^{(5)}_5  &\equiv &    (m \,  G \tilde{G}) = 
 \textrm{Tr} \left[ \mathcal M  t^a \right] \,
  \frac{1}{2} \epsilon^{\mu \nu \alpha \beta}  G^b_{\mu \nu}  G^b_{\alpha \beta}
\\
O^{(5)}_6  &\equiv&   (m  \, \partial \cdot A)_1 = 
\textrm{Tr}\left[\mathcal M t^a\right]  \partial_\mu  \left( \bar\psi\gamma^\mu\gamma_5   \psi \right)
\\
O^{(5)}_7  &\equiv&   (m  \, \partial \cdot A)_2 = 
\frac{1}{2} \partial_\mu  \left(  \bar\psi\gamma^\mu\gamma_5   \left\{  \mathcal M, t^a \right\} \psi \right) 
- \frac{1}{3} \textrm{Tr}\left[\mathcal M t^a\right]  \partial_\mu  \left( \bar\psi\gamma^\mu\gamma_5   \psi \right)
\\
O^{(5)}_8   &\equiv &    (m^2  P)_1 = 
\frac{1}{2}  \, \bar\psi i\gamma_5    \left\{ \mathcal M^2, t^a \right\}   \psi
\\
O^{(5)}_9   &\equiv &    (m^2  P)_2 = 
\textrm{Tr} \left[\mathcal M^2 \right]  \ \bar\psi i\gamma_5 t^a  \psi
\\
O^{(5)}_{10}   &\equiv &    (m^2 P)_3 = 
\textrm{Tr} \left[\mathcal M t^a \right]  \ \bar\psi i\gamma_5  \mathcal M  \psi
\\
%%%%%%%%%%%%%%%%%%%%%%%%%%%%%%%%%%%%%%%%%%%%%%%%%
O^{(5)}_{11} &\equiv&    P_{EE}  =  i\bar\psi_E\gamma_5 t^a \psi_E 
\\
O^{(5)}_{12} &\equiv&    \partial \cdot A_E  
=  \partial_\mu[\bar\psi_E\gamma^\mu\gamma_5 t^a  \psi+\bar\psi\gamma^\mu\gamma_5 t^a  \psi_E]
\\ 
O^{(5)}_{13}  &\equiv&  A_\partial  = 
  \bar\psi  \gamma_5  \slashed{\partial} t^a   \psi_E   \ -   \bar {\psi}_E   \overleftarrow{\slashed{\partial}}  \gamma_5   t^a  \psi  
  \\ 
O^{(5)}_{14} &\equiv &   A_{A^{(\gamma)}}  =
\frac{ i  e}{2} \left( \bar\psi    \{ Q, t^a \}  \Aslash^{(\gamma)}  \gamma_5\psi_E - \bar\psi_E  \{ Q, t^a \}  \Aslash^{(\gamma)}  \gamma_5\psi \right)~.
\eea

With a flavor blind breaking of chiral symmetry,   the CEDM can mix with additional dimension-5 operators, namely:
\bea
&{\rm Tr} \left[ {\cal M}  \right]    \partial_\mu  ( \bar\psi\gamma^\mu\gamma_5 t^a \psi), 
\qquad 
\textrm{Tr}\left[\mathcal M \right]   \bar\psi  i\gamma_5     \mathcal M  t^a  \psi , 
\qquad  
\left( {\rm Tr}{ \cal M} \right)^2   \  \bar\psi i\gamma_5  t^a \psi , &
\nn\\
&\textrm{Tr}\left[\mathcal M^2  t^a\right]   \bar\psi i\gamma_5    \psi ,  \qquad  
\textrm{Tr}\left[\mathcal M \right]   \textrm{Tr}\left[\mathcal M  t^a\right]    \bar\psi i\gamma_5    \psi ~. &
\eea

In the perturbative analysis presented below, we will use dimensional regularization. 
For $\gamma_5$ we will present results for both the na\"{\i}ve  anti-commuting  scheme  known 
as na\"{\i}ve dimensional regularization (NDR)  and the consistent 
't Hooft-Veltman (HV) scheme (see~\cite{Collins} and references therein). 
It is important that the regulator does not break the hermiticity of the operator basis: 
when considering operator insertions in the dimensionally regulated theory,  
care must be taken to  ensure that the operators remain Hermitean
for  arbitrary space-time dimension $d$. This is essential in order to obtain correct results for the finite 
parts of the diagrams.  In what follows we will need to insert  $O_1^{(5)}$ in loop diagrams, so we provide 
in Eq.~\eqref{eq:CEDMdef}  the explicit Hermitean  form of  $O_1^{(5)}$, valid both in HV and NDR schemes.

\subsection{Mixing structure and Renormalization Conventions} 
\label{sec:RCmixing}

The relation between renormalized operators ($O_i$) in any given scheme 
and bare operators   ($O^{(0)}_i$)    (expressed in terms of the bare fields)  can be written as: 
\be
O_i^{(0)}     \ =  \  Z_{ij} \  O_j ~.
\label{eq:ren1}
\ee 
%
%\be
%O_i^{(5)}  \big\vert_{\rm bare}    \ =  \  Z_{ij} \  O_j^{(5)} \big\vert_{\rm ren} ~.
%\label{eq:ren1}
%\ee 
%
The renormalization mixing matrix $Z_{ij}$ is 
scheme-dependent and has the general structure given in
Table~\ref{tab:mixing} \footnote{Working to first order in insertions of the new physics operator, 
each sector labeled by the diagonal flavor structure $t^a$ ($a=0,3,8$) renormalizes independently,
so that the renormalization matrix has a block-diagonal form in flavor space.}.
This structure is dictated by several considerations, including (i) power-counting (some operators are
effectively of dimension three and four with either factors of masses
or external derivatives and cannot mix with genuinely dimension-5
operators); (ii) BRST invariance~\cite{Deans:1978wn}; (iii) vanishing
by EOM or at zero four-momentum injection.  Indicating the
gauge-invariant operators that do not vanish on using the EOM
($O^{(5)}_i$ for $i=1,..., 10$) by $O_\alpha$ and the ``nuisance''
operators ($O^{(5)}_i$ for $i=11,..., 14$) by $N_\alpha$, the
renormalization matrix has the block-structure 
\be
\left(\begin{array}{c} O^{(0)} \\ N^{(0)} 
\end{array}\right)
%\end{array}\right)_{\rm ren}
= 
\left(\begin{array}{cc}
Z_O  &  Z_{ON}\\
0 & Z_N 
\end{array}\right)
\, 
\left(\begin{array}{c}
O \\ N
\end{array}\right)~~.
%\end{array}\right)_{\rm bare}~~.
\label{eq:Zstructure}
\ee
The divergent part of $Z_O$ (proportional to $1/(d-4)$ in dimensional regularization or 
$\log \Lambda^2$ in a cutoff theory), controlling the physical  anomalous dimension, 
is  independent of the gauge-fixing  choice~\cite{Deans:1978wn}. 

In  the following  we will provide $Z_O$ in the $\overline{\rm MS}$ scheme and in 
a momentum subtraction scheme to one loop order. 
We will perform the calculations in dimensional regularization ($d=4-2\epsilon$) 
and will  present results in both the HV and NDR schemes~\cite{Collins} for $\gamma_5$ and the $\gamma$-matrix algebra
(we use the definition of  $\gamma_5$ and $\epsilon^{\mu \nu \alpha \beta}$  given in Ref.~\cite{Peskin:1995ev}).  
To extract the operator renormalization matrix we define the 
field, coupling, and mass  renormalization constants as 
\begin{subequations}
\begin{align}
\psi^{(0)}   =\ &  \sqrt{Z_\psi}  \, \psi 
\\
A_\mu ^{(0)}  =\ &  \sqrt{Z_G}  \, A_\mu 
\\
g^{(0)}  =\ & Z_g \, g  \, \mu_{\overline{\rm MS}}^\epsilon \qquad \qquad 
\mu_{\overline{\rm MS}} \equiv \mu \ \frac{e^{\gamma_E/2}}{(4 \pi)^{1/2}}~
%\bar{\mu} \equiv \mu \, \frac{e^{\gamma_E/2}}{(4 \pi)^{1/2}}~
\\
m^{(0)}  =\ &  Z_m \, m~.
\end{align}
\label{eq:ren3}
\end{subequations}
Here, as usual,  $\mu$ denotes an arbitrary parameter with dimensions of mass,  introduced 
to keep the renormalized coupling $g$ dimensionless ($[g]=0$), while $[m]=1$, $[\psi] = 3/2 - \epsilon$, and $[A_\mu] = 1- \epsilon$. 
Note that $g$  and $\alpha_s \equiv g^2/(4 \pi)$   depend on both $\mu$ and $\epsilon$, so that 
$d \alpha_s/ d (\log \mu) = - 2 \epsilon \alpha_s + O(\alpha_s^2)$.  

Finally, let us discuss different conventions for the renormalization factors 
for  fields, couplings, masses, and  operators, generically denoted by  $Z$.
Our definitions in Eqs.~(\ref{eq:ren1}) and (\ref{eq:ren3}) follow the notation typically used in 
the perturbative QCD literature  (see for example \cite{Buras:1998raa}).
However, we warn the reader that the lattice community typically uses a different convention 
(fleshed out explicitly in Ref.~\cite{Sturm:2009kb}), which is related to the one followed here 
%by  the replacement $Z \to Z^{-1}$.  
by   replacing everywhere  $Z \to Z^{-1}$.

%\psi^{(0)}  & = &  \sqrt{Z_\psi}  \, \psi 
%\\
%A_\mu ^{(0)}  & = &  \sqrt{Z_A}  \, A_\mu 
%\\
%g^{(0)} &=& Z_g \, g  \, \mu_{\overline{\rm MS}}^\epsilon \qquad \qquad 
%\mu_{\overline{\rm MS}} \equiv \mu \ \frac{e^{\gamma_E/2}}{(4 \pi)^{1/2}}~ \\
%m^{(0)} &=&  Z_m \, m~.

%%%%%%%%%%%%%%%%%%%%%%%%%%%%%%%%%%%%%%%%%%%%%%%%
\begin{table}[htb]
\begin{center}
\setlength{\tabcolsep}{3pt}
\begin{tabular}{|c||c|c|c|c|c|c|c|c|c|c|||c|c|c|c|}
\hline
  			&	&	   		&	& 				& 			  	& 	  			& 	&			&		&  		&		 &   	 &	      	&			\\[-1\jot]
    			&  \scriptsize{$C$}	& \scriptsize{$\partial^2 P$}  	&  \scriptsize{$E$} 	& \scriptsize{$m F \tilde F$} 		& \scriptsize{$m G \tilde G$} 		& \scriptsize{$(m \partial \cdot A)_1$}  	& \scriptsize{$(m \partial \cdot A)_2$}  	&\scriptsize{ $(m^2  P)_1$}  	& \scriptsize{$(m^2 P)_2 $} &  \scriptsize{$(m^2 P)_3 $}  
 &\scriptsize{$P_{EE}$} 	 & 	 \scriptsize{$\partial \cdot A_E$}  & \scriptsize{$ A_\partial$} & \scriptsize{$A_{A^{(\gamma)}}$} 	 
\\ [1\jot]
\hline
\hline
%\\[1\jot]
\scriptsize{  $C$}	& x & x		 	& x	& x		 	& x			  	& x				& x 				& x	& x 	& x	& x		& x	 &  x & x
\\
%\\[1\jot]
\hline
\scriptsize{ $\partial^2 P$}		&  	& x 		& 	&  				& 			  	& 	 			& 		&		&  		&   	 	&   		&    &    & 
\\
%[1\jot]
\hline
\scriptsize{  $E$}			&  	&  			& x	& 			 	& 			  	& 		  		& 		&		&  		&   	 	&    	&  &  &  
\\
%[1\jot]
\hline
\scriptsize{  $m F \tilde{F}$}	&  	&  			& 	& x	& 			  	& 		  		& 	&			&  		&   	 	&   		&   &  &  
\\%[1\jot]
\hline
\scriptsize{$m\, G \tilde{G}$}	&  	&  			& 	&  				& x	& x		  		& 	&			&  		&    	&   		& &  & 
\\%[1\jot]
\hline	
\scriptsize{ $(m \partial \cdot A)_1$}  &  	&  			& 	&  				& 			  	& x	& 	&			&  	  	&   	 	&   		& &  & 
 \\%[1\jot]
 \hline
 \scriptsize{$(m  \partial \cdot A)_2$}  &  	&  			& 	&  				& 			  	& 	& x	&			&  	  	&   	 	&   		& &  & 
 \\%[1\jot]
 \hline
\scriptsize{$ (m^2 P)_1$}		&  	&  			& 	&  				& 			  	& 		  		& 			&  x &		&         	&   		& &  & 
	 \\%[1\jot]
\hline
\scriptsize{ $ (m^2 \hat P)_2$}	&  	&  			& 	&  				& 			  	& 		  	& 			&   		& x  &        	&   		& &  & 
	 \\%[1\jot]
	 \hline
\scriptsize{ $ (m^2 \hat P)_3$}	&  	&  			& 	&  				& 			  	& 		  	& 			&   		&     & x   	&   		& &  & 
	 \\%[1\jot]
\hline
\hline
\hline
 \scriptsize{$P_{EE}$}		&  	& 		 	& 	& 			 	& 			  	& 		  		& 		&		&  		&   	  	&  x 		& x& x & 
\\%[1\jot]
\hline
\scriptsize{ $\partial \cdot A_E$}	&  	&  			& 	& 	 			& 			  	& 		  		& 		&		&  		&   	      	&   		& x&  & 
\\%[1\jot]
\hline
\scriptsize{$A_{\partial}$}	&  	&  			& 	&  				& 		  		& 		  		& 	    &			&  		&   	      	&  x		&  x& x &  x
\\%[1\jot]
\hline
\scriptsize{$A_{A^{(\gamma)}}$}	&  	&  			& 	&  				& 			  	& 		  		& 		&	 	&  		&   	      	&   		&  &  & x
 \\%[1\jot]
\hline
\hline	 
\end{tabular}
\end{center}
\caption{Mixing structure of the dimension-5 operators, with ``x'' representing  non-zero entries.  
Throughout, we neglect effects proportional to the electroweak coupling $\alpha_{EW}$. \label{tab:mixing}}
\end{table}
%%%%%%%%%%%%%%%%%%%%%%%%%%%%%%%%%%%%%%%%%%%%%%%%

\section{Green's function calculations} 
\label{sect:4}

\begin{figure}[!t]
\centering\includegraphics[width=0.65\textwidth]{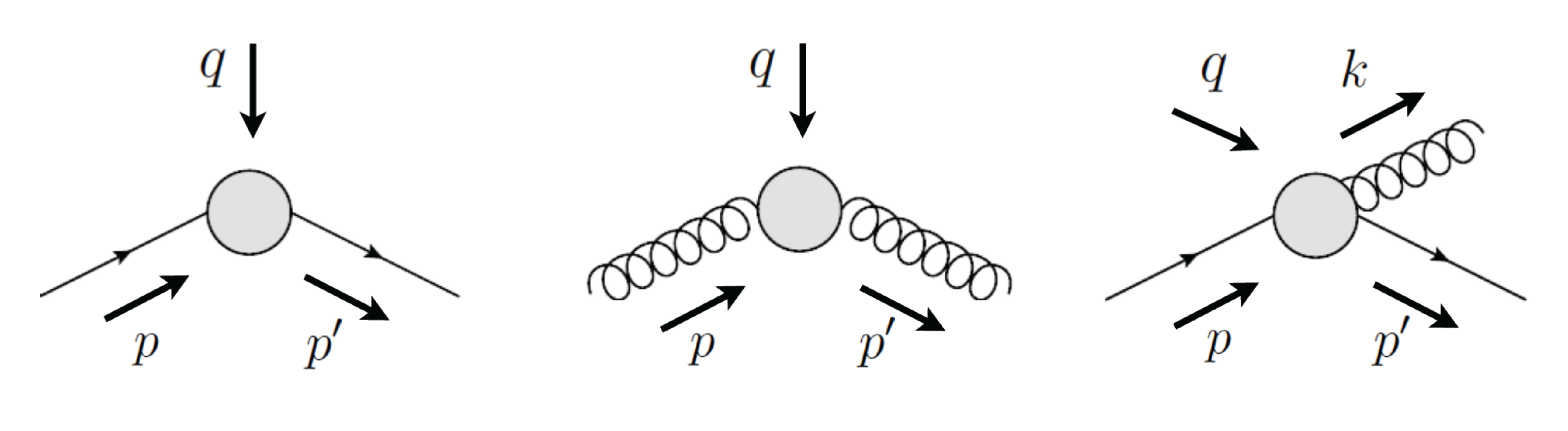}
\caption{Momentum flow of generic diagrams contributing to the
  quark-quark, gluon-gluon, and quark-quark-gluon Green's functions
  with operator insertion.  The shaded blob represents the operator
  insertion with incoming 4-momentum $q$ and higher order
  corrections. In the four-point function, the gluon (photon) momentum is labeled by $k$. }
\label{fig:fig0}
\end{figure}

In order to determine $Z_{ij}$ and the relation between $\overline{\rm MS}$ and the 
RI-$\tilde{\rm S}$MOM scheme to be defined in Section~\ref{sect:MStoRISMOM} below,  
we will study amputated two- and three-point functions\footnote{Since the terminology of
lattice simulations also counts the points at which the operator is inserted, these correspond to
three- and four- point functions in that terminology.} with operator insertion. These 
are shown in Fig.~\ref{fig:fig0} and defined as follows:
\bea
  \int  d^4x  \  e^{ - i q \cdot x}   \  \langle   g (p', \epsilon^{*'})  |   \,  O(x)  \,   |   g(p, \epsilon)  \rangle 
 &=&  (2 \pi)^4  \  \delta^{(4)}  (q + p - p') \  \, 
\epsilon_\mu^{*'} (p')  \ \Gamma_O^{\mu \nu} (p,p') \ \epsilon_\nu (p)
\label{eq:ggO}
\\ 
\int  d^4x  \  e^{- i q \cdot x}   \  \langle q (p')  |   \,  O(x)  \,   |   q  (p)  \rangle       &=&  (2 \pi)^4  \ \delta^{(4)}  (q + p - p') \ \,   \bar{u} (p')    \Gamma_O^{(2)} (p,p')   u (p)  
\label{eq:qqO}
\\
\int  d^4x  \,  e^{ - i q \cdot x}   \  \langle q (p'),   g (k,\epsilon^*)   |   \,  O(x)  \,   |   q(p)  \rangle       &=&  (2 \pi)^4  \  \delta^{(4)}  (q + p - p' - k) \  \, \bar{u} (p')    \Gamma_O^{(3)}(p,p',k)  u (p). \ \ \ \ \ 
\label{eq:gggO}
\eea
To minimize notational clutter in the above equations and throughout
the paper we will suppress the color indices, which can be restored as
follows.
The gluon two-point function $\Gamma^{\mu \nu}_O$ carries the color structure $\delta^{c c'}$,
where $c,c'$ are the  octet color indices labeling the two amputated gluon external legs.
The quark two-point function $\Gamma^{(2)}_O$ carries the  color structure $\delta_{ij}$,
where $i,j$ are the color indices labeling the two amputated quark external legs. 
The quark-quark-gluon three-point function $\Gamma^{(3)}_O$ carries the color structure $T^b_{ij}$,
where $b$ is the octet color index labeling the amputated gluon external leg and  
$i,j$ are the color indices labeling the amputated quark external legs. 
Moreover, in our  notation  $\Gamma^{(3)}_O$ is linear in the gluon polarization vector, {\it i.e.}
 $\Gamma^{(3)}_O =  \epsilon_{\mu}^{b*} (k)  \ \Gamma^{(3) \mu}_O $.  
Analogous definitions exist for the photon two-point function and the
quark-quark-photon three-point function, which we will denote by
$\Gamma_O^{\mu \nu (\gamma)} (p,p')$ and $
\Gamma_O^{(3,\gamma)}(p,p',k)$ (the latter carries color structure
$\delta_{ij}$).  The assignment of momentum flow in these two- and
three-point functions with operator insertion is shown in
Fig.~\ref{fig:fig0}.

In any scheme, the renormalization factors  $Z_{ij}$ introduced in Eq.~(\ref{eq:ren1}) can then be determined by imposing conditions on the 
two- and three-point  functions defined above.  
Working to first order in $\alpha_s$,  the needed Green's functions  with insertion of $O_1^{(5)}  \equiv  C$ read:
\bea
\Gamma_C^{(2)}   &=&    \Gamma_C^{(2)}  \Big \vert_{\rm 1-loop}  \ + \   
\sum_{j \neq 1} \, (Z^{-1})_{1j}  \,  \Gamma_{O_j^{(5)}}^{(2)}  \Big  \vert_{\rm tree}   
\\
\Gamma_C^{(3)}   &=&    \Gamma_C^{(3)}  \Big \vert_{\rm 1-loop}  \ + \   
 \Big(  Z_\psi  Z_g  \sqrt{Z_G}  \, (Z^{-1})_{11} - 1 \Big)  \,    \Gamma_C^{(3)} \Big \vert_{\rm tree}  \ + \   
\sum_{j \neq 1} \, (Z^{-1})_{1j}  \,  \Gamma_{O_j^{(5)}}^{(3)}  \Big  \vert_{\rm tree}   
\label{eq:3pt1loop}
\\
\Gamma_C^{\mu \nu}  &=&    \Gamma_C^{\mu \nu}  \Big \vert_{\rm 1-loop}  \ + \   
\sum_{j \neq 1} \, (Z^{-1})_{1j}  \,  \Gamma_{O_j^{(5)}}^{\mu \nu}  \Big  \vert_{\rm tree}   ~. 
\eea
The simplest perturbative scheme is $\overline{\rm MS}$, in which one  
determines the $(Z^{-1})_{1j}$ by requiring  cancellation of the poles in $\epsilon = (4 - d)/2$. 
Similar relations involving insertions of $O^{(5)}_{i \neq1}$ allow one to determine 
the remaining entries $Z_{ij}$ of the renormalization matrix. 
This program requires computing the tree-level and one-loop results  for the 
two- and three-point functions, to which we turn next.

\subsection{Tree level matrix elements} 

In this section we give tree-level results for the 
gluon two-point functions   $\Gamma_O^{\mu \nu} (p,p')$, 
the quark two-point functions $\Gamma_O^{(2)} (p,p')$, 
the gluon-quark-quark three-point functions  $ \Gamma_O^{(3)}(k,p,p')$, 
and the photon-quark-quark three-point functions  $ \Gamma_O^{(3,\gamma)}(k,p,p')$, 
for all  the relevant operators $O_i^{(d)}$.

The only operator with non-zero  two-gluon matrix element at tree level  
is $O^{(5)}_5  \equiv m G \tilde{G}$:  
\bea
\Gamma_{O_5^{(5)}}^{\mu \nu}  (p,p') =
\textrm{Tr} \left[ \mathcal M  t^a \right] \times 
 4  \, \epsilon^{\mu \nu \alpha \beta}  p_\alpha  p_\beta' ~. 
\label{eq:ggt}
\eea
An analogous result holds for the photon two-point function  $\Gamma_{O_4^{(5)}}^{\mu \nu (\gamma)} (p,p')$.

In Tables~\ref{tab:tree}, \ref{tab:tree2}, and \ref{tab:tree3}, we
give the tree-level 1-particle irreducible (1PI) matrix elements
$\Gamma_O^{(2)} (p,p')$, $\Gamma_O^{(3)}(k,p,p')$, and
$\Gamma_O^{(3\gamma)}(k,p,p')$ for each operator. Throughout, we use
the notation: \bea \sigma (a,b) &\equiv & a_\mu \sigma^{\mu \nu} b_\nu
\nn \\ \epsilon_\mu (a,b,c) &\equiv & \epsilon_{\alpha \beta \rho \mu}
\, a^\alpha b^\beta c^\rho ~.  \eea

\begin{figure}[!tp]
\centering\includegraphics[width=0.65\textwidth]{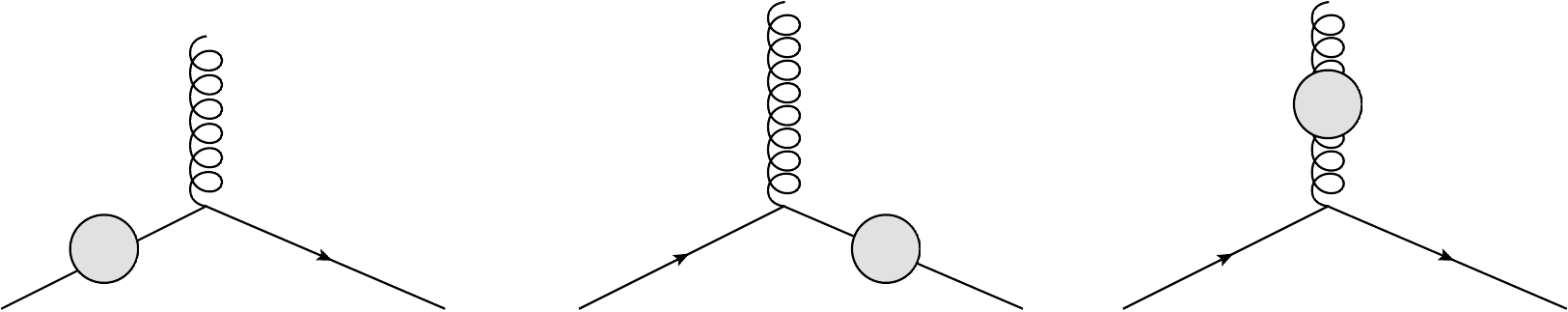}
\caption{Non-1PI diagrams contributing to the quark three-point function. The shaded blob represents 
the 1PI contribution to the relevant two-point function with  operator insertion.} 
\label{fig:non1PI}
\end{figure}

Finally, for a given operator $O$, non-1PI tree level contributions to
the three-point functions (see Fig.~\ref{fig:non1PI}) can be expressed
in terms of quark and gluon two-point functions as follows \bea
\Gamma_{O}^{(3)} (p,p',k) &=& - g \slashnext{\epol} \,
\frac{\slashed{k} + \slashed{p'} + m}{s - m^2} \, \Gamma_{O}^{(2)} (p,
k + p') - g \Gamma_{O}^{(2)} (p- k, p') \, \frac{\slashed{p}-
  \slashed{k} + m}{u - m^2} \, \slashnext{\epol} \nonumber \\ &-&
\frac{g}{t} \, \gamma_\mu \, \Gamma_O^{\mu \nu} (p - p',k) \,
\epol_\nu ~,
\label{eq:non1PI}
\eea
where  $s = (p'+k)^2$,  $u = (p-k)^2$, $t = (p' - p)^2$.

\begin{table}[ht]
\begin{center}
\setlength{\tabcolsep}{3pt}
\begin{tabular}{|c|c|}
\hline
 &  \\[-1\jot]
$O$   &  $ \Gamma_O^{(2)}$      \\[1\jot]
\hline
\hline
\strut\vphantom{\({}^2\)}\( O^{(3)} =  \  P  \)&\(i\gamma_5 t^a\) \\[1\jot]
\hline
\strut\vphantom{\({}^2\)}\( O^{(5)}_{2} = \partial^2 P \)&\(   - i  q^2 \gamma_5 t^a  \)\\[1\jot]
\strut\vphantom{\({}^2\)}\( {O}^{(5)}_{6} =   (m \partial \cdot A)_1   \)&   \(   \, \textrm{Tr}\left[ \mathcal M t^a\right]  \,  i \qslash     \gamma_5  \)      \\[1\jot]
\strut\vphantom{\({}^2\)}\( O^{(5)}_{7} =   (m \partial \cdot A)_2 \)&   \(  \left( \frac{1}{2}\, \left\{ \mathcal M, t^a\right\} - \frac{1}{3} \textrm{Tr}\left[ \mathcal M t^a
\right] \right) \,  i  \qslash     \gamma_5  \)      \\[1\jot]
\strut\vphantom{\({}^2\)}\( O^{(5)}_{8} =  \ ( m^2 P)_1  \)&     \(     \frac{1}{2}  \left\{ \mathcal M^2, t^a\right\}  i \gamma_5    \)     \\[1\jot]
\strut\vphantom{\({}^2\)}\( {O}^{(5)}_{9} =  \ (m^2 P)_2 \)&\(  \textrm{Tr} \left[ \mathcal M^2 t^a\right] i \gamma_5\)\\[1\jot]
\strut\vphantom{\({}^2\)}\( {O}^{(5)}_{10} =  \ (m^2 P)_3 \)&\(  \textrm{Tr} \left[ \mathcal M t^a\right] i  \mathcal M \gamma_5\)\\[1\jot]
\strut\vphantom{\({}^2\)}\(  O^{(5)}_{11} =  P_{EE} \)&\(
-i \Big[ p\cdot p' \,  t^a - \frac{1}{2} \left\{ \mathcal M^2, t^a\right\}    + i \sigma (p,p')  \, t^a
 + \frac{1}{2} \left\{ \mathcal M, t^a \right\}\qslash
 \Big] \gamma_5
\) \\[1\jot]
\strut\vphantom{\({}^2\)}\(O^{(5)}_{12}=   \partial \cdot A_{E}    \)&\(
i    \Big[ q^2  \, t^a   - \left\{\mathcal M, t^a\right\}\qslash   -       2i  \sigma(p, p') \, t^a  \Big] \gamma_5
\)\\[1\jot]
\strut\vphantom{\({}^2\)}\( 
O^{(5)}_{13} =  A_{\partial}  \) &\(
-i \Big[  (p^2 + p'^2) t^a -  \frac{1}{2} \left\{\mathcal M, t^a \right\}\qslash     \Big] \gamma_5
\)\\[1\jot]
\hline
\end{tabular}
\end{center}
\caption{Non-vanishing tree-level   2-point functions with operator insertion. 
For notational conventions and momentum flow, see discussion below 
Eq.~\eqref{eq:gggO}. 
 \label{tab:tree}}
\end{table}

\begin{table}[htp]
\begin{center}
\setlength{\tabcolsep}{3pt}
\begin{tabular}{|c|c|}
\hline
  & \\[-1\jot]
$O$      &   $ \Gamma_O^{(3)} $  (1PI)   \\[1\jot]
\hline
\hline
\strut\vphantom{\({}^2\)}\( O^{(5)}_{1}= C \) &\( 2g  \, \sigma(\epol,k) \gamma_5 t^a  \)\\[1\jot]
\strut\vphantom{\({}^2\)}\(  O^{(5)}_{11} =   P_{EE}   \)&\(- i g \Big[ \epol \cdot (p+p')   -  i \sigma (\epol, p-p')  \Big] \gamma_5 t^a
  \)\\[1\jot]
\strut\vphantom{\({}^2\)}\(O^{(5)}_{12}=   \partial \cdot A_{E}    \) &\( 2  g \ \sigma(\epol,q) \gamma_5  t^a
\)\\[1\jot]
\strut\vphantom{\({}^2\)}\( 
O^{(5)}_{13} =  A_{\partial}  \) &\(- i g \Big[ \epol \cdot (p+p')   +  i \sigma (\epol, p-p' - 2k)  \Big] \gamma_5 t^a
\)\\[1\jot]
\hline
\end{tabular}
\end{center}
\caption{Non-vanishing tree-level 1PI   quark-quark-gluon  3-point functions.  
For notational conventions and momentum flow, see discussion below 
Eq.~\eqref{eq:gggO}. 
\label{tab:tree2}}
\end{table}

\begin{table}[htp]
\begin{center}
\setlength{\tabcolsep}{3pt}
\begin{tabular}{|c|c|}
\hline
  & \\[-1\jot]
$O$      &   $ \Gamma_O^{(3, \gamma)} $  (1PI)   \\[1\jot]
\hline
\hline
\strut\vphantom{\({}^2\)}\( O^{(5)}_{3}= E \) &\( e  \{Q,t^a\}  \   \sigma(\epol,k) \gamma_5   \)\\[1\jot]
\strut\vphantom{\({}^2\)}\(  O^{(5)}_{11} =   P_{EE}   \)&\(- \frac{i e}{2}  \{Q,t^a\}   \Big[ \epol \cdot (p+p')   -  i \sigma (\epol, p-p')  \Big] \gamma_5  
  \)\\[1\jot]
\strut\vphantom{\({}^2\)}\(O^{(5)}_{12}=   \partial \cdot A_{E}    \) &\(   e  \{Q,t^a\}   \ \sigma(\epol,q) \gamma_5 
\)\\[1\jot]
\strut\vphantom{\({}^2\)}\( 
O^{(5)}_{13} =  A_{\partial}  \) &\(-\frac{ i e}{2}  \{Q,t^a\}    \Big[ \epol \cdot (p+p')   +  i \sigma (\epol, p-p' - 2k)  \Big] \gamma_5  
\)\\[1\jot]
\strut\vphantom{\({}^2\)}\( 
O^{(5)}_{14} =  A_{A^{(\gamma)}}  \) &       \(  - \frac{i e}{2}  \{Q,t^a\}   \Big[ \epol \cdot (p+p')   -  i \sigma (\epol, p-p')  \Big] \gamma_5  
\)\\[1\jot]
\hline
\end{tabular}
\end{center}
\caption{Non-vanishing tree-level 1PI   quark-quark-photon 3-point functions. 
For notational conventions and momentum flow, see discussion below 
Eq.~\eqref{eq:gggO}.   \label{tab:tree3}}
\end{table}

\subsection{One-loop Green's functions with  CEDM insertion}
\label{sect:loop1}

\begin{figure}[!tp]
\centering
\includegraphics[width=0.65\textwidth]{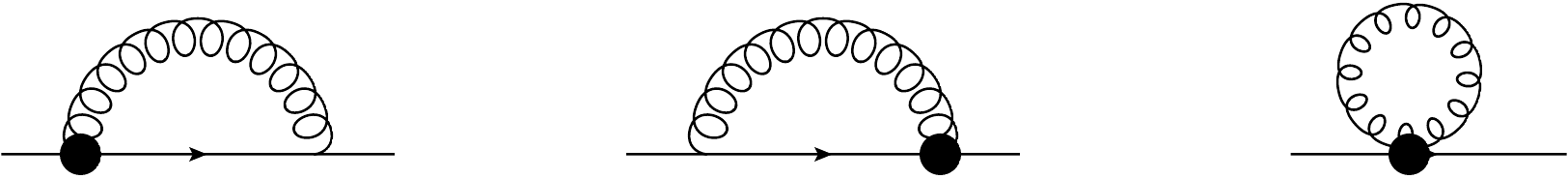}
\caption{Diagrams contributing to the quark two-point function. The dot denotes the insertion of the CEDM operator.} 
\label{fig:qqC}
\end{figure}

At one-loop level, we regulate the diagrams with dimensional
regularization, following the notation introduced in Section~\ref{sec:RCmixing}.
Working in general covariant gauge (with gauge fixing
parameter $\xi$) \footnote{
Feynman gauge corresponds to $\xi=1$, while Landau gauge corresponds to $\xi=0$.}, 
we have computed both the divergent and finite parts
of the Green's functions at generic kinematic points, before
specializing to non-exceptional momentum configurations needed to
define the operators in the RI-$\tilde{\rm S}$MOM scheme (see
Section~\ref{sect:MStoRISMOM}).  Specifically, for the two-point
functions (with $p + q = p'$) we work at the symmetric point $p^2 =
p'^2 =q^2 = - \Lambda^2$.  For the three-point functions ($p +q = p' +
k$) we work at the non-symmetric point $\tilde{S}$ characterized by
$p^2 = p'^2 = k^2 =q^2 = s = u = t/2 = - \Lambda^2$.  We will provide
the motivation behind this choice in Section~\ref{sect:MStoRISMOM}.

Throughout this work we will denote the $SU(N_C)$ color factors as follows:
\be
C_F = \frac{N_C^2 - 1}{2 N_C},   \qquad 
 C_A = N_C, \qquad  
T_F = \frac{1}{2}~.
\label{eq:colorfactors}
\ee

\subsubsection{Quark two-point function}

At one loop,  $\Gamma^{(2)}_{C}  (p,p')$ receives contributions from the diagrams in Fig.~\ref{fig:qqC} 
and reads:
\bea
\Gamma^{(2)}_{C}  (p,p')   &=& 
\frac{i \alpha_s}{4 \pi}  
\Bigg\{
\left(p^2 + p'^2 \right) \gamma_5 t^a  \left[  3 C_F  \left(\frac{1}{\epsilon} + \log \frac{\mu^2}{\Lambda^2}\right)   + f_0 \right]
\nn \\
&+& \{ \mathcal M, t^a \} \qslash \gamma_5  \left[  - \frac{3 C_F}{2}   \left(\frac{1}{\epsilon} + \log \frac{\mu^2}{\Lambda^2}\right)  + f_1   + O \left( \frac{m_q^2}{\Lambda^2} \right)  \right] 
\nn \\
&+& \{ \mathcal{M}^2, t^a \} \gamma_5  \left[  - 6 C_F   \left(\frac{1}{\epsilon} + \log \frac{\mu^2}{\Lambda^2}\right)  + f_2  + O \left( \frac{m_q^2}{\Lambda^2} \right)    \right]  \ \Bigg\}~,
\label{eq:q2pt}
\eea
where
\bea
f^{HV}_0  &=&    \frac{22}{9}   \times 3 C_F~,  \qquad
f^{NDR}_0  =    \frac{4}{3} \times 3 C_F  
\\
f^{HV}_1  &=&  - 3 C_F~,   \qquad 
f^{NDR}_1 =  -\frac{2}{3}  \times 3 C_F
\\
f^{HV}_2 &=&  -\frac{10}{3} \times 3 C_F~,   \qquad
f^{NDR}_2  =    -\frac{2}{3} \times 3 C_F ~.
\eea

\subsubsection{Gluon two-point function}

\begin{figure}[!t]
\centering\includegraphics[width=0.65\textwidth]{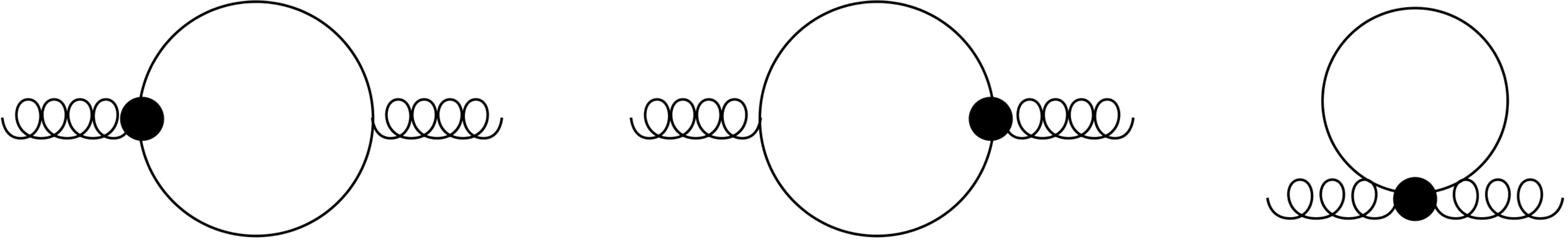}
\caption{Diagrams contributing to the gluon  two-point function. The dot denotes the insertion of the CEDM operator.} 
\label{fig:ggC}
\end{figure}

As illustrated in Fig.~\ref{fig:ggC},  at  one loop level three diagrams contribute to the  gluon two-point function with insertion of the CEDM operator,  $\Gamma_C^{\mu\nu} (p,p')$,  defined in Eq.~\eqref{eq:ggO}. 
The third diagram vanishes  due to  the anti-symmetry of $\sigma_{\mu \nu}$, while the other two contribute 
\be
\Gamma_{C}^{\mu \nu}  (p,p')   =   \frac{\alpha_s}{4 \pi} \,   \textrm{Tr}\left[ \mathcal M t^a \right]  \, 
\Gamma_{G \tilde{G}}^{\mu \nu}  (p,p')  \,   \left[  2 \left( \frac{1}{\epsilon} + \log \frac{\mu^2}{\Lambda^2} \right) + 4   
+ O \left( \frac{m_q^2}{\Lambda^2} \right)  \right]~,
\label{eq:g2pt}
\ee
where
$\Gamma_{G \tilde{G}}^{\mu \nu}  (p,p') = 4  \, \epsilon^{\mu \nu \alpha \beta}  p_\alpha  p_\beta' $ (see  Eq.~\eqref{eq:ggt}).
This result allows us to identify the mixing between the CEDM operator 
$C$   and   the operator $ O_5^{(5)} =  m G \tilde{G} $.

\subsubsection{Quark-quark-gluon three-point function}

\begin{figure}[!t]
\centering
\includegraphics[width=0.12\textwidth]{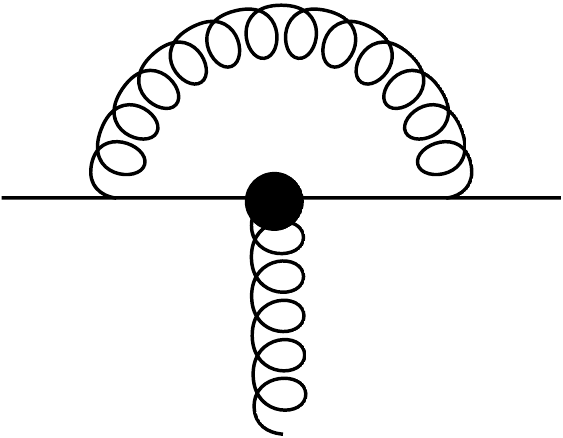}
\hspace{0.2in}
\includegraphics[width=0.30\textwidth]{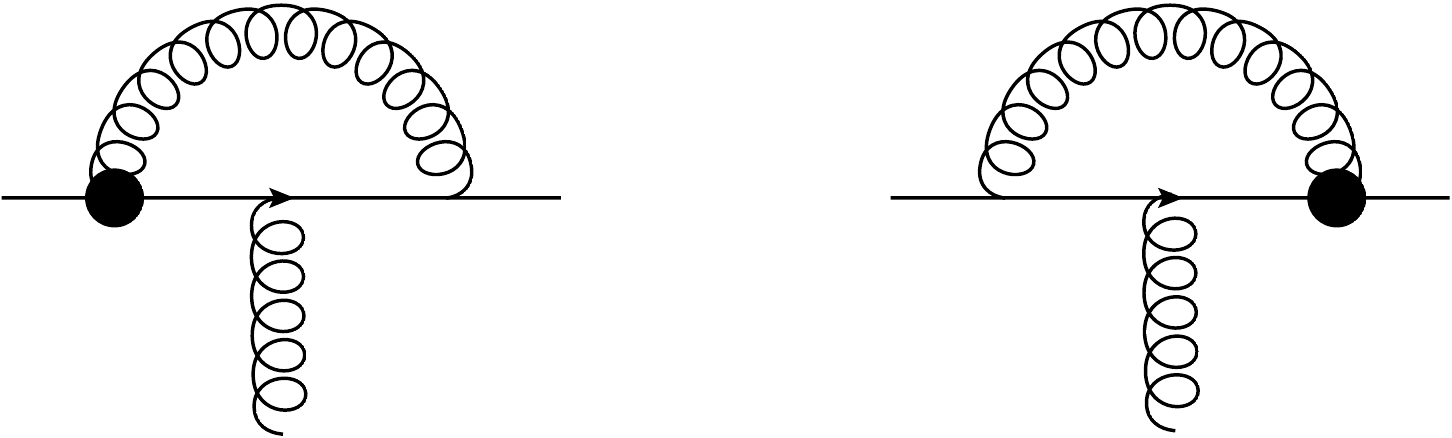}
\hspace{0.2in}
\includegraphics[width=0.30\textwidth]{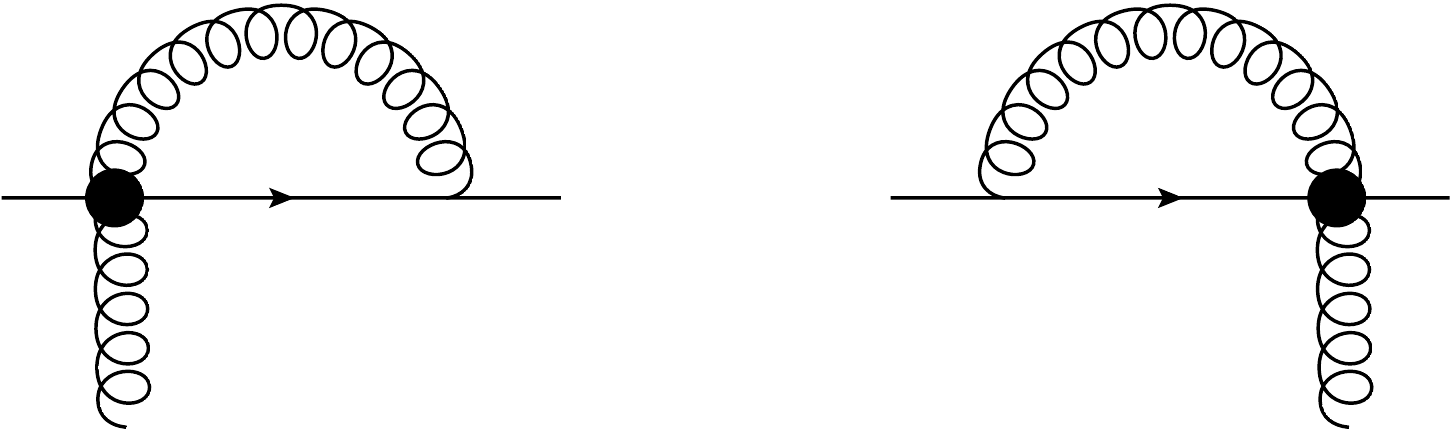}
\hspace{0.2in}
\\
\includegraphics[width=0.12\textwidth]{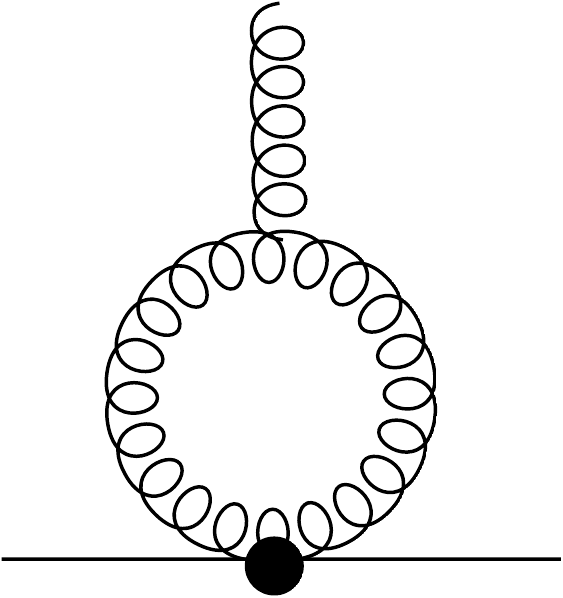}
\hspace{0.2in}
\includegraphics[width=0.3\textwidth]{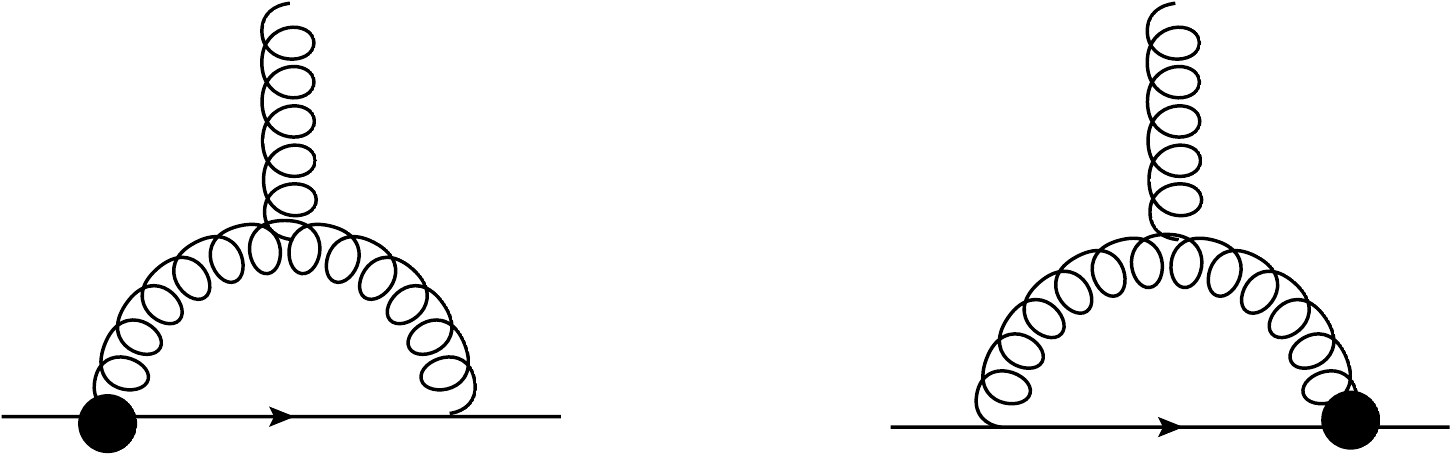}
\caption{1PI diagrams contributing to the quark three-point function. The dot denotes the insertion of the CEDM operator.} 
\label{fig:qqgC}
\end{figure}

We now turn to the  quark-quark-gluon  three-point function with insertion of the chromo-electric operator, 
$ \Gamma_C^{(3)}  (p,p',k)$,   defined in Eq.~\eqref{eq:gggO}.  
In all diagrams we chose to eliminate the four-momentum $q^\mu$ in favor of $(k + p'-p)^\mu$. 
The amputated  three-point function receives contributions from 1PI diagrams  (see Fig.~\ref{fig:qqgC}),   non-1PI diagrams (see Fig.~\ref{fig:non1PI}), 
and quark and gluon wave-function renormalization.  
In this section we summarize our results for the 1PI  diagrams 
and note that  the non-1PI contributions of   Fig.~\ref{fig:non1PI} 
are determined by  the one-loop results for the quark and gluon two-point functions 
$\Gamma_{C}^{(2)}$ and $\Gamma_{C}^{\mu \nu}$  
presented in  Eqs.~\eqref{eq:q2pt} and \eqref{eq:g2pt}, 
as detailed in Eq.~\eqref{eq:non1PI}. 
As we will  discuss in Section~\ref{sect:MStoRISMOM},   we  can choose 
a kinematic point and appropriate conditions so that the non-1PI diagrams are not 
needed to determine the RI-$\tilde{\rm S}$MOM  renormalization constants. 

$ \Gamma_C^{(3)}  (p,p',k)$  can be decomposed in terms of sixteen spinor structures, 
and is  characterized by sixteen scalar coefficients $c_{1, ..., 16}$\footnote{Hermiticity 
of the operator implies constraints among the various coefficients, such as  $c_{12} = - c_{13}$, which we have 
used to check our calculation.}  
\bea
\Gamma^{(3)}_{C} &=&   \Big[
c_1  \gamma_5  +
c_2 \epsilon (\epsilon^*,k,p,p') +
c_3 \slashed{\epsilon}^* \gamma_5 + 
c_4 \slashed{k} \gamma_5 + 
c_5 \slashed{p} \gamma_5 + 
c_6 \slashed{p}' \gamma_5 
 \\
&+& c_7 \epsilon_\mu (\epsilon^*,k,p)  \gamma^\mu+ 
c_8 \epsilon_\mu (\epsilon^*,k,p')  \gamma^\mu + 
c_9 \epsilon_\mu (\epsilon^*,p,p')  \gamma^\mu + 
c_{10}  \epsilon_\mu (k,p,p')  \gamma^\mu 
\nonumber \\
& +&   
c_{11} \sigma (\epsilon^*, k) \gamma_5  +  
c_{12} \sigma (\epsilon^*, p) \gamma_5  +  
c_{13} \sigma (\epsilon^*, p') \gamma_5  \nonumber\\
& +&
c_{14} \sigma (k, p) \gamma_5  +  
c_{15} \sigma (k, p') \gamma_5  +  
c_{16}  \sigma (p, p') \gamma_5  \Big]  ~.
\nonumber 
\eea
The coefficients $c_i$ are functions of the invariants $p^2$,
$p^{\prime\,2}$, $k^2$, $q^2$, $s$, $t$, $u$, 
and $\epsilon^* \cdot (p \pm p')$.
The $c_i$'s   can be expressed in
terms of triangle and bubble scalar integrals and their derivatives
with respect to the invariants they depend on.  For a generic
kinematic configuration, the result involves logarithms and
dilogarithms of ratios of invariants, and logarithms of ratios of the
invariants to the renormalization scale $\mu$.  Working at the
RI-$\tilde{\rm S}$MOM kinematic point $p^2 = p^{\prime\, 2} = k^2 =
q^2 = s = u = t/2 = -\Lambda^2$, and in the massless limit, greatly
simplifies the integrals, reducing them to single-scale integrals. At
this point, the triangle scalar integrals collapse to constants, and
contribute in two forms. First, triangles that are functions of three
invariants that become equal at the renormalization point, like $p^2$,
$k^2$ and $s$, or $p^{\prime 2}$, $k^2$ and $u$, are proportional to
the constant
\begin{eqnarray}
\psi & = &   \frac{2}{3} \left(  \psi^{(1)} \left(\frac{1}{3}\right) - \frac{2}{3} \pi^2\right)~,
\label{eq:psi}
\end{eqnarray}
called $C_0$ in Ref.~\cite{Sturm:2009kb}.  Here $\psi^{(1)}$ denotes
the first derivative of the Digamma function.  Second, triangles that depend
on the invariants $p^2, p^{\prime\,2}$ and $t$ are proportional to the
Catalan constant, which can also be expressed in terms of the first
derivative of the Digamma function
\begin{eqnarray}
K &= &  \frac{1}{8} \left( \psi^{(1)}\left(\frac{1}{4}\right)-  \pi^2\right).
\label{eq:K}
\end{eqnarray}
The only other non-rational number occurring in the result is $\log(2)$, which originates from  the choice $t = - 2 \Lambda^2$.

Next, we give the UV divergent parts of the diagrams in
Fig.~\ref{fig:qqgC}, and the finite pieces of those Dirac structures
that give non-vanishing contributions to the projections used to
define the quark CEDM operator in the RI-$\tilde{\rm S}$MOM scheme
(see Section~\ref{sect:MStoRISMOM}).  The quark-quark-gluon
three-point function is
\begin{eqnarray}
\label{eq:qqgFinite}
\Gamma_C^{(3)}(p, p^{\prime}, k) &=&  g \   \frac{\alpha_s}{4\pi}  
\left\{ 
2 \, \sigma(\varepsilon^{*}, k)  \gamma_5 \left[   
\left( C_F  \left( \xi-2 \right) + C_A  \left(\frac{11}{4}+ \frac{\xi}{4}\right)  \right) \left( \frac{1}{\varepsilon} +  \log \frac{\mu^2}{\Lambda^2} \right) 
+ k_1\right] \nonumber  \right.
\\ &+& \left.   \sigma(\varepsilon^* , p - p^\prime )  \gamma_5 \left[ 
 - \frac{3 C_A}{4} \left( \frac{1}{\varepsilon} +  \log \frac{\mu^2}{\Lambda^2} \right) + k_2
  \right]\right.  \nonumber \\ 
&+ & 
\left.  i (p+p^\prime)\cdot \varepsilon^* \gamma_5 
\left[  \left(6 C_F - \frac{3}{4} C_A\right)\left(\frac{1}{\varepsilon} +  \log \frac{\mu^2}{\Lambda^2} \right) 
+ k_3
\right] \right\} \, t^a + \ldots
\end{eqnarray}
where, here and later, $\ldots$ denotes the contribution of the Dirac structures that are not relevant to defining the quark CEDM 
in RI-$\tilde{\rm S}$MOM scheme.
The constants $k_1$, $k_2$ and $k_3$ depend on the definition of $\gamma_5$ in $d$-dimension. In the 't Hooft-Veltman scheme, they are given by
\begin{eqnarray}
k^{HV}_1 &=& -2 C_F \left(  2 -  \xi \right)
+     \frac{33  - \xi^2}{4} C_A
+ \left( \frac{2}{3} C_F  - \frac{ 5 + 2 \xi}{3} C_A \right) \psi
+  (2 C_F -  C_A )(1-\xi) K\\
k^{HV}_2 &=&    C_F (2-\xi)  - 
 C_A \left(  \frac{13 - 2 \xi}{4}      -\frac{\xi}{6}\psi^{} \right)  
 + (2 C_F - C_A)(1-\xi) \left( \frac{1}{2} \log(2) - K \right)
 \\
k^{HV}_3 &=& 
  \frac{44}{3} C_F - 2 C_A     + \left( - 4 C_F + \frac{3+2\xi}{6} C_A \right)\psi + (2 C_F - C_A) \frac{(3+\xi)}{2}\log(2),
\end{eqnarray}
while in NDR
\bea
k^{NDR}_1 &=&   k^{HV}_1  -2 \left( C_F  + C_A \right)
\\
k^{NDR}_2 &=&   k^{HV}_2  + \frac{1}{2}  C_A
\\
k^{NDR}_3 &=&   k^{HV}_3  - \frac{20}{3}  C_F  + \frac{1}{2} C_A~.
\eea

\subsubsection{Quark-quark-photon three-point functions}

The quark-quark-photon three-point function with insertion of the quark CEDM gives
\begin{eqnarray}\label{eq:qqgammaFinite}
\Gamma_C^{(3,\gamma)} & =  & 
\frac{e}{2}  \{ Q, t^a \} \
 \frac{\alpha_s}{4\pi}\left\{   2 \,  \sigma(\varepsilon^*, k)  \gamma_5  \left[ - C_F   \left( \frac{1}{\varepsilon}+ \log \frac{\mu^2}{\Lambda^2} \right)+
k_1^{(\gamma)}\right]  +   C_F \, \sigma(\varepsilon^*,   p - p^{\prime})  \gamma_5    
\nonumber  \right. 
\\ &+ & \left.
 i (p+p^\prime) \cdot \varepsilon^* \gamma_5    \left[  6 C_F  \left( \frac{1}{\varepsilon} + \log \frac{\mu^2}{\Lambda^2} \right)  + k_3^{(\gamma)}\right]  \right\}
+ \ldots
\end{eqnarray}
with
\begin{eqnarray}
k_1^{(\gamma) HV} &=& C_F \, \left(- 2  + \frac{2}{3} \psi^{} \right)  \qquad  k_1^{(\gamma) NDR} = C_F \, \left( - 4  + \frac{2}{3} \psi^{} \right)  
\\
k_3^{(\gamma) HV} &=& C_F  \, \left( \frac{44}{3} - 4 \psi^{} \right) \qquad  k_3^{(\gamma) NDR} = C_F \, \left( 8 - 4 \psi^{} \right)~.
\end{eqnarray}

\subsection{One-loop Green's functions with insertions of
 \texorpdfstring{$E$}{E}, 
 \texorpdfstring{$P$}{P}, 
         \texorpdfstring{$\partial \cdot A$}{\textpartial \textcdot A} and 
         \texorpdfstring{$G \tilde{G}$}{GG\textcomposetilde}}
\label{sect:loop2}

The determination of the physical   block $Z_O$ of the mixing matrix in Eq.~\eqref{eq:Zstructure} requires the calculation of quark and/or  gluon two-point functions with   insertions of the operators $E$, $\partial^2 P$, $(m^2 P)_{1,2,3}$, $(m\partial \cdot A)_{1,2}$ and $mG \tilde G$.
The renormalization of the pseudoscalar and tensor densities,  and axial current has been studied 
in many papers,  and  the conversion between $\overline{\textrm{MS}}$-NDR  and RI-SMOM to one loop was addressed  
in Ref.~\cite{Sturm:2009kb}. The renormalization of $G \tilde G$ in $\overline{\textrm{MS}}$ was studied in Ref.~\cite{Espriu:1982bw}.
Here we provide one-loop 1PI results for the Green's functions  in $\overline{\rm MS}$-HV  and  $\overline{\rm MS}$-NDR.

The relevant projection of the quark-quark-photon 1PI  three-point function 
(this is essentially a quark-quark function)  with insertion of the quark EDM operator, 
evaluated at the symmetric point gives 
\begin{eqnarray}\label{eq:qqgammaEFinite}
\Gamma_E^{(3,\gamma)} & =  & 
- e  \{ Q, t^a \} \  \frac{\alpha_s}{4\pi}     \sigma(\varepsilon^*, k )  \gamma_5   \, 
\left[  (1-\xi)   C_F  \left( \frac{1}{\varepsilon} + \log \frac{\mu^2}{\Lambda^2} \right)   + k_T
\right] + \ldots~
\\
k_T &=& C_F  \, (1 - \xi)  \left(  2 - \frac{5}{6} \psi \right)~, 
\label{eq:kT}
\end{eqnarray}
both in HV and NDR.

At one loop, the 1PI quark two-point functions with insertions of the operators $\partial^2 P$ and $(m^2 P)_{1,2,3}$, 
evaluated at the symmetric point, are given by 
\begin{eqnarray}
\Gamma^{(2)}_{\partial^2 P,\, (m^2 P)_{1,2,3}}&=&  i \gamma_5 
\left\{ -q^2 t^a , \frac{1}{2} \left\{\mathcal M^2, t^a \right\} ,
 \textrm{Tr}\left[\mathcal M^2 t^a\right] {\mathbbm 1} , \
\textrm{Tr}\left[\mathcal M  t^a\right] \, \mathcal M 
\right\} 
\nonumber \\
& \times &  \frac{\alpha_s}{4\pi}\left[   k_P  + (3+\xi) C_F \left(
\frac{1}{\varepsilon} + 
\log \frac{\mu^2}{\Lambda^2} 
% - \frac{1}{2} \psi^{}
\right) 
 + C_F\,  \frac{1-\xi}{3 \Lambda^2} i \sigma^{\alpha \beta} p_{\alpha} p_{\beta}^{\prime}
 + O \left( \frac{m_q}{\Lambda} \right)
\right]~,  \qquad \label{eq:2pP}
\end{eqnarray}
where  $k_P$ depends on the $d$-dimensional definition of $\gamma_5$, namely
\begin{equation}
k_P^{HV} = C_F \left[ 2 ( 6 +  \xi)  - \frac{3+\xi}{2} \psi  \right], \qquad 
k_P^{NDR}  = k_P^{HV} - 8 C_F ~. 
%2 (2 + \xi).
\end{equation}
The gluon and photon two-point functions with insertions of $P$ are finite, and not needed for renormalization. Eq.~\eqref{eq:2pP} is in agreement with the result of Ref.~\cite{Sturm:2009kb}, where the calculation was carried out in the NDR scheme.

The 1PI quark two-point function with insertion of the operators proportional to the  divergence of the axial current is  
\begin{eqnarray}\label{eq:2ppartialA}
\Gamma^{(2)}_{ ( m\partial \cdot A)_1, ( m\partial \cdot A)_2}  &=&
 i  \slashed{q} \gamma^5 
\left\{ 
\textrm{Tr} \left[\mathcal M t^a \right]  {\mathbbm 1} \ , \  
\frac{1}{2} \,  \left\{\mathcal   M, t^a \right\} - \frac{1}{3} \textrm{Tr}\left[\mathcal M t^a\right]  {\mathbbm 1} 
 \right\} 
\frac{\alpha_s}{4\pi} \left[  k_A   + C_F  \xi \left( \frac{1}{\varepsilon} + \log \frac{\mu^2}{\Lambda^2} \right)\right]
\nonumber\\
k_A^{HV} &=& C_F \left (\xi + 4 \right) \qquad \qquad k_A^{NDR} =   C_F \, \xi~. 
\end{eqnarray}

The gluon two-point function with insertion of the operator  $(m \partial \cdot A)_1$ is finite, and we find,
\begin{eqnarray}
\Gamma^{(2)\, \mu \nu}_{ (m \partial \cdot{A})_1}  =   \textrm{Tr}\left[\mathcal M t^a\right] \frac{\alpha_s n_F}{4\pi}  4 e^{\mu \nu \alpha \beta} p_{\alpha} p^{\prime}_{\beta} ,
\end{eqnarray}
where $n_F = 3$ is the number of flavors we are considering.
The insertion of the operator $(m \partial \cdot A)_2$ vanishes.

Finally, the gluon  and quark  two-point functions with insertion of $(m G \tilde{G})$ are  given by 
\begin{eqnarray}
\label{eq:2gGGtilde}
\Gamma^{(2)\, \mu \nu}_{(m G \tilde{G})} &=&  4 \,   \textrm{Tr}\left[\mathcal M t^a\right]\varepsilon^{\mu \nu \alpha \beta} p_{\alpha} p^{\prime}_{\beta}   \frac{\alpha_s}{4\pi} \left\{
C_A   \frac{3 + \xi}{2}
\left( \frac{1}{\varepsilon} + \log \frac{\mu^2}{\Lambda^2} \right) 
+ k_G  \right\}
\\
\label{eq:2qGGtilde}
\Gamma^{(2)}_{(m G \tilde{G})} &=&   i \textrm{Tr}[\mathcal M t^a]
\,  \slashed{q} \gamma_5
\  \frac{\alpha_s}{4\pi}   \left[   6 C_F  \left( \frac{1}{\varepsilon} + \log \frac{\mu^2}{\Lambda^2} \right)   + \tilde{k}_G  \right]
+ \mathcal O \left( {\cal M}^2 \right)   
\\
k_G &=& \frac{C_A }{2} \left( 17 - \xi^2    - \frac{ 4}{3} (3+\xi)   \psi^{} \right) 
\\
\tilde{k}_G &=& C_F \, \left(16 - 4 \psi \right)~.
\end{eqnarray}
These results determine the self-renormalization of 
 $(m G \tilde G)$  and its mixing with  $(m \partial\cdot A)_1$.

\section{Renormalization matrix in \texorpdfstring{$\overline{\rm MS}$}{M\textcomposemacron S\textcomposemacron} scheme} 
\label{sect:5}

In this section we provide one-loop results for the $Z_O$ block 
%the first row 
of the renormalization matrix given in Eq.~\eqref{eq:Zstructure}.
At various stages of the calculation we need the  one-loop results for the mass, couplings, and field 
renormalization constants  in general covariant gauge (recall $d = 4 - 2 \epsilon$): 
\bea
Z_m &=&   1 - \frac{1}{\epsilon}   \frac{\alpha_s}{4 \pi}    \ 3  \, C_F
\\
Z_q &=& 1 - \frac{1}{\epsilon}   \frac{\alpha_s}{4 \pi}    \  \xi  \, C_F
\\
Z_G &=& 1 +    \frac{1}{\epsilon} \frac{\alpha_s}{4 \pi} 
\left[  - \frac{4}{3} n_F  T_F   + C_A  \left(\frac{13}{6} - \frac{\xi}{2} \right) \right]
\\
Z_g &=& 1 -   \frac{1}{\epsilon}   \frac{\alpha_s}{4 \pi}
\  \frac{11 C_A - 4 T_F  n_F}{6} ~, 
\eea
where the color factors are $C_F$, $C_A$, and $T_F$ are 
given in Eq.~\eqref{eq:colorfactors}.
We will also need the renormalization constants for the pseudoscalar  $\bar \psi \gamma_5 \psi$ and  tensor
$\bar \psi \sigma_{\mu \nu} \psi$   densities, defined by $O_\Gamma^{(0)} = Z_\Gamma O_\Gamma$: 
\bea
Z_P &=& Z_m^{-1}  =   1 +  \frac{1}{\epsilon}   \frac{\alpha_s}{4 \pi}    \ 3  \, C_F 
\\
Z_T &=&   1 -   \frac{1}{\epsilon}   \frac{\alpha_s}{4 \pi}    \   C_F.
\eea
For the mixing of dimension-5 operators, 
specializing Eq.~\eqref{eq:ren1} to the $\MSbar$  scheme at one loop 
one finds
\be
O_i^{\MSbar} \ = \ (Z^{-1})^{\overline{\rm MS}}_{ij} \ O_j^{(0)}~,
\qquad Z^{\overline{\rm MS}}_{ij} \equiv \delta_{ij} \ -
\ \frac{1}{\epsilon} \frac{\alpha_s}{4 \pi} \ z_{ij}~.  
\label{eq:renmsbar}
\ee
Note that in the above expressions  $\alpha_s$ denotes the $d$-dimensional renormalized  coupling 
defined in Section~\ref{sec:RCmixing}, satisfying $d \alpha_s/ d (\log \mu) = - 2 \epsilon \alpha_s + O(\alpha_s^2)$. 
So to $O(\alpha_s)$ the anomalous dimension matrix $\gamma  \equiv d (\log Z)/d (\log \mu)$ can be immediately read off Eq.~(\ref{eq:renmsbar}): 
$\gamma_{ij} = 2  \alpha_s/(4 \pi) \, z_{ij}$. 
%so that the anomalous dimension can be immediately read off:   $\gamma = d (\log Z)/d (\log \mu)$ 

The various entries of the renormalization matrix are determined as follows:

\begin{itemize} 

\item Finiteness of the quark two-point function $\Gamma^{(2)}_C$,
  gluon two-point function $\Gamma_C^{\mu \nu}$, quark-quark-gluon
  $\Gamma^{(3)}_C$ and quark-quark-photon $\Gamma^{(3,\gamma)}_C$
  three-point functions implies a set of conditions for $z_{1n}$,
  $n=1,..., 14$.  Note that only the results for $n=1,...,10$ affect
  physical observables, the rest are given for completeness.

\item The operator $O_{2}^{(5)} = \partial^2 P$ renormalizes
  diagonally with constant $Z_P$.

\item The quark EDM operator $O_{3}^{(5)} \equiv E$ renormalizes diagonally  
(to zeroth order in the fine structure constant)  in the same way as the 
 tensor quark bilinear, $i.e.$, $(Z^{-1})_{33} = Z_T^{-1}$. 

\item To zeroth order in the electromagnetic couplings, $O_{4}^{(5)} =
  m F \tilde F$ renormalizes diagonally with the renormalization constant
  $(Z^{-1})_{44} = Z_m^{-1}$.

\item The subset of  operators  $O_{5,6,10}^{(5)}$ related to the axial anomaly 
renormalize, to one-loop, as follows~\cite{Espriu:1982bw} (recall $Z_m  Z_P = 1$): 
\be
\left(\begin{array}{c}
m  \, G \tilde G  \\
(m \,  \partial  \cdot A)_1 \\
(m^2 P)_3
\end{array}\right)^{\overline{\rm MS}}
= 
\left(\begin{array}{ccccc}
Z_m^{-1}  Z_g^2  &    &  - \frac{1}{\epsilon} \frac{\alpha_s}{4 \pi} \ 6 C_F &     & 0   \\
0 &  & Z_m^{-1} & & 0 \\
0 &  & 0 &  & Z_m^{-1}  
\end{array}\right)
\, 
\left(\begin{array}{c}
m \, G \tilde G \\
(m  \,  \partial \cdot A)_1 \\
(m^2 P)_3
\end{array}\right)^{(0)} ~~.
\label{eq:Zstructure2}
\ee
To explicitly check Eq.~\eqref{eq:Zstructure2} at one
loop use Eqs.~\eqref{eq:2pP}, \eqref{eq:2ppartialA}, \eqref{eq:2gGGtilde}
and \eqref{eq:2qGGtilde}. 

\item Finally,  $O_7^{(5)} = (m \partial \cdot A)_2$ renormalizes as  $(m \partial \cdot A)_1$ 
and $O_{8,9}^{(5)}= (m^2 P)_{1,2}$ renormalize as $(m^2 P)_3$, 
thus leading to   $(Z^{-1})_{77,88,99} = Z_m^{-1}$. 

\end{itemize}
 
In summary, the entries in the first row in Eq.~\eqref{eq:Zstructure} are:
\begin{subequations}
\begin{align}
z_{11} &= 5 C_F - 2 C_A
\\
z_{12} &= 0 
\\
z_{13} &=  4 C_F
\\
z_{14} &= 0 
\\
z_{15} &= - 2
\\
z_{16} &=  C_F - \frac{1}{4} C_A
\\
z_{17} &= 3 C_F - \frac{3}{4} C_A
\\
z_{18} &=  6 C_F + \frac{3}{2} C_A
\\
z_{19} &=  0
\\
z_{1,10} &= 0
\\
z_{1,11} &=  6 C_F - \frac{3}{2} C_A
\\
z_{1,12}  &=- 3 C_F +  \frac{3}{4} C_A 
\\
z_{1,13}  &= \frac{3}{4} C_A
\\
z_{1,14}  &=  \frac{3}{4} C_A~.
\end{align}
\end{subequations}
For the remaining non-zero entries we have:
\begin{subequations}
\begin{align}
z_{22} &= -3 C_F 
\\
z_{33} &=  C_F
\\
z_{44} &= 3 C_F
\\
z_{55} & =  - \frac{11 C_A - 4 T_F  n_F}{3}  + 3 C_F
\\
z_{56} & =  -6 C_F 
\\
z_{66} & = z_{77} = z_{88} = z_{99} = z_{10,10}=  3 C_F.
\end{align}
\end{subequations}

The submatrix $z_{11}$, $z_{13}$ and $z_{33}$ agrees with the original calculation of Refs.
 \cite{Wilczek:1976ry,Vainshtein:1976nd,Ellis:1976uz,Ciuchini:1993fk,Degrassi:2005zd}.

\section{Definition of  RI-\texorpdfstring{$\tilde{\bf S}$}{S\textcomposetilde}MOM operators and matching to 
\texorpdfstring{$\overline{\bf MS}$}{M\textcomposemacron S\textcomposemacron}} 
\label{sect:MStoRISMOM}

A consistent phenomenological analysis of BSM-induced CP violation in
hadronic systems requires computation of the effect of the CP-odd
operators in Eq.~\eqref{eq:Leff1} on couplings at the hadronic scale,
such as the nucleon EDM and the T-odd $\pi N N$ couplings.  This is an
intrinsically non-perturbative problem.
The  first step in this program  involves  {\it defining}  UV finite operators in a suitable renormalization  scheme, 
whose matrix elements can be then  computed non-perturbatively  within lattice QCD.  
Here we will define finite operators  within a class of regularization-independent (RI)  
momentum subtraction (MOM) schemes~\cite{Martinelli:1994ty,Sturm:2009kb}.
Next, one converts the 
matrix elements in the RI-MOM scheme to the $\MSbar$ scheme, 
commonly adopted to  compute the Wilson coefficients  
and their renormalization-group evolution  
down to the hadronic scale, using continuum perturbation theory.

In this section we address the following issues:
\begin{enumerate}
\item We provide a set of regularization independent normalization
  conditions for the amputated Green's functions $\Gamma_{O_i^{(5)}}$
  that subtract all the UV divergences and fix the finite parts of the
  renormalization constants for the gauge-invariant CP-odd operators
  $O^{(5)}_{1, ...,10}$.  Since we will use subtraction conditions for
  the three-point functions at a non-symmetric momentum point, we call
  this scheme RI-$\tilde{\rm S}$MOM, as opposed to
  RI-SMOM~\cite{Sturm:2009kb}.
\item We provide the finite matching matrix that relates the
  RI-$\tilde{\rm S}$MOM and $\MSbar$ operators to one-loop in
  perturbation theory: 
\be O_i^{\rm RI-\tilde{\rm S}MOM} = C_{ij} \,
  O_j^{\MSbar}~.
\label{eq:match1}
\ee 
In practice this amounts to finding a linear combination of
$\MSbar$ operators $O_i^{\MSbar}$ such that the Green's functions with
insertions of $O_i^{\rm RI-\tilde{\rm S}MOM}$ satisfy the
normalization conditions that define the scheme (see item 1. above).
\end{enumerate}

\subsection{Defining the RI-\texorpdfstring{$\tilde{\bf S}$}{S\textcomposetilde}MOM scheme}
\label{sect:RISMOM}

We follow the strategy outlined in Refs.~\cite{Martinelli:1994ty,Sturm:2009kb}, 
with appropriate modifications related to the operators we are dealing with.
The content of this scheme can be summarized as follows:
\begin{itemize}

\item We require that the quark and gluon two-point functions with insertion of the quark CEDM 
operator $\Gamma^{(2)}_C (p,p')$ and $\Gamma^{(2) \mu \nu}_C (p,p')$ 
 vanish at the symmetric kinematic  point $S$ 
defined by $p^2 = p'^2 =q^2 = - \Lambda^2$. 

\item We require that certain projections of the three-point functions
  with quark CEDM insertion $\Gamma_C^{(3)}$ and
  $\Gamma_C^{(3,\gamma)}$ take the tree-level value at the
  non-symmetric kinematic point $\tilde{\rm S}$ (involving only
  non-exceptional momenta) characterized by $p^2 = p'^2= k^2 =q^2 = s
  = u = t/2 = - \Lambda^2$.

With this choice, and by virtue of the normalization condition imposed on
$\Gamma_{C}^{(2)} (p,p')$, the non-1PI diagrams (see
Eq.~\eqref{eq:non1PI}) contributing to the three-point function with
insertion of $O_1^{(5)} = C$ on the quark external legs vanish.  In
other words, the amputated Green's function coincide with the 1PI
Green's functions up to a non-1PI term arising from operator
insertion on the gluon external leg.  This non-1PI term does not
project on the spin/Lorentz structures that we use to impose the
normalization conditions, so for all practical purposes the
renormalization conditions can be imposed on the 1PI vertices.

\item 
We require that  the gluon  and quark  two-point functions with insertion 
$O_5^{(5)}  = (m G \tilde G)$  take their tree-level value at the symmetric  point $S$ given by  $p^2 = p'^2 =q^2 = - \Lambda^2$. 
The condition  on the gluon two-point function involves overall factors of the quark masses.  
While one can  use  quark masses in any scheme, 
we choose to use the quark masses in the $\overline{\rm MS}$ scheme. 
This leads to the simplest matching factors, and  corresponds to imposing the 
subtraction conditions on the operator $G \tilde G$, ignoring the mass factors.

\item  The remaining operators 
are  related to quark bilinears: 
 $O^{(5)}_{2,8-10}$ 
 are related to the  pseudoscalar density,  
 $O^{(5)}_{3}$ is related to the tensor density,
 and  $O^{(5)}_{6-7}$ are related to the divergence of the axial current. 
We exploit this factorized structure and 
impose the ``standard'' RI-SMOM conditions~\cite{Sturm:2009kb} 
on the quark bilinear part.  
The  subtraction condition for   $O^{(5)}_{6-10}$  
involves again overall factors of the quark masses, 
for which we choose the $\overline{\rm MS}$ values. 
This is equivalent to imposing the conditions on the quark bilinears, ignoring the 
overall quark mass factors. 

\end{itemize}

Throughout, we impose the normalization conditions in the chiral limit $m_q \to 0$.  
This is achieved as follows:  (i) we expand two- and three-point  Green's functions in spin-flavor structures, keeping 
explicit powers of the quark mass.   (ii) Through appropriate projections we then isolate the coefficients of the various 
spin-flavor structures,  which are defined for any value of the quark mass. 
(iii) We finally impose normalization coefficients on these coefficient functions in the chiral limit. 
This procedure defines  a mass-independent renormalization scheme. 

This RI scheme, defined in terms of gauge fixed correlation functions
of quark and gluon states in the deep Euclidean region, serves as a
useful intermediary for converting non-perturbative results to those
required for phenomenology. In this work, we only discuss the matching
of this RI scheme to the perturbative {\MSbar} scheme in covariant
gauges.  To complete the program of connecting the {\MSbar} to a
lattice scheme, we also need to calculate the matching between lattice
and this RI scheme. Among the covariant gauges, the Landau gauge is the
most convenient for lattice calculations. The calculation of the
corresponding matrix elements on the lattice can be done either using
lattice perturbation theory, or non-perturbatively.  In fact, matrix
elements with quark external states are used extensively nowadays for
renormalizing lattice operators~\cite{Martinelli:1994ty}.
However,
renormalization of the CEDM operator needs extension of such
calculations to include gluon external states.  Even though gluonic
correlators have long been studied on the
lattice~\cite{Bernard:1990dn}, they are typically noisy.  In addition,
the matrix elements with two quarks and a gluon external state gives
rise to ``four-point'' functions\rlap,\footnote{It is conventional in
  the lattice literature to count the point of operator insertion.}
and there is little experience with calculating these in the lattice
community.

Apart from these difficulties, however, the non-perturbative
evaluation of the matrix elements is theoretically straightforward.
The large number of off-shell operators does not pose a significant
challenge either. In particular, since these operators explicitly
involve the equation of motion, an \(n+1\)-point
function involving them is straightforwardly related to a \(n\)-point
function obtained by exactly canceling an external propagator using
the equation of motion.  With such reductions, the number of
correlation functions that need to be evaluated non-perturbatively are
much fewer than the number of operators in the basis.

\subsubsection{Subtraction conditions on the  quark CEDM}
\label{sect:sub1}

We now give explicitly  the fourteen conditions needed to determine $Z^{\rm RI-\tilde{\rm S}MOM}_{1n}$\rlap.\footnote{This
is in addition to the condition for eliminating  possible power divergences (see Section~\ref{sect:dim3}).} 
We begin with the conditions on the two-point functions with external gluons and photons: 
\begin{subequations}
\bea
\epsilon_{\mu \nu \alpha \beta}  p^\alpha p'^\beta   
\  \Gamma_C^{\mu\nu}   (p,p') \Big\vert_{S} &=& 0
\\
\epsilon_{\mu \nu \alpha \beta}  p^\alpha p'^\beta   \ 
\Gamma_C^{\mu\nu (\gamma)}  (p,p')  \Big\vert_{S} &=& 0~.
\eea
\end{subequations}

The quark-quark Green's function has the following spin-flavor structures, 
\bea
\Gamma_{C}^{(2)} &=&   \alpha_1  \gamma_5 t^a    
+ \alpha_2  \sigma (p, p') \gamma_5 t^a 
 +  \alpha_3 \,    \mathcal M  t^a  \slashed q \gamma_5    
 +  \alpha_4 \,  {\rm Tr} \left[  \mathcal M   t^a  \right]   \slashed q \gamma_5   
 \nn \\
 &+&  \alpha_5 \,   \mathcal M^2  t^a   \gamma_5      
 +  \alpha_6 \,  {\rm Tr} \left[  \mathcal M^2     \right]   \gamma_5     t^a 
 + \alpha_7 \,  {\rm Tr} \left[  \mathcal M   t^a  \right]   \mathcal M  \gamma_5  ~, 
\label{alphas}
\eea
where the $\alpha_i$ are functions of the kinematic invariants.  We impose the
RI-$\tilde{\rm S}$MOM condition that all the $\alpha_i$ vanish at the
symmetric kinematic point $S$ in the chiral limit $m_q \to0$.  This
can be achieved with the following projections (traces are over color,
spin, and flavor indices):
\begin{subequations}
\bea
{\rm Tr}  \left[\Gamma_C^{(2)} \, \gamma_5  \, t^a \right]_{S} &=& 0
\\
  {\rm Tr}  \left[\Gamma_C^{(2)} \, \gamma_5   \sigma_{\mu \nu}  t^a  \right]_{S} &=& 0 
\\
M_2^{-1}
\left(
\begin{array}{c}
{\rm Tr}  \left[ \Gamma^{(2)}_C \, \gamma_5 \slashed q  \,  \mathcal M t^a \right]
\label{mixwithax}
\\
{\rm Tr}  \left[ \Gamma^{(2)}_C \, \gamma_5 \slashed q  \right]   \, {\rm Tr} \left[ \mathcal M t^a \right]
\end{array}
\right)_{S}   &=&0 
\\
M_3^{-1}
\left(
\begin{array}{c}
{\rm Tr}  \left[ \Gamma^{(2)}_C \, \gamma_5     \,  \mathcal M^2  t^a \right]
\\
{\rm Tr}  \left[ \Gamma^{(2)}_C \, \gamma_5   t^a  \right]   \, {\rm Tr} \left[ \mathcal M^2 \right]
\\
{\rm Tr}  \left[ \Gamma^{(2)}_C \, \gamma_5   \mathcal M  \right]   \, {\rm Tr} \left[ \mathcal M t^a \right]
\end{array}
\right)_{S}   &=&0 ~,
\eea
\end{subequations}
where the matrices  $M_2$ and $M_3$  are given by (here ${\rm Tr}_F$ denotes the 
trace over flavor indices only):
\be
M_2 = 
\left(
\begin{array}{cc}
{\rm Tr}_F  \left[ (\mathcal M t^a)^2 \right]   &   \left( {\rm Tr}_F \left[ \mathcal M t^a  \right] \right)^2   \\  
\left(  {\rm Tr}_F \left[ \mathcal M t^a  \right] \right)^2    & \ \  n_F \left(  {\rm Tr}_F \left[ \mathcal M t^a  \right] \right)^2 
\end{array}
\right)
\label{m2}
\ee
and 
\be
\label{m3}
M_3 = 
\left(
\begin{array}{ccc}
{\rm Tr}_F  \left[ (\mathcal M^2  t^a)^2 \right]   
&   
\ \ {\rm Tr}_F  \mathcal M^2 \,   {\rm Tr}_F \left[  \mathcal M^2 (t^a)^2   \right] 
& 
\ \ {\rm Tr}_F  \left[ \mathcal M    t^a \right] \,   {\rm Tr}_F \left[  \mathcal M^3   t^a   \right] 
\\  
{\rm Tr}_F  \mathcal M^2 \,   {\rm Tr}_F \left[  \mathcal M^2 (t^a)^2   \right] 
&
\ \ \left({\rm Tr}_F   \mathcal M^2\right)^2  {\rm Tr}_F \left[ t^a t^a \right]  
& 
\ \ {\rm Tr}_F  \mathcal M^2 \,  \left( {\rm Tr}_F \left[  \mathcal M  t^a \right]  \right)^2   
\\
{\rm Tr}_F  \left[ \mathcal M    t^a \right] \,   {\rm Tr}_F \left[  \mathcal M^3  t^a   \right] 
&
\ \ {\rm Tr}_F  \mathcal M^2 \,  \left( {\rm Tr}_F \left[  \mathcal M  t^a   \right]  \right)^2  
&
\ \ {\rm Tr}_F  \mathcal M^2 \,  \left( {\rm Tr}_F \left[  \mathcal M  t^a   \right]  \right)^2  
\end{array}
\right)~.
\ee
The above projections  work for non-degenerate  quark masses  ($m_u \neq m_d \neq m_s$). 
In the isospin limit $m_u = m_d$ the matrices $M_2$ and $M_3$ become singular. 
In Appendix~\ref{app:projections} we describe the projections needed in this case. 

To express the subtraction  conditions on  the quark-quark-gluon three-point function we restore the color 
and flavor indices  of these objects. Recalling that 
$ \Gamma_C^{(3)}$ is proportional to $t^a T^b$ where
$t^a$ is a matrix in flavor space \footnote{We use $t^0 = 1/\sqrt{6} I_{3 \times 3}$ so that  ${\rm Tr}_F  (t^a t^a) = 1/2$ for $a=0,3,8$.}
while $T^b$ is a color generator,  we will use the notation $\Gamma_C^{(3)} \to  \Gamma_C^{(3), ab}$. 
The conditions then read (there is no summation over $a$ and $b$)
\begin{subequations}
\bea
\frac{1}{   i \epsilon (\epsilon^*, k, p, p') }\, 
{\rm Tr} \left[ \Gamma_C^{(3),ab} \, \sigma(p,p')  t^a  T^b \right]_{\tilde S}  &=& 2 g^{\overline{\rm MS}}
\\
{\rm Tr} \left[ \Gamma_C^{(3),ab} \, \sigma(k,p+p')  t^a  T^b \right]_{\tilde S}  &=& 0 
\\
{\rm Tr} \left[ \Gamma_C^{(3),ab} \,  \gamma_5  t^a  T^b \right]_{\tilde S}  &=& 0 ~.
\eea
\end{subequations}
Note that in the first condition above, we could have used the renormalized value 
of the strong coupling constant in any renormalization scheme. 
The use of  $g^{\overline{\rm MS}}$ makes the connection between RI-$\tilde{\rm S}$MOM 
and  $\overline{\rm MS}$ schemes simpler. 

Finally,  we impose the following conditions on the quark-quark-photon three-point function:
\begin{subequations}
\bea
{\rm Tr} \left[ \Gamma_C^{(3,\gamma)} \, \sigma(p,p') Q  t^a   \right]_{\tilde S}  &= & 0 
\\
{\rm Tr} \left[ \Gamma_C^{(3,\gamma)} \,  \gamma_5  Q t^a  \right]_{\tilde S}  &=& 0 ~.
\eea
\end{subequations}

\subsubsection{Subtraction conditions on the remaining operators} 
\label{sect:sub2}

We give here the subtraction conditions needed to determine the remaining entries of 
$Z^{\rm RI\hbox{-}\tilde{\rm S}MOM}_{ij}\!$.
For the operator $O_{5}^{(5)} = (m G \tilde G)$ we prescribe 
\begin{subequations}
\begin{align}
- \frac{1}{6 \Lambda^4} \epsilon_{\mu \nu \alpha \beta}  p^\alpha p'^\beta   
\  \Gamma_{O_5^{(5)}}^{\mu\nu}  (p,p') \Big\vert_{S} & =    \, {\rm Tr}_F \left[ \mathcal M^{\overline{\rm MS}} t^a \right]
\label{CondGGtilde}   \\
{\rm Tr} \left[  \Gamma_{O_5^{(5)}}^{(2)}   \gamma_5 \slashed q \right]_{S} & =  0~.
\end{align}

The remaining operators  $O^{(5)}_{2,3,6-10}$   are  related to quark bilinears, 
and we wish to impose the ``standard'' RI-SMOM conditions~\cite{Sturm:2009kb}. 
$O^{(5)}_{2,6-10}$  have a simple factorized form, and 
the normalization conditions of Ref.~\cite{Sturm:2009kb} are equivalent to: 
\begin{align}
\frac{i}{6 q^2} {\rm Tr} \left[   \Gamma_{O_2^{(5)}}^{(2)} \gamma_5 t^a \right]_{S} &= 1
\label{CondP} \\
\frac{1}{12 q^2} {\rm Tr} \left[  \Gamma_{O_6^{(5)}}^{(2)}   \gamma_5 \slashed q \right]_{S} & =  
n_F  {\rm Tr}_F \left[ \mathcal M^{\overline{\rm MS}} \,  t^a\right]
\label{CondA0}\\
\frac{1}{12 q^2} {\rm Tr} \left[  \Gamma_{O_7^{(5)}}^{(2)}   \gamma_5 \slashed q  \, \mathcal M \right]_{S} & = 
{\rm Tr}_F   \left[  \left(  \mathcal M^2 \right)^{ \overline{\rm MS}}  \, t^a
- \frac{1}{3}  \mathcal M^{\overline{\rm MS}} \,   {\rm Tr}_F \left[ \mathcal M^{\overline{\rm MS}} \,  t^a\right] \right] 
\label{CondA}\\
\frac{1}{12 i } {\rm Tr} \left[  \Gamma_{O_{8}^{(5)}}^{(2)}   \gamma_5 \right]_{S}  & =
  {\rm Tr}_F  \left[   \left( \mathcal M^2 \right)^{\overline{\rm MS}}  t^a \right] 
\\
\frac{1}{12 i } {\rm Tr} \left[  \Gamma_{O_{9}^{(5)}}^{(2)}   \gamma_5  t^a  \right]_{S}  & =
\frac{1}{2} \, {\rm Tr}_F  \left[   \left( \mathcal M^2 \right)^{\overline{\rm MS}}     \right] 
\\
\frac{1}{12 i } {\rm Tr} \left[  \Gamma_{O_{10}^{(5)}}^{(2)}   \gamma_5 \right]_{S}  & =
  {\rm Tr}_F  \left[   \mathcal M^{\overline{\rm MS}}  t^a  \right]  \ {\rm Tr}_F 
  \left[  \mathcal M^{\overline{\rm MS}}  \right] ~.
\end{align}
\end{subequations}

The operator  $O^{(5)}_{3}$  is related to the tensor density, 
but contains an explicit photon field strength.  
One would be tempted to impose the following condition on the quark-quark-photon matrix element
\be
\label{eq:vcT1}
\frac{1}{12 \, i \,  \epsilon (\epsilon^*, k, p, p') {\rm Tr}  \left[ (Q t^a)^2\right] } 
 \,   {\rm Tr} \left[   \Gamma_{O_3^{(5)}}^{(3, \gamma)} \  \sigma (p,p') \  Q t^a \right]_{S}  = 2 e^{\overline{\rm MS}}~,
\ee
which effectively fixes the  projection on the structure $\sigma (\epsilon^*, k) \gamma_5$  to its tree-level value. 
However,  in terms of matrix elements of the tensor density, this prescription 
corresponds to 
\be
\label{eq:vcT2}
{\rm Tr} \left[  \Gamma^{\mu \nu}_T (p,p')  \, \gamma_5 \sigma (p,p') \right]_{S} =  12 i \, \epsilon^{\mu \nu \alpha \beta} p_\alpha p'_\beta ~, 
\ee
which differs   from the standard one \cite{Sturm:2009kb}
 \be
 \label{eq:Taoki}
{\rm Tr} \left[  \Gamma^{\mu \nu}_T (p,p')  \, \sigma_{\mu \nu}  \right]_{S} =  144~
\ee
and would lead to a finite difference in the renormalization factors \footnote{
Note that in the free theory ($\Gamma_T^{\mu \nu} \to \sigma^{\mu \nu}$) both 
Eq.~\eqref{eq:vcT2} and Eq.~\eqref{eq:Taoki} hold.  However,  when including interactions 
a difference arises:   the projection Eq.~\eqref{eq:vcT2}  selects the $\sigma^{\mu \nu}$ component of 
$\Gamma_T^{\mu \nu}$, while 
the projection Eq.~\eqref{eq:Taoki} picks up not only $\sigma^{\mu \nu}$  but also 
additional terms in $\Gamma_T^{\mu \nu}$,  such as 
$\sigma^{\alpha  \beta} p_\alpha p'_\beta (p^\mu p'^\nu - p^\nu p'^\mu)$. 
}. In our analysis we stick to the standard normalization condition Eq.~\eqref{eq:Taoki}.
This can be obtained by imposing Eq.~\eqref{eq:vcT1} 
while performing  a finite shift $\delta k_T$  in the loop factor $k_T$ 
given in Eq.~\eqref{eq:kT},  namely 
\be
\delta k_T=  C_F (1 - \xi)  \left(\frac{2}{3} - \frac{1}{3} \psi \right)~. 
\ee

\subsection{Matching
  RI-\texorpdfstring{$\tilde{\bf S}$}{S\textcomposetilde}MOM and
  \texorpdfstring{$\overline{\bf MS}$}{M\textcomposemacron
    S\textcomposemacron} operators}

We now determine the conversion matrix appearing in Eq.~\eqref{eq:match1}
\be
C_{ij}  =  \left(  \left(Z^{\rm RI-\tilde{\rm S}MOM}\right)^{-1}  \cdot  \ Z^{\overline{\rm MS}}  \right)_{ij} ~
\ee
to first order in $\alpha_s$. 
Denoting  field renormalization and renormalized amputated Green functions  of  any  operator $O$ 
in the  RI-$\tilde{\rm S}$MOM scheme  with $\tilde{Z}_{q,G}$ and   $\tilde{\Gamma}_O$, respectively, 
and the corresponding quantities  in the $\overline{\rm MS}$ scheme with 
$Z_{q,G}$  and ${\Gamma}_O$,  
the matching conditions take the form:
\begin{subequations}
\bea
\tilde{\Gamma}^{(2)}_{O_i}   &=&  \  \frac{\tilde{Z}_q}{Z_q}  \  \sum_j  \ C_{ij}  \   {\Gamma}^{(2)}_{O_j}  
\\
\tilde{\Gamma}^{\mu \nu}_{O_i}   &=&  \  \frac{\tilde{Z}_G}{Z_G}  \  \sum_j  \ C_{ij}  \   {\Gamma}^{\mu \nu}_{O_j}  
\\
\tilde{\Gamma}^{(3)}_{O_i}   &=&  \  \frac{\tilde{Z}_q \tilde{Z}_G^{1/2}}{Z_q Z_G^{1/2}}  \  \sum_j  \ C_{ij}  \   {\Gamma}^{(3)}_{O_j}  
\\
\tilde{\Gamma}^{(3, \gamma)}_{O_i}   &=&  \  \frac{\tilde{Z}_q}{Z_q}  \  \sum_j  \ C_{ij}  \   {\Gamma}^{(3, \gamma)}_{O_j}  ~.
\eea
\end{subequations}
When one imposes that the $\tilde{\Gamma}_{O_i}$ satisfy the RI-$\tilde{\rm S}$MOM
subtraction conditions given in subsections~\ref{sect:sub1} and \ref{sect:sub2}, 
one obtains a system  of  linear equations for the $C_{ij}$ matching factors. 

Using the explicit one-loop results of Sections~\ref{sect:loop1} and \ref{sect:loop2}
and the ratios of wave-function renormalization factors, 
\bea
 \frac{\tilde{Z}_q}{Z_q}  
 &\equiv & 1+  \frac{\alpha_s}{4 \pi} r_q   
= 1  - \frac{\alpha_s}{4 \pi} \ C_F \xi  \left[ 1 + \log \frac{\mu^2}{\Lambda^2} \right] 
 \\
  \frac{\tilde{Z}_G}{Z_G}
 &\equiv & 1+  \frac{\alpha_s}{4 \pi} r_G   
 \nn \\
&=&    1  +  \frac{\alpha_s}{4 \pi} \left[ C_A \left( \frac{97}{36} + \frac{\xi}{2}  +\frac{\xi^2}{4} \right) 
   - \frac{20}{9}   n_F T_F 
+   \left[   C_A  \left(\frac{13}{6} - \frac{\xi}{2} \right)      - \frac{4}{3} n_F  T_F   \right] \, \log \frac{\mu^2}{\Lambda^2}    \right], \ \ \ 
\eea
we solve for the $C_{ij}$. 

To  $O(\alpha_s)$ the matching coefficients have the structure
\be
C_{ij} \equiv  \delta_{ij}   \ + \ \frac{\alpha_s}{4\pi} \left[ c_{ij}  +  z_{ij}  \, \log \frac{\mu^2}{\Lambda^2}  \right] ~, 
\ee
corresponding to the   RI-$\tilde{\rm S}$MOM renormalization matrix 
\be
 Z_{ij}^{\rm RI-\tilde{\rm S}MOM}   = \delta_{ij}   -  \frac{\alpha_s}{4 \pi}  \, \left[
 z_{ij} \left(\frac{1}{\epsilon}  + \log \frac{\mu^2}{\Lambda^2} \right)
 %\frac{z_{ij}}{\epsilon} +  z_{ij}  \, \log \frac{\mu^2}{\Lambda^2}  
  + c_{ij}     \right] ~.
\ee
We have given the pole terms  $z_{ij}$ in Section~\ref{sect:5}, while the constants $c_{ij}$  can be expressed 
in terms of the loop factors  $r_{q,G}$ defined above 
and  $f_{0,1,2}$, $k_{1,2,3}$, $k_{1,3}^{(\gamma)}$, $k_G$, $\tilde{k}_G$, and $k_{A,P,T}$ 
defined in Sections~\ref{sect:loop1} and \ref{sect:loop2}.
We find for the first row $c_{1n}$:
\begin{subequations}
\bea
c_{11} &=&  - k_1 - \frac{1}{2} k_2  + \frac{1}{2} k_3  - r_q - \frac{1}{2} r_G
\\
c_{12} &=& 2 f_0  + k_2 - k_3 
\\
c_{13} &=&  - k_1^{(\gamma)}  + \frac{k_3 - k_2}{2} 
\\
c_{14} &=& 0 
\\
c_{15} &=& - 4 
\\
c_{16} &=&   \frac{k_2 - 2\, f_1}{3} 
\\
c_{17} &=& k_2 - 2\,  f_1
\\
c_{18} &=& - 2 f_2 - k_2  - k_3
\\
c_{19} &=& 0 
\\
c_{1,10} &=& 0 
\\
c_{1,11} &=& k_2 + k_3 
\\
c_{1,12} &=&  - \frac{k_2 + k_3}{2}
\\
c_{1,13} &=& - k_2 
\\
c_{1,14} &=& k_3^{(\gamma)} - k_3~.
\eea
For the remaining non-zero entries of $c_{ij}$ we find:
\bea
c_{22} &=& - k_P - r_q
\\
c_{33} &=& - k_T - \delta k_T - r_q
\\
c_{55} &=&  - \kappa_G - r_G
\\
c_{56} &=& - \tilde{\kappa}_G
\\
c_{66} &=& c_{77}= - k_A - r_q
\\ 
c_{88} &=& c_{99} = c_{10,10} = - k_P - r_q  ~.
\eea
\end{subequations}

In Appendix~\ref{app:Matching}, we report explicit results for the
matching coefficients $c_{ij}$ using both the HV and NDR prescriptions
for $\gamma_5$.

\section{Renormalization and the  axial  Ward Identities}
\label{sect:WI}

In the previous section we have imposed a set of subtraction conditions 
on  CP-odd operators of dimension five, some of  which  are related to 
the axial current ($O_{6-7}^{(5)}$),
 the pseudoscalar density ($O_{8-10}^{(5)}$), 
 and $G \tilde G$   ($O_{5}^{(5)}$). 
So far we have not  discussed  whether the resulting finite operators 
satisfy the non-singlet and singlet axial Ward Identities (WIs). 
In particular, the normalization conditions on 
the singlet $A$, $P$, and $G \tilde G$ may be inconsistent with the 
singlet WIs. For the non-singlet case,  RI-SMOM subtraction conditions 
have been shown to be consistent with the WIs~\cite{Sturm:2009kb}. 

In general one can  obtain   properly normalized  symmetry currents  
through the Ward Identity method,   discussed in Refs.~\cite{Luscher:1996sc,Bhattacharya:2005rb}.
Moreover,  in Ref.~\cite{Sturm:2009kb} the  RI-SMOM conditions were suitably chosen so that 
they are consistent with the non-singlet axial WIs.   
Here  we take a different point  of view: 
we  discuss  how to define  renormalized  (singlet and non-singlet) axial current  
and pseudoscalar density operators  that satisfy the axial WIs, 
starting from an arbitrary  subtraction scheme,  such as $\overline{\rm MS}$ or the RI-$\tilde{\rm S}$MOM 
scheme defined in Section~\ref{sect:RISMOM}.   
We put forward a  two-step approach:
\begin{enumerate}
\item Using any regulator and any subtraction scheme,  
define renormalized  (finite)  axial  ($A_\mu$), pseudoscalar ($P$) and $G \tilde G$ operators.  
\item  Starting from any of the above schemes, perform   
a finite renormalization that  leads to  operators 
$A_\mu$, $P$ and $G \tilde G$ that obey properly normalized WIs.  
The resulting $A_\mu$  is the ``symmetry current''  associated with axial transformations. 
We may call this  new scheme the ``WI scheme''. 
\end{enumerate}
In the case of $\overline{\rm MS}$ and the RI-$\tilde{\rm S}$MOM scheme defined in the previous section 
we provide  the explicit matching  factors  to the WI scheme to $O(\alpha_s)$.  
We will also  describe the procedure to obtain non-perturbative matching factors 
connecting the RI-$\tilde{\rm S}$MOM and WI schemes. 

Our discussion is inspired by  the  analysis  of Refs.~\cite{Collins,Bos:1992nd} for a dimensionally regulated theory 
and of  Refs.~\cite{Bochicchio:1985xa,Testa:1998ez} for a lattice regulated theory. 
While we give details pertaining to the dimensionally regulated theory,  
our aim is to point out that the general features of the analysis are ``RI\rlap,'' {\it i.e.}, regularization independent. 
Therefore, we will draw parallels with discussions of the axial current in various 
lattice QCD formulations~\cite{Karsten:1980wd,Bochicchio:1985xa,Furman:1994ky} in appropriate places.

\subsection{PCAC relation in terms of bare operators}

We focus on the singlet axial current for concreteness. 
A discussion of the  non-singlet current in the context of dimensional regularization 
and minimal subtraction is presented in  Ref.~\cite{Collins}, 
and the relevant  results are a special case of the  analysis  presented below.  
%
%can be inferred from our  analysis  presented below.  
%
In terms of suitably  regularized operators, the PCAC relation takes the form 
\be
\partial \cdot A  =  2  (m P)  +   P_E    +  X~.
\label{eq:A1}
\ee
In dimensional regularization  the bare operators take the form
$A_\mu = \bar{\psi} (1/2)  [\gamma_\mu, \gamma_5] \psi$, 
$(m P) \equiv    \bar{\psi} {\cal M}  i \gamma_5  \psi$, 
$P_E  =  \bar{\psi}_E  i \gamma_5  \psi +    \bar{\psi}   i \gamma_5  \psi_E$. 
$X$ is the anomaly operator,   whose tree-level insertions in Green's functions vanish as one removes 
the regulator ($d \to 4$ in dimensional regularization  or $a \to 0$ in the lattice theory).
In the dimensionally regulated theory, with HV prescription for the $\gamma_5$, one has
\be
X =  \frac{1}{2}   \bar{\psi}  \left\{ \gamma_5 ,   \overrightarrow{\slashed{D}}   -   \overleftarrow{\slashed{D}}  \right\}   \psi~, 
\ee
which clearly vanishes at the classical level  in $d=4$ due to the anti-commutation properties of $\gamma_5$.  
For $d \neq 4$ this operator is non-vanishing and through divergent quantum corrections it can leave a finite remnant  
in Green's functions,  including anomalous terms in the axial current conservation equation. 
In NDR, $X$  always  vanishes. For this reason, NDR does not ``see'' the axial anomaly, and cannot consistently be used for the discussion of the singlet axial current.
In the lattice theory with Wilson fermion discretization $X$ is the  variation of the Wilson term 
under axial transformation~\cite{Karsten:1980wd,Bochicchio:1985xa}, and its properties are
similar to those of $X$ in the HV scheme. 

The anomalous term $X$ can be expressed as a linear combination of  other regulated operators 
with same quantum numbers  and an evanescent operator $\bar{X}$, whose 
insertions in Green's functions with arbitrary number of fields vanish at the quantum level   
as one removes the regulator. 
To perform the projection on non-evanescent operators one defines~\cite{Bochicchio:1985xa}
\be
\bar{X}    =   X    \ + \  \alpha  \ \partial \cdot A    \ + \ \beta  \  2  (m P)    \ + \   \gamma \ G \tilde{G} ~, 
\label{eq:Xeps}
\ee
and determines  the coefficients $\alpha, \beta, \gamma$  
perturbatively or non-perturbatively by requiring that 
appropriate projections of matrix elements of $\bar{X}$ in  quark and gluon states  
%insertions 
% the insertions of $\bar{X}$ 
%in  quark states 
(and their derivative with respect to the mass)  
%and gluon states 
vanish \footnote{Insertions of $X$
can give non-vanishing results only in Green's functions with positive superficial degree of divergence. 
Of these,  one needs to analyze only the one with two quarks and the one with  two gluons, 
as the others do not provide independent information.}
\bea
\langle q |    \bar{X}   | q \rangle \Big \vert_{\slashed{q} \gamma_5} & = &  0 \\
\frac{\partial}{\partial m}  \langle q |    \bar{X}  | q \rangle \Big \vert_{ \gamma_5} & = &  0 \\
\langle g |  \bar{X} | g \rangle  & = &  0 ~.
\eea
Analyzing Green's functions of $X$ with two quarks and with two gluons to one-loop in perturbation theory
in the $\overline{\rm MS}$-HV scheme,  we find
\bea
\Gamma^{(2)}_{X}  &=&   \frac{\alpha_s}{4 \pi}    \ 4 C_F \,   \bigg[  i \slashed{q} \gamma_5   \ 
   \ - \   4 \, i  {\cal M}   \gamma_5  \bigg] 
 \\
\Gamma^{(2) \mu \nu}_{X}  &=& n_F  \frac{\alpha_s}{4 \pi}   \  4 \epsilon^{\mu \nu \alpha \beta}    p_\alpha p_\beta^\prime~, 
\eea
leading to 
\bea
\alpha &=& - 4 C_F  \frac{\alpha_s}{4 \pi}  \label{alpha}
\\
\beta  &=&  8 C_F  \frac{\alpha_s}{4 \pi}  \label{beta}
\\
\gamma  &=&  - n_F  \frac{\alpha_s}{4 \pi}  ~. \label{gamma}
\eea
Using Eq.~\eqref{eq:Xeps} into Eq.~\eqref{eq:A1} one  gets the final result
\be
(1 + \alpha)  \partial \cdot A =  2 \,  (1 - \beta)   (m P) +      P_E  -   \gamma G \tilde G   +  \bar{X} ~,  
\label{eq:A2}
\ee
which is still expressed in terms of bare operators and couplings. 
The non-singlet case is now straightforward: 
one finds the same values of $\alpha$ and $\beta$,  and  the non-singlet anomalous operator $\bar X^{a}$ does not have  a $G \tilde G$ component.

\subsection{PCAC relation in terms of renormalized operators}

We next express the PCAC relation in terms of renormalized operators $[O]_i$, 
related to bare operators $O_j$ via  (note that in this section we use a 
different notation compared to Eq.~(\ref{eq:ren1}))
\be
%[O]_i   =  (Z^{-1})_{ij}  \, O_j
O_i   =  Z_{ij}  \, [O]_j
\ee
with $Z_{ij}$ given in an arbitrary scheme. 
For the operators of interest,    we have the mixing structure\footnote{This is valid in schemes in which 
$m_{u,d,s}$ are multiplicatively renormalized with the same constant $Z_m$.}
\be
\left(\begin{array}{c}
\, G \tilde G  \\  \,  \partial \cdot A  \\
(m  P)
\end{array}\right)
= 
\left(\begin{array}{ccccc}
 Z_{G \tilde G}   &    & Z_{G\tilde G, \partial A}  &     & 0   \\
0 &  &  Z_A   & & 0 \\
0 &  & 0 &  &    Z_m   Z_P
\end{array}\right)
\, 
\left(\begin{array}{c}
 \left[ G \tilde G \right] \\
\left[  \partial  \cdot A \right]   \\
\left[ m P \right]
\end{array}\right)~~. 
\label{eq:Zstructure3}
\ee
Using Eqs.~\eqref{alpha}, \eqref{beta}, \eqref{gamma}, and \eqref{eq:A2} leads to  the renormalized PCAC relation
\bea
\overline{C}_1 (g^2)  [\partial \cdot A]  &=& 
 \overline{C}_2 (g^2)  \ 2   [m P]   +  \, 
\overline{C}_3 (g^2)  \frac{n_F}{16 \pi^2}  \, [g^2 \, G \tilde G]   +    P_E  + \bar{X}~, 
\label{eq:A3}
\eea
with  coefficients, in terms of the bare coupling $g$, 
\begin{subequations}
\label{eq:Cbar}
\bea
\overline{C}_1 (g^2)  &=& Z_A (1 + \alpha)  + \gamma  \, Z_{G \tilde G, \partial A} 
\\ 
\overline{C}_2 (g^2)  &=&   Z_P Z_m  \, (1 -   \beta )
\\
\overline{C}_3 (g^2)  &=&  
- \frac{ 16 \pi^2 \ \gamma}{n_F \, g^2} \,  Z_g^2  \, Z_{G \tilde G}
\eea
\end{subequations}
satisfying $\overline{C}_{1,2,3} (0) = 1$.   
As a consequence  of the finiteness of the EOM operator, $[P_E] = P_E$, and of the independence of the operators $[\partial \cdot A]$, $[m P]$ and $[G \tilde G]$,
$\overline{C}_{1,2,3} (g^2)$ must be finite.

\subsection{Finite renormalization and WIs}

Eq.~\eqref{eq:A3} shows that  in a given renormalization scheme
({\it i.e.}  a  choice of  $Z_{A,P}$, $Z_{m}$, $Z_{G \tilde G}$, $Z_{G \tilde G, \partial \cdot A}$ that makes the 
operator insertions finite)  the renormalized quantities do not necessarily satisfy 
properly normalized (anomalous) WIs:  $\overline{\rm MS}$-HV scheme is one example. 
However, given the scheme-dependent $\overline{C}_{1,2,3} (g^2)$,  
through a finite renormalization one can restore the WIs,  as seen from Eq.~\eqref{eq:A3}. 
Operators in the  ``WI scheme''  are defined by:
\bea
\left[A^\mu \right]_{\rm WI} &=&   \overline{C}_1 (g^2) \ [A^\mu]   
\\
\left[m P\right]_{\rm WI} &=&   \overline{C}_2 (g^2) \ [m P] 
\\
\left[g^2  G \tilde G \right]_{\rm WI} &=&   \overline{C}_3 (g^2) \ [g^2 G \tilde G] ~.
\eea
Applying the operator 
% $\mu d/d \mu$ 
 $d/d (\log \mu)$ 
to both sides of
Eq.~\eqref{eq:A3}, using the finiteness of $P_E$ and the independence of the remaining operators, 
one obtains a set of differential equations for $\overline{C}_{1,2,3} (g^2)$. 
The solution  reveals that the coefficients $\overline{C}_{2,3} (g^2)$ 
are such that  $[m P]_{\rm WI}$ and $[g^2 G \tilde G]_{\rm WI}$ have vanishing diagonal anomalous dimension to all orders. 
On the other hand, $\overline{C}_{1} (g^2)$ is such that 
$[A_\mu]_{\rm WI}$ has an anomalous dimension starting at $O(g^4)$,  
related to the off-diagonal 
anomalous dimension 
$\gamma_{G \tilde G, \partial \cdot A} =  - (Z^{-1}  \, dZ/d (\log \mu) )_{G \tilde G, \partial \cdot A}$, 
namely  $\gamma_{A_{WI}} = \gamma_{G \tilde G, \partial \cdot A}   \cdot \alpha_s /(4 \pi)$.  
The rescaled operators satisfy the properly normalized PCAC relation: 
\bea
  \partial  \cdot [A]_{\rm WI}  &=& 
  \ 2  [m P]_{\rm WI}   +  \, 
\frac{n_F}{16 \pi^2}  \, [g^2 \, G \tilde G]_{\rm WI}   + i P_E  + \bar{X}~.
\eea
The coefficients needed to  reach  the ``WI'' scheme from the 
 $\overline{\rm MS}$-HV scheme,  to $O(g^2)$  are\footnote{To determine $\overline{C}_3$ we rely on the two-loop calculations of Ref.~\cite{Bos:1992nd}.}
\be
\overline{C}_1 = 1  - 4 C_F \ \frac{\alpha_s}{4 \pi}  \qquad \qquad 
\overline{C}_2  =     1  - 8 C_F \ \frac{\alpha_s}{4 \pi}    \qquad \qquad 
\overline{C}_3 = 1 + O(\alpha_s^2) ~. 
\ee

In the case of the  RI-$\tilde{\rm S}$MOM scheme defined in Sec. \ref{sect:RISMOM},
the perturbative values of $\alpha$, $\beta$ and $\gamma$ are still given by Eqs.~\eqref{alpha}, \eqref{beta} and \eqref{gamma}.
In HV, the conditions given in Eqs.~\eqref{CondP}, \eqref{CondA0}, and \eqref{CondA}, which are the equivalent to the RI-SMOM condition of Ref.~\cite{Sturm:2009kb},   
give 
\begin{eqnarray}
Z_A &=& 1 + \frac{\alpha_s }{4\pi} (4 C_F) \label{AxialRIMOM}\\
Z_P Z_m  &=& 1 +  \frac{\alpha_s }{4\pi} (8 C_F)  \label{PseudoRIMOM}~, 
\end{eqnarray}
where we used the value of $Z_m$  obtained in Ref.~\cite{Sturm:2009kb}
\begin{equation}
Z_m = 1 - \frac{\alpha_s C_F}{4\pi} \left( 4 + \xi - (3 + \xi) \frac{\psi}{2}\right).
\end{equation}
This leads to $\bar{C}_1(g^2) = 1 + \mathcal O(g^4)$ and $\bar C_2(g^2) = 1 + \mathcal O(g^4)$, 
thus showing that singlet and non-singlet axial currents and  pseudoscalar densities are already correctly normalized, up to corrections of $\mathcal O(g^4)$. 
The RI-$\tilde{\rm S}$MOM condition Eq.~\eqref{CondGGtilde} leads to a $G \tilde G$ which is not correctly normalized. However, once 
a definition of  $Z_g$ is given, for example by fixing the three-gluon or quark-gluon vertex at the symmetric point to its tree-level value   \cite{Celmaster:1979km,Braaten:1981dv},
Eq.~\eqref{eq:Cbar} allows one  to define $[G \tilde G]_{\rm WI}$.

Eqs.~\eqref{AxialRIMOM} and \eqref{PseudoRIMOM} differ by a finite piece from the results in
Ref.~\cite{Sturm:2009kb}, which are obtained using NDR and found $Z_A = Z_P Z_m =1$.
The finite pieces in $Z_A$ and $Z_P Z_m$ are crucial in compensating the anomalous dimension 
of the axial current and pseudoscalar density arising from divergences in the $\overline{\textrm{MS}}$-HV two-loop calculation, as can be explicitly verified from the results of Ref.~\cite{Larin:1993tq}.
For the non-singlet axial current,  the cancellation is exact, and the RI-$\tilde{\rm S}$MOM axial current does not have anomalous dimension at $\mathcal O(\alpha_s^2)$. 
In the singlet case, the $\mathcal O(\alpha_s)$ finite piece ensures that the relation  $\gamma_A = \gamma_{G \tilde G, \partial \cdot A} \alpha_s/(4\pi)$ is respected.

While  we have  given explicit  results in perturbation theory
within the $\overline{\rm MS}$-HV and  RI-$\tilde{\rm S}$MOM scheme,  the above 
discussion provides  the  steps needed
to determine the coefficients
$\alpha, \beta, \gamma$ starting from any regulator 
and any scheme. These, in turn, in combination 
with the renormalization factors  of Eq.~\eqref{eq:Zstructure3} 
determine  the finite rescaling factors $\overline{C}_{1,2,3}$
in Eqs.~\eqref{eq:Cbar} needed to obtain renormalized operators that satisfy the axial Ward identities.

\section{Relation to the $\Delta S=1$ chromomagnetic operator}
\label{sect:comparison}

In a recent article~\cite{Constantinou:2015ela}, the renormalization of the strangeness changing 
quark chromo-magnetic dipole moment (CMDM) operator has been studied. In our notation 
the P- and CP-even operator studied in~\cite{Constantinou:2015ela} reads
\be
O_{CM}  =   g\,  \bar\psi t^{\Delta S}   \sigma^{\mu\nu} G_{\mu\nu} \psi~,   \qquad \qquad 
t^{\Delta S} = 
\left(
\begin{array}{ccc}
0 & 0 & 0 \\
0 & 0 & 1\\
0 & 1 & 0 \\
\end{array}
\right) ~.
\ee
Ref.~\cite{Constantinou:2015ela}  
studies the mixing of $O_{CM}$ with lower-dimensional operators non-perturbatively on the lattice, 
and the mixing of  $O_{CM}$  with other dimension-5 operators in perturbation theory 
both in  the lattice and in $\overline{\rm MS}$ schemes. Clearly, a number of  common issues arise in our study and in 
Ref.~\cite{Constantinou:2015ela},  so a closer comparison of  operator basis and mixing results is desirable. 

\subsection{Operator basis}
\label{sect:opbasis}

First, let us focus on the operator basis.  Note that our operator basis was constructed  
assuming diagonal flavor structures, so  that $[\mathcal M, t^a]=  [Q, t^a]=0$ and ${\rm Tr} (\mathcal M t^a) \neq 0$. 
In the case of flavor off-diagonal generators, such as $t^{\Delta S}$,  a number of new operators appears at dimension-5, 
while all the operators involving ${\rm Tr} (M t^a)$ vanish. 
In what follows we provide (i)  the basis of  dimension-5 operators mixing with the  P- and CP-odd CEDM operator  $O_{CE} \equiv C$ defined in 
Eq.~(\ref{eq:Cdef}),  with off-diagonal flavor structure  $t^a \to t^{\Delta S}$;  (ii) the corresponding  
basis for  the P- and CP-even sector (mixing with $O_{CM}$), to be compared with Ref.~\cite{Constantinou:2015ela}. 

For ease of comparison with Ref.~\cite{Constantinou:2015ela}  we omit the operators involving the electromagnetic field, i.e. 
$O^{(5)}_{3}$ and $O^{(5)}_{14}$  of Section~\ref{sect:dim5basis}.  
With this in mind,  for the $\Delta S=1$ sector we find ten independent  CP-odd operators.
In the notation of Section~\ref{sect:dim5basis} (with $t^a \to t^{\Delta S}$),  we find  
$C, \partial^2 P, (m\, \partial \cdot A)_2,  (m^2 P)_{1,2}, P_{EE}, \partial \cdot A_E, A_\partial$ and 
two new structures that vanish for diagonal flavor generators, namely 
$(m^2 P)_4 = \bar\psi i\gamma_5  \left[ \mathcal M,  \left[ \mathcal M,  t^{\Delta S} \right] \right]  \psi$ and 
$ (m P_E)_1 =   \bar\psi_E   \ i \gamma_5  \left[ \mathcal M, t^{\Delta S} \right]  \psi   -  
\bar\psi     \ i \gamma_5  \left[ \mathcal M, t^{\Delta S} \right]  \psi_E  $. 
In order to match more closely the operator basis of Ref.~\cite{Constantinou:2015ela} 
we can trade the operator $(m \,  \partial \cdot A)_2$ involving derivatives of the axial current in favor 
of $ (m P_E)_2 =    \bar\psi_E   \ i \gamma_5  \left\{ \mathcal M, t^{\Delta S} \right\}  \psi   +
\bar\psi     \ i \gamma_5  \left\{ \mathcal M, t^{\Delta S} \right\}  \psi_E $,  via the relation
\be
2 (m \, \partial \cdot A)_2  =   4 (m^2 P)_1 - (m^2 P)_4  + (m P_E)_2~.
\label{eq:mpcac}
\ee
So we end up with the dimension-5 basis in the P- and CP-odd sector reported in the left column of Table~\ref{tab:bases}.
The  corresponding P- and CP-even sector operators can be obtained from the above ones with the substitution $i \gamma_5 \to 1$ 
and are given explicitly in the right  column of Table~\ref{tab:bases}.

\begin{table}[th]
\begin{center}
\setlength{\tabcolsep}{3pt}
\begin{tabular}{|c|c|}
\hline
&  \\[-1\jot]
CP-odd   &   CP-even   \\[1\jot]
\hline
\hline 
&  \\[-1\jot]
 %%%%%% %%%%%% %%%%%% %%%%%% %%%%%% %%%%%% %%%%%% %%%%%%
$
  O_{CE} = ig\,  \bar\psi  \sigma^{\mu\nu} \gamma_5 \, G_{\mu\nu}  t^{\Delta S}   \psi 
$
 &       
 $
O_{CM} = g\,  \bar\psi  \sigma^{\mu\nu} \, G_{\mu\nu}  t^{\Delta S}   \psi 
 $
 \\[1\jot]
 %%%%%% %%%%%% %%%%%% %%%%%% %%%%%% %%%%%% %%%%%% %%%%%%
$
 \partial^2 P =   \partial^2  \left(   \bar\psi i\gamma_5  t^{\Delta S} \psi \right) 
$
&
$
 \partial^2 S =   \partial^2  \left(   \bar\psi  t^{\Delta S} \psi \right) 
$
 \\[1\jot]
 %%%%%% %%%%%% %%%%%% %%%%%% %%%%%% %%%%%% %%%%%% %%%%%%
$
 (m^2  P)_1 = \frac{1}{2}  \, \bar\psi i\gamma_5    \left\{ \mathcal M^2, t^{\Delta S} \right\}   \psi
$
&
$
 (m^2  S)_1 = \frac{1}{2}  \, \bar\psi   \left\{ \mathcal M^2, t^{\Delta S} \right\}   \psi
$
 \\[1\jot]
 %%%%%% %%%%%% %%%%%% %%%%%% %%%%%% %%%%%% %%%%%% %%%%%%
$
  (m^2  P)_2 = \textrm{Tr} \left[\mathcal M^2 \right]  \ \bar\psi i\gamma_5 t^{\Delta S}  \psi
$
&
$
  (m^2  S)_2 = \textrm{Tr} \left[\mathcal M^2 \right]  \ \bar\psi  t^{\Delta S}  \psi
$
 \\[1\jot]
 %%%%%% %%%%%% %%%%%% %%%%%% %%%%%% %%%%%% %%%%%% %%%%%%
 $
    (m^2 P)_4 =  \bar\psi i\gamma_5  \left[ \mathcal M,  \left[ \mathcal M,  t^{\Delta S} \right] \right]  \psi
$
&
$
    (m^2 S)_4 =  \bar\psi  \left[ \mathcal M,  \left[ \mathcal M,  t^{\Delta S} \right] \right]  \psi
$
 \\[1\jot]
 %%%%%% %%%%%% %%%%%% %%%%%% %%%%%% %%%%%% %%%%%% %%%%%%
$
 P_{EE}=  i\bar\psi_E\gamma_5 t^{\Delta S} \psi_E 
$
&
$
 S_{EE}=  \bar\psi_E t^{\Delta S} \psi_E 
$
 \\[1\jot]
 %%%%%% %%%%%% %%%%%% %%%%%% %%%%%% %%%%%% %%%%%% %%%%%%
 $
(m P_E)_1  =    \bar\psi_E  i \gamma_5  \left[ \mathcal M, t^{\Delta S} \right]  \psi   -  
\bar\psi      i \gamma_5  \left[ \mathcal M, t^{\Delta S} \right]  \psi_E   
$
&
$
(m  S_E)_1  =    \bar\psi_E     \left[ \mathcal M, t^{\Delta S} \right]  \psi   -  
\bar\psi       \left[ \mathcal M, t^{\Delta S} \right]  \psi_E   
$
 \\[1\jot]
 %%%%%% %%%%%% %%%%%% %%%%%% %%%%%% %%%%%% %%%%%% %%%%%%
$
(m P_E)_2   =    \bar\psi_E  i \gamma_5  \left\{ \mathcal M, t^{\Delta S} \right\}  \psi   +
\bar\psi      i \gamma_5  \left\{ \mathcal M, t^{\Delta S} \right\}  \psi_E   
$
&
$
(m S_E)_2   =    \bar\psi_E   \left\{ \mathcal M, t^{\Delta S} \right\}  \psi   +
\bar\psi       \left\{ \mathcal M, t^{\Delta S} \right\}  \psi_E   
$
 \\[1\jot] 
 %%%%%% %%%%%% %%%%%% %%%%%% %%%%%% %%%%%% %%%%%% %%%%%%
$
   \partial \cdot A_E  
=  \partial_\mu[\bar\psi_E\gamma^\mu\gamma_5 t^{\Delta S}  \psi    -   \bar\psi \gamma_5  \gamma^\mu t^{\Delta S}  \psi_E]
$
&
$
   \partial \cdot V_E  
=  i \partial_\mu[\bar\psi_E\gamma^\mu    t^{\Delta S}  \psi    -   \bar\psi  \gamma^\mu t^{\Delta S}  \psi_E]
$
 \\[1\jot]
 %%%%%% %%%%%% %%%%%% %%%%%% %%%%%% %%%%%% %%%%%% %%%%%%
$
A_\partial  =
  \bar\psi  \gamma_5  \slashed{\partial} t^{\Delta S}   \psi_E   \ -   \bar {\psi}_E   \overleftarrow{\slashed{\partial}}  \gamma_5   t^{\Delta S}  \psi  
$
&
$
V_\partial  =
  \bar\psi  i  \slashed{\partial} t^{\Delta S}   \psi_E   \ -   \bar {\psi}_E  i  \overleftarrow{\slashed{\partial}}    t^{\Delta S}  \psi  
$
 \\[1\jot]
 %%%%%% %%%%%% %%%%%% %%%%%% %%%%%% %%%%%% %%%%%% %%%%%%
\hline
\end{tabular}
\end{center}
\caption{Operator basis in the CP-odd and CP-even sectors. 
 \label{tab:bases}}
\end{table}

We can now compare our basis to the one in Ref.~\cite{Constantinou:2015ela},  which consists of 
ten dimension-5 operators ${\cal O}_{1, ..., 10}$: 

\begin{itemize}

\item 
For the gauge-invariant operators that do not vanish by the EOM,  
after converting the operators in \cite{Constantinou:2015ela}
from Euclidean to Minkowski metric, we find the correspondence: 
${\cal O}_1 =  O_{CM}$, 
${\cal O}_2 =  2 (m^2 S)_1$, 
${\cal O}_3 =  (m^2 S)_1 -  1/2 (m^2 S)_4$, 
${\cal O}_4 = - \partial^2 S$.
%
%${\cal O}_1  \leftrightarrow  O_{CM}$, 
%${\cal O}_2  \leftrightarrow  2 (m^2 S)_1$, 
%${\cal O}_3  \leftrightarrow  (m^2 S)_1 -  1/2 (m^2 S)_4$, 
%${\cal O}_4  \leftrightarrow  \partial^2 S$.

\item For the operators vanishing by the EOM  we find: 
${\cal O}_5  =   S_{EE}$,  
${\cal O}_{7} =  1/2 [ (m S_E)_1  - ( m S_E)_2] $, 
${\cal O}_{8} =  - 1/2 [ (m S_E)_1  +  ( m S_E)_2 ]$, 
${\cal O}_9 = -  ( \partial \cdot V_E   +   V_\partial)$, 
${\cal O}_{10} =   V_\partial$.

% Give equations here

\item There is no operator in our basis corresponding to ${\cal O}_6$ in  \cite{Constantinou:2015ela}.
The CP-odd counterpart  of ${\cal O}_6$ is 
$\tilde{P}_{EE} =  \bar \psi  i \gamma_5 t^{\Delta S} \psi_{EE} + \bar{\psi}_{EE} i \gamma_5 t^{\Delta S} \psi$,  
and it can be expressed in terms of operators already present in the basis, via:
\be
\partial \cdot A_E =  \tilde{P}_{EE} + 2 P_{EE} + (m P_E)_2~,
\ee
A similar  linear dependence relation holds in the P- and CP-even sector.  
Ref.~\cite{Constantinou:2015ela} finds  at one loop that  ${\cal O}_6$ is not needed to renormalize $O_{CM}$. 
This is consistent with our finding that ${\cal O}_6$ is not linearly independent. 

\item  In Ref.~\cite{Constantinou:2015ela}  the   operator  $(m^2 S)_2 = (m_u^2+m_d^2 + m_s^2)  \bar{s} d$  is absent. 
This operator is allowed by the symmetries of the problem. In perturbation theory it can  mix with $O_{CM}$ 
starting at two-loop order, so its omission does not affect the results of Ref.~\cite{Constantinou:2015ela}.
However, the operator should be included in non-perturbative  renormalization treatments.  

\end{itemize}

\subsection{One-loop renormalization factors}

Using the CP-odd operator  basis of Table~\ref{tab:bases}, 
we have extended our  analysis of the two- and  three-point functions to include  off-diagonal flavor  structures 
and have found the mixing to the additional operators   (in the ${\overline {\rm MS}}$ scheme)
\be
Z_{O_{CE}, (m^2 P)_4} = -Z_{O_{CE}, (m P_E)_1} =  \frac{1}{\epsilon} \frac{\alpha_s}{4 \pi} \frac{3}{8} \left (C_A - 4 C_F \right)~.
\label{eq:newmixing}
\ee
Using (i)  the results  given in Section~\ref{sect:5}  for the operator mixing in our original basis 
(extended to the new structures through  Eq.~(\ref{eq:newmixing})); 
and (ii)  the change of basis implied by  Eq.~(\ref{eq:mpcac}),    we have computed the renormalization  matrix 
relevant to  the CP-odd operators  in Table~\ref{tab:bases}.

In order to compare to Ref.~\cite{Constantinou:2015ela}, we need the  relation between the divergence structure of the 
CP-even and CP-odd sectors.  At one loop we have verified that  the divergences of two- and three-point functions 
with insertion of $O_{CM}$ and $O_{CE}$ are related by a simple operation $ \hat{\tau}$: 
\be
\Gamma_{O_{CM}}    =    \hat{\tau} \left[  \Gamma_{O_{CE}}    \right]~, \qquad 
\hat{\tau}:  \left\{   i \gamma_5 \to 1~, \ \   t^{\Delta S} {\cal M}^n \to  (-1)^n  t^{\Delta S} {\cal M}^n  \ (n=0,1,2)     \ \right\}~. 
\ee
Similarly,  the tree-level insertions of the CP-even (${O}_+$) and CP-odd  ($O_-$)  operators appearing in 
each line of Table~\ref{tab:bases}  are related by $\Gamma_{O_{+}}    =    \hat{\tau} \left[  \Gamma_{O_{-}} \right]$, 
except for the following cases:  
% $\Gamma_{(m S_E)_{1,2}}    =    \hat{\tau} \left[  \Gamma_{(mP_E)_{2,1}} \right]$
\begin{subequations}
\label{eq:relationsdiv}
\bea
\Gamma_{(m S_E)_{1,2}}   & = &   \hat{\tau} \left[  \Gamma_{(mP_E)_{2,1}} \right]
\\
\Gamma_{S_{EE}}   & = &   - \hat{\tau} \left[  \Gamma_{P_{EE}} \right]
\\
\Gamma_{\partial \cdot V_{E}}   & = &   - \hat{\tau} \left[  \Gamma_{\partial \cdot A_E} \right]
\\
\Gamma_{V_{\partial}}   & = &   - \hat{\tau} \left[  \Gamma_{A_\partial} \right]
\eea
\end{subequations}
From the  renormalization matrix  in the CP-odd sector and the relations   (\ref{eq:relationsdiv}),   
we have computed the renormalization factors in the CP-even sector,  in the basis 
of Table~\ref{tab:bases}.  Finally,  converting to the basis ${\cal O}_{1, ..., 10}$ of Ref~\cite{Constantinou:2015ela} 
(using the relations given in  Section~\ref{sect:opbasis}), we  find our   results for the renormalization coefficients  
to agree with Eqs. (66-75) of Ref.~\cite{Constantinou:2015ela}.

\section{Conclusions}
\label{sect:conclusion}

In this work we have studied the off-shell renormalization and 
mixing of CP-odd dimension-5 
operators in QCD in both the $\MSbar$ and RI-$\tilde{\rm S}$MOM schemes 
(the latter amenable to implementation in lattice QCD), 
providing the matching matrix between operators in 
RI-$\tilde{\rm S}$MOM and $\MSbar$ to $O(\alpha_s)$. 

We have  paid special attention to the definition of a finite 
quark CEDM operator in the RI-$\tilde{\rm S}$MOM scheme, 
identifying all the needed subtractions.
This is the first step towards  a lattice QCD calculation of the impact of
the quark CEDM  on the nucleon EDM, 
which is currently afflicted by one order of magnitude uncertainty. 
This paper sets the stage to perform non-perturbative renormalization of 
the CEDM. The next steps in the program involve 
(i) performing exploratory computations of the needed CEDM quark and gluon Green's functions on the lattice, 
and comparing  this method to lattice perturbation theory;
(ii) performing exploratory calculation of the CEDM insertion in the neutron state, correlated with 
the electromagnetic current or in external electric field~\cite{Bhattacharya:2012bf}.

Besides inducing  nucleon EDM, the quark CEDM induces T-odd P-odd  pion-nucleon 
couplings that are  a key input in the computation of  EDMs of both light and heavy nuclei. 
Chiral symmetry  implies that   the T-odd pion-nucleon coupling induced by the quark CEDM 
can be extracted (up to  chiral corrections)   by calculating the baryon mass splittings 
induced by the quark chromo-magnetic dipole moment (CMDM) operator~\cite{deVries:2012ab}. 
In a future publication we will explore the 
non-perturbative  renormalization and mixing in the 
flavor-diagonal CMDM sector and its relation to the CEDM. 

Finally,  a desirable extension of this work involves  studying  the non-perturbative renormalization
and mixing structure of CP-odd  dimension-6 operators,  such as
Weinberg's operator~\cite{Weinberg:1989dx} and  four-quark operators. 
\\

\noindent {\bf Aacknowledgements}  We acknowledge support by the US  DOE Office of Nuclear Physics  and Office of High Energy Physics, 
and by the LDRD program at Los Alamos National Laboratory. 
We thank T. Blum, T.  Izubuchi,  C. Lehner, and  A. Soni for useful discussions. 

\appendix

\section{CP transformation}
\label{app:CPtransformation}

In this Appendix we review  the definition and properties of the CP transformation. 
On the fermion fields \(\psi\), CP is defined as the linear operator ${\cal CP}$:
\begin{eqnarray}
 {\cal CP}^{-1}\psi  {\cal CP} &\equiv & 
\psi^{CP} = i e^{i\phi} \gamma_2 \gamma_0 \psi^* \nonumber\\
          &=& -i e^{i\phi} (\bar\psi\gamma_2^*)^T =   i  e^{i\phi}     \gamma_2  \bar\psi^T \ , 
\label{eq:CP1}
\end{eqnarray}
where \(\psi^* \equiv \psi^{\dagger T}\),   \(\phi\) is an arbitrary phase, and
we are using the convention that \(\gamma_2\) is an antihermitean
matrix\rlap.\footnote{Whenever we need an explicit representation for the $\gamma$ matrices, 
we use that one provided in Ref.~\cite{Peskin:1995ev}.}
Note that the CP transformation for the $U(1)$-transformed  fermion field 
$\tilde \psi =  e^{i \theta} \psi$  looks like Eq.~\eqref{eq:CP1} with 
$\psi \to \tilde \psi$ and $\phi \to \phi + 2 \theta$.  

In addition to this, the CP transformation
changes all vector operators \(v^\mu\) to \(v_\mu\) in the metric with
signature (\(\mathord{+}\mathord{-}\mathord{-}\mathord{-}\)), and  
changes every charge generator $T^a$ to  $(T^{a})^T$.  

Let \(\Gamma_M\)  denote a gamma structure with \(M\)  Lorentz indices 
and  \(O^N\) denote   an operator involving derivatives  with \(N\) Lorentz indices.  Then
\begin{eqnarray}
(\bar\chi \Gamma_M O^N \psi)^{CP} &=& 
  -(\psi^T O^N \Gamma_M^T \gamma_0^T \chi^*)^{CP} \nonumber\\
                                &=&
  -\left[ \left(-i e^{i\phi_\psi} \bar\psi \gamma_2^*\right)
          O_N \Gamma_M^T \gamma_0^T 
          \left(-i e^{-i\phi_\chi} \gamma_2^* \gamma_0^* 
               \chi\right) \right] \nonumber\\
                                &=& e^{i\Delta\phi}
  \left[\bar\psi \left(\gamma_0^\dagger \gamma_2^\dagger \gamma_0 
                       \Gamma_M \gamma_2^\dagger\right)^T O_N 
                       \chi \right]\nonumber\\
                                &=& e^{i\Delta\phi}
  \left[\bar\psi \left(-\gamma_2 \Gamma_M \gamma_2\right)^T
                       O_N\chi\right]\nonumber\\
                                &=& e^{i\Delta\phi}
  \bar\psi \Gamma_M^{CP} O_N \chi\,,
\end{eqnarray}
where \(\Gamma_M^{CP} \equiv (-\gamma_2 \Gamma_M \gamma_2)^T\) and we
have used the hermiticity of \(\gamma_0\), antihermiticity of
\(\gamma_2\), and \(\gamma_0^2 = 1\).  Using  \(\gamma_2^T =
\gamma_2^{\dagger*} = \gamma_2\), we can now write the simpler expression
\(\Gamma_M^{CP} \equiv -\gamma_2 \Gamma_M^{\dagger*} \gamma_2\).  For
the sixteen Clifford  matrices, we then have
\begin{subequations}
\begin{eqnarray}
   1^{CP} &=& 1 \\
   \gamma_5^{CP} &=& -\gamma_5 \\
   \gamma_\mu^{CP} &=& -\gamma^\mu \\
   (\gamma_\mu\gamma_5)^{CP} &=& \gamma_5\gamma^\mu 
                              = -\gamma^\mu\gamma_5 \\
   \sigma_{\mu\nu}^{CP} &=& \sigma^{\nu\mu} 
                       =-\sigma^{\mu\nu}~.
\end{eqnarray}
\end{subequations}

For the equation of motion field  \(\psi_E = (i D^\mu \gamma_\mu - m) \psi\),  the transformation is 
\begin{eqnarray}
\psi_E^{CP} &\equiv& 
    (i {D^\mu}^{CP} \gamma_\mu - m) \psi^{CP} \nonumber\\
            &=& 
    i e^{i\phi} (i D_\mu^* \gamma_\mu - m) \gamma_2 \gamma_0 \psi^* \nonumber\\
            &=& 
    i e^{i\phi} \gamma_2 (- i D_\mu^* \gamma_\mu^* - m) \gamma_0 \psi^* \nonumber\\
            &=& 
    i e^{i\phi} \gamma_2 \gamma_0 (-i D_\mu^* \gamma_\mu^T - m) \psi^* \nonumber\\
            &=& 
    i e^{i\phi} \gamma_2 \gamma_0 (-i D_\mu^* \gamma^{\mu*} - m) \psi^* \nonumber\\
            &=& 
    i e^{i\phi} \gamma_2 \gamma_0 (-i D_\mu^* \gamma^{\mu*} - m) \psi^* \nonumber\\
            &=& 
    i e^{i\phi} \gamma_2 \gamma_0 \left[(i D_\mu \gamma^\mu - m) \psi \right]^* \nonumber\\
            &=& 
    i e^{i\phi} \gamma_2 \gamma_0 \psi_E^*\,,
\end{eqnarray}
where the conjugate of \(D_\mu = \partial_\mu - i g A_\mu^a T^a\) is
defined as \(D_\mu^* = \partial_\mu + i g A_\mu^a T^{a*}\) to take
into account the opposite gauge charge of the antiparticle.  One way
to state this result is that the CP phase is the same for the fields
\(\psi\) and \(\psi_E\), {\it i.e.,} \(\phi_{\psi_E} = \phi_{\psi}\).

Finally, note that the CP transformation on chiral fields $\psi_{L,R} = (1 \mp \gamma_5)/2  \,  \psi$ 
% is given by (up to a phase)
\begin{eqnarray}
  {\cal CP}^{-1}\psi_L \, {\cal CP} &=& i
    e^{i \phi}
    \gamma_2 \bar{\psi_L}^T\,\nonumber\\
  {\cal CP}^{-1}\psi_R \, {\cal CP} &=&  i 
    e^{i \phi}
   \gamma_2   \bar{\psi_R}^T\, ~. 
\end{eqnarray}

\section{BRST symmetry and operator basis}
\label{app:BRST}

A given gauge invariant operator $O$  mixes under renormalization with  
two classes of operators of same (or lower) dimension~\cite{Deans:1978wn,Collins}: (i) ghost-free gauge-invariant operators 
with the same symmetry properties of $O$ that do not vanish by the equations of motion (EOM);  
(ii) ``nuisance'' operators allowed by the solution to the Ward Identities  
associated with the BRST symmetry. These include non-gauge-invariant operators. 
For completeness,  we sketch below  the procedure to obtain the ``nuisance'' operators,  paraphrasing Ref.~\cite{Deans:1978wn}. 

The gauge and fermion Lagrangian density for the $SU(3)_C \times U(1)_{EM}$ group is 
expressed in terms of   physical fields
\(A_\mu^a\), \(A_\mu^\gamma\), \(\psi\), \(\bar\psi\), 
the dynamical ghosts \(c^a\), \(\bar{c}^a\), \(c^\gamma\), \(\bar{c}^\gamma \), and the non-propagating sources for BRST
transformations \(M\), \(\bar M\), \(J_\mu^a\), \(K^a\), \(J_\mu^\gamma \), 
whose properties are summarized  in Table~\ref{tab:BRST}. 
This  Lagrangian is:
\bea
{\cal L}_0 &=&  -\frac{1}{4} G_{\mu \nu}^a G^{a \mu \nu}   -  \frac{1}{2 \xi} \left( \partial \cdot A^a \right)^2
- (J_\mu^a - \partial_\mu \bar{c}^a)  \, D^{\mu, ab}  c_b 
+ \frac{1}{2} g f^{abc}  K^a  c^b c^c
\\
&-& \frac{1}{4} F_{\mu \nu} F^{\mu \nu}   -  \ \frac{1}{2 \xi_\gamma} \left( \partial \cdot A^\gamma \right)^2
- (J_\mu^\gamma - \partial_\mu  \bar{c}^\gamma)  \, \partial^{\mu}  c^\gamma
\\
&+&  \bar{\psi}   \left( i \slashed{D} - m \right) \psi  + \bar{M} \left( -i g  c^a T^a   - i e c^\gamma \right) \psi
+  \bar{\psi} \left( -i g  c^a T^a   - i e c^\gamma \right) \, M ~,
\eea
where 
\bea
G_{\mu \nu}^a &=&  \partial_\mu A_\nu^a - \partial_\nu A_\mu^a + g f^{abc} A_\mu^b A_\nu^c  
\\
F_{\mu \nu}^a &=&  \partial_\mu A^\gamma _\nu - \partial_\nu A^\gamma_\mu
\\
D_\mu^{ab} c^b &=&   \partial_\mu  c^a  + g f^{abc}  A_\mu^b  c^c
\\
D_\mu \psi &=&  \left( \partial_\mu    - i g A_\mu^a T^a   - i e Q A_\mu^\gamma  \right) \, \psi~.
\eea
The action $S$  obtained by adding to the Lagrangian density 
a  set of infinitesimal sources $\Phi$ for   gauge-invariant ghost-free operators $O$   
\be
S =   \int d^4x {\cal L}_0 (x)    + \int d^4x   \ \Phi(x)  O(x)    \equiv  S_0 +  \Phi \cdot O~, 
\ee
is invariant  under the BRST transformations given by: 
\bea
\Delta A_\mu^a &=&  - \frac{\delta S}{\delta J_\mu^a}  \, \delta \lambda
\qquad \qquad  \Delta A_\mu^\gamma =  - \frac{\delta S}{\delta J_\mu^\gamma}  \, \delta \lambda
\\
\Delta c^a &=&   \frac{\delta S}{\delta K^a}  \, \delta \lambda
\qquad \qquad  \ \  \Delta c^\gamma =   0 
\\
\Delta \bar{c}^a &=&   \frac{1}{\xi}  \partial \cdot A^a  \, \delta \lambda
\qquad \ \  \  \ \Delta \bar{c}^\gamma =   \frac{1}{\xi_\gamma}  \partial \cdot A^\gamma  \, \delta \lambda
\\
\Delta \psi_i   &=&   \frac{\delta S}{\delta \bar{M}_i}     \, \delta \lambda
\\
\Delta \bar{ \psi}_i   &=&   \frac{\delta S}{\delta{M}_i}     \, \delta \lambda~, 
\eea
with $\delta \lambda$ an anti-commuting infinitesimal parameter. This invariance leads to the Ward identities for the 
generating functional of 1PI Green's functions, that in particular  imply the following  identity for 
$\hat S\equiv S  +  \int  d^4x  [  1/(2 \xi)  (  \partial \cdot A^a)^2  +  1/(2  \xi_\gamma) ( \partial \cdot A^\gamma)^2  ]$:
\be
\int d^4x \left( 
\frac{\delta\hat S}{\delta A_\mu^\gamma}   \frac{\delta \hat S}{\delta J_\mu^\gamma}  
+ \frac{\delta\hat S}{\delta A_\mu^a}   \frac{\delta \hat S}{\delta J_\mu^a}  
+ \frac{\delta \hat S}{\delta  c^a}   \frac{\delta \hat S}{\delta K^a}  
+ \frac{\delta \hat S}{\delta  \psi_i}   \frac{\delta \hat S}{\delta \bar{M}_i}  
+ \frac{\delta \hat S}{\delta  \bar{\psi}_i}   \frac{\delta \hat S}{\delta M_i}  
\right) = 0 ~.
\label{eq:WIS}
\ee
While $S = S_0 + \Phi \cdot O$ satisfies the Ward identity Eq.~\eqref{eq:WIS}, 
the general solution involves  additional terms.  Writing the general solution
symbolically as 
\be
S = S_0 + \Phi \cdot O  + \Phi \cdot N ~, 
\ee
and working to first order in the external sources (one operator insertion), 
one finds that the nuisance operators $N$ must satisfy:
\bea
\hat{W}  \ \left( \Phi \cdot N  \right) &= & 0 ~, 
\eea
with the operator 
\bea
\hat{W} &=& 
\frac{\delta \hat{S}_0}{\delta A_\mu^\gamma}   \frac{\delta    }{\delta J_\mu^\gamma}  
+  \frac{\delta  \hat{S}_0  }{\delta J_\mu^\gamma}  \frac{\delta }{\delta A_\mu^\gamma}   
+ \frac{\delta \hat{S}_0}{\delta A_\mu^a}   \frac{\delta    }{\delta J_\mu^a}  
+  \frac{\delta  \hat{S}_0  }{\delta J_\mu^a}  \frac{\delta }{\delta A_\mu^a}   
+ \frac{\delta \hat{S}_0}{\delta c^a}   \frac{\delta    }{\delta K^a}  
+  \frac{\delta  \hat{S}_0  }{\delta K^a}  \frac{\delta }{\delta c^a}   
\nn \\
&+&  
 \frac{\delta \hat{S}_0}{\delta \psi_i}   \frac{\delta    }{\delta \bar{M}_i}  
+  \frac{\delta  \hat{S}_0  }{\delta \bar{M}_i}  \frac{\delta }{\delta \psi_i}   
+  \frac{\delta \hat{S}_0}{\delta \bar{\psi}_i}   \frac{\delta    }{\delta  {M}_i}  
+  \frac{\delta  \hat{S}_0  }{\delta  {M}_i}  \frac{\delta }{\delta  \bar{\psi}_i}   ~.
\eea
Since $\hat{W} \hat{W} = 0$, it turns out that 
\be
\Phi \cdot N  =  \hat{W}  \left (\Phi \cdot F \right) 
\ee
where $F$ is a set of operators with 
the same Lorentz property of $O$, same dimension, and ghost number
\(-1\).   After acting with $\hat W$ one sets the sources  \(M,\bar M,K\)   to
zero, and \(J_\mu\) to \(-\partial_\mu \bar c\).

\begin{table}[!t]
\begin{center}
\tabcolsep=1pt
\begin{tabular}{||l||c|c|c|c||c|c|c|c|c||c|}
\hline
         &\(M\)&\({\bar M}^{\strut}\)&\(J_\mu - \partial_\mu \bar c\)&\(K\)&\(\psi\)&\(\bar\psi\)&\(A_\mu\)&\(c\)&\(\bar c\)&\(\partial_\mu\)\\[1\jot]
\hline
Comm.    & +   &     +    &   $ -$    &  +  &   $-$    &     $-$      &    +    &  $- $ &     $-$    &      +         \\[1\jot]
Lorentz  & \(\frac12\) & \(\bar{\frac12}\)  &  1  & 0 & \(\frac12\) & \(\bar{\frac12}\) & 1 & 0 & 0 & 1\\[1\jot]
Color    & \(3\) & \(3^*\) & \(8\)  &\(8\)& \(3\)  & \(3^*\)    & \(8\)   &\(8\)&   \(8\)  & \(0\) \\[1\jot]
Ghost    &\(-1\)& \(-1\)  & \(-1\)  &\(-2\)& \(0\) & \(0\)      & \(0\)   &\(1\)&   \(-1\)  & \(0\) \\[1\jot]
Dim.     &\(\frac52\)&\(\frac52\)&\(3\)&\(4\)&\(\frac32\)&\(\frac32\)&\(1\)&\(0\)& \(2\)  & \(1\) \\[1\jot]
\hline
\end{tabular}
\end{center}
\caption{\label{tab:BRST} Properties for dynamical fields and BRST sources.  
The first row indicates whether the variable is commuting ($+$) or anti-commuting ($-$). The second and third 
row list the transformation under  Lorentz  and color groups. The fourth row gives the ghost number assignments and 
the fifth row lists the mass-dimension.}
\end{table}

We are now ready to classify  the $F$ operators and resulting 
nuisance operators $N$:
\begin{itemize} 

\item At dimension five, the only Lorentz scalars of ghost number \(-1\)
that we can write down are:  
$\bar\psi\chi_\pm \slashnext A M$, 
$\bar\psi\chi_\pm \slashnext A^\gamma M$, 
$\bar M \slashnext A \chi_\pm\psi$, 
$\bar M \slashnext A^\gamma \chi_\pm\psi$, 
$\bar\psi\chi_\pm \slashnext \partial M$,
$\bar M \slashnext \partial \chi_\pm\psi$, 
$\bar M c M$, 
$\bar M c^\gamma M$,
where \(\chi_\pm =  (1 \pm \gamma_5)/2 \) is a chiral projector. 
Acing on these structures with $\hat W$ 
produces the terms
$\bar{\psi}_E   \slashed{A} \chi_\pm  \psi$, 
$\bar{\psi}_E   \slashed{A}^\gamma  \chi_\pm  \psi$, 
$\bar{\psi}_E  \slashed{\partial} \chi_\pm  \psi$, 
$\bar{\psi}   \slashed{A} \chi_\pm  \psi_E$, 
$\bar{\psi}   \slashed{A}^\gamma \chi_\pm  \psi_E$, 
$\bar{\psi}   \slashed{\partial} \chi_\pm  \psi_E$.

In addition, we have the
gauge-invariant ghost-free terms that are not zero by equations of
motion in the massless limit:~$\bar\psi \sigma^{\mu\nu} G_{\mu\nu}\chi_\pm\psi$,
$\bar\psi \sigma^{\mu\nu} F_{\mu\nu}\chi_\pm\psi$,
$ \partial^2  (\bar \psi \chi_\pm \psi)$, 
$\partial_\mu (\bar\psi \sigma^{\mu\nu}D_\nu^*\chi_\pm\psi)$, 
$\partial_\mu (\bar\psi \sigma^{\mu\nu}D^{\vphantom{*}}_\nu\allowbreak\chi_\pm\allowbreak\psi)$.

\item At dimension four, the only Lorentz scalars of ghost number \(-1\)
are:     $\bar M \chi_\pm\psi$, $ \bar\psi\chi_\pm M$, 
$J^{a \mu} A^a_\mu$, 
$J^{\gamma  \mu} A^\gamma_\mu$, and 
$K c$.  The variation of these produce 
$\bar\psi_E  \chi_\pm\psi$, $\bar\psi\chi_\pm  \psi_E$, 
$(D_\nu G^{\nu\mu} A_\mu + g \bar\psi \slashnext A \psi) - g [\partial_\mu\bar c,c]A^\mu$,  
$(\partial_\nu D^{\nu\mu} A^\gamma_\mu + e \bar\psi \slashnext A^\gamma \psi)$,  
$(\partial^\mu \bar c) D_\mu c$, 
$(D_\mu \partial^\mu \bar c) c$.  

The only gauge-invariant
ghost-free operators not zero by equation of motion in the mass-less
limit are: \(G_{\mu\nu}G^{\mu\nu}\) and \(G_{\mu\nu}\allowbreak\tilde
G^{\mu\nu}\).

\item At dimension three and below, there are no ghost number \(-1\) scalars, so
the only operators we need to consider are the gauge-invariant
ghost-free operators that are not zero by the  massless equation of motion.
The only possible such terms are \(\bar\psi\chi_\pm\psi\).
\end{itemize}
Selecting the T-odd and P-odd structures, including 
gauge-invariant ghost-free operators that do not vanish by the EOM, 
and eliminating linearly dependent operators\footnote{
We used the relation 
$\partial_\mu ( \bar\psi\tilde\sigma^{\mu\nu}   \overleftrightarrow D_\nu\psi) = 
-    ( \partial^2  + 4 m^2)  \, P   -    \partial \cdot  A_E    - 2  m P_E$, 
to eliminate one T-odd, P-odd structure. Moreover, there are no T-odd and P-odd 
operators containing the ghost fields up to dimension five.}
we arrive at the basis presented in Section~\ref{sect:basis}.

\section{Axion Mechanism}
\label{app:axion}

A very elegant way to dynamically set $\bar\theta$ to zero is the 
Peccei-Quinn (PQ) mechanism \cite{Peccei:1977hh},
which predicts the existence of a new light particle, the axion \cite{Weinberg:1977ma,Wilczek:1977pj}.
We follow here the discussion of the PQ mechanism in the EFT framework of  Ref.~\cite{Georgi:1986df}.
A common feature of axion models is the existence of a $U_{\textrm{PQ}}(1)$ symmetry, which is spontaneously broken at high energy.
The axion is the Goldstone boson of the symmetry, and, under $U_{\textrm{PQ}}(1)$, it changes by an additive constant, $a \rightarrow a + c$,
while the SM fields are chosen to be invariant. 
At low energy, around the QCD scale, the Lagrangian includes derivative couplings of the axion to the quarks, which respect $U_{\textrm{PQ}}(1)$. 
Furthermore, the symmetry is explicitly broken by the anomalous coupling to  $G \tilde G$ \cite{Georgi:1986df}, so that the quark-axion Lagrangian has the form  
\begin{eqnarray}\label{eq:axion1}
\mathcal L & = &  \bar\psi i \slashed{D} \psi + \frac{1}{2}\partial_{\mu} a \, \partial^{\mu} a + \bar \psi \left(C_{0} + C_{1} \tau_3 \right) \gamma^5 \gamma^{\mu} \psi\, \partial_{\mu} \frac{a}{f_a} 
-c_{a \gamma \gamma}\frac{e^2}{32 \pi^2}   \frac{a}{f_a}    F  \tilde{F} 
\nonumber \\ & &
-\frac{g^2}{32 \pi^2} \left( \theta +  \frac{a}{f_a} \right)     G  \tilde{G} 
-  e^{i \rho} \bar\psi_L  \mathcal{M} \psi_R - e^{-i \rho} \bar\psi_R  \mathcal M  \psi_L
,
\end{eqnarray} 
where  $f_a$ is the axion decay constant. 
The couplings $C_{0,1}$ and $c_{a\gamma \gamma}$ are model dependent, while the coupling to gluons is fixed by the $U_A(1)$  anomaly. 

As in Section~\ref{Sec2}, the $G \tilde G$ term can be eliminated in favor of a complex mass term, 
with the difference that the $U_A(1)$ rotation depends on the axion field. 
The rotation has the effect of modifying the couplings $ C_{0,1}$ and $c_{a\gamma \gamma}$, and, more importantly, affects the mass sector. 
The discussion of vacuum alignment of Section~\ref{Sec2} can be immediately generalized, by replacing $\theta$ with 
$\bar\theta + a/f_a$. In this context, vacuum alignment achieves the diagonalization of the pion-axion mass term.

After imposing the vacuum alignment condition, the quark-axion Lagrangian becomes 
\begin{eqnarray}\label{eq:axion2}
\delta \mathcal L &=&  - \bar\psi   \left[  \mathcal M  -  \mathcal M^{-1}  \frac{m^2_*}{2} \left( \bar\theta + \frac{a}{f_a}\right)^2 \right]  \psi
+ \bar\psi i \gamma_5 \psi m_* \left( \bar\theta + \frac{a}{f_a}\right),
\end{eqnarray}
where we have kept terms quadratic in $\bar\theta + a/f_ a$.
When chiral symmetry is spontaneously broken,  $\bar \psi \psi$ acquires
a non-vanishing  vacuum expectation value,  $- (m_u + m_d) \langle \bar\psi \psi \rangle = 3 m_{\pi}^2 f^2_{\pi}$, and the CP-even quark mass term in Eq.~\eqref{eq:axion2}
generates an axion potential
\begin{eqnarray}\label{eq:axion3}
V_0\left(\bar\theta + \frac{a}{f_a}\right) &=& \frac{1}{3}\langle \bar\psi \psi \rangle \textrm{Tr} \left[  \mathcal M  -  \mathcal M^{-1}  \frac{m^2_*}{2} \left( \bar\theta + \frac{a}{f_a}\right)^2 \right] \nonumber \\
& =&  - \frac{m^2_\pi f^2_{\pi}}{( m_u + m_d)} \left( (m_s + m_d + m_s - m_* \frac{1}{2} \left(\bar\theta + \frac{a}{f_a}\right)^2	   \right).
\end{eqnarray}
$V_0$  is an  even function of $\bar\theta + a/f_a$, and it is minimized by 
\begin{equation}
\bar\theta + \frac{\langle a \rangle}{f_a} = 0,
\end{equation}
thus canceling the CP-violating effects of the $\bar\theta$ term.
Oscillations around the minimum  determine the axion mass in terms of the pion mass and decay constant, and of the axion decay constant 
\begin{equation}
m_a^2 = \frac{m^2_{\pi} f^2_{\pi}}{f^2_a} \frac{m_u m_d}{(m_u + m_d)^2},
\end{equation}
where we neglected small corrections $\sim m_{u,d}/m_s$.

The presence of additional, chiral symmetry breaking sources of CP violation has the effect of shifting the minimum of the axion potential, inducing a residual $\bar\theta$ term, proportional to the amount of CP violation. As an example, we discuss the case of CP violation from a quark CEDM.
Performing vacuum alignment, as discussed in Section~\ref{Sec2}, induces a CP even axion-quark Lagrangian of the form
\begin{eqnarray} \label{eq:axion32}
\delta \mathcal L &=& - \frac{g}{2}  m_*  \left(\bar\theta + \frac{a}{f_a} \right)  \   \bar\psi \sigma^{\mu \nu} G_{\mu \nu} \mathcal M^{-1} [d_{CE}]  \,  \psi 
\nn \\ 
&+ & \frac{r}{2} m_*    \left(\bar\theta + \frac{a}{f_a} \right)  \ \bar \psi  \
\Big\{  {\cal M}^{-1}  [d_{CE}]  -  m_*   {\cal M}^{-1} \, {\rm Tr} \left[ {\cal M}^{-1} d_{CE} \right]  \Big\}   \, \psi
+ O(\bar{\theta}^2)~.
\label{eq:axion4}
\end{eqnarray}
When chiral symmetry is broken, the isoscalar components in Eq.~\eqref{eq:axion32} 
give a correction to the axion potential Eq.~\eqref{eq:axion3}.
Up to terms of $[d_{CM}] \times (\bar\theta + a/f_a)^2$ 
that  affect the value of the axion mass but do not change the minimum of the potential, 
the shifted potential reads:
\begin{eqnarray}\label{eq:axion4dup}
V\left(\bar\theta + \frac{a}{f_a}\right) &=&
\frac{1}{6} m_*   \ \left( \bar\theta + \frac{a}{f_a}\right)^2  \  \langle \bar \psi \,  \psi \rangle 
\nn\\
&-& \frac{1}{6} m_*   \ \left( \bar\theta + \frac{a}{f_a}\right) \ {\rm Tr}  \left[ {\cal M}^{-1}  [d_{CE}] \right]  
 \  \langle  \bar \psi \sigma^{\mu \nu} g G_{\mu \nu} \psi \rangle~. 
\end{eqnarray}
The term proportional to $[d_{CE}]$, odd in $\bar \theta + a/f_a$, causes 
the potential to be minimized at a non-zero value of the $\bar\theta$ angle,
\begin{equation}
\bar\theta + \frac{\langle a \rangle}{f_a} = \bar\theta_{\textrm{ind}} = 
\frac{r}{2}  \textrm{Tr} \left[ \mathcal M^{-1} \left[d_C\right]\right], \qquad
 r = \frac{\langle   \bar \psi \sigma^{\mu \nu} g G_{\mu \nu} \psi \rangle }{\langle  \bar \psi  \psi \rangle} ,
\end{equation}
where,  by chiral symmetry,  $r$ is the same ratio defined in Section~\ref{Sec2}.

\section{Projections in the isospin limit}
\label{app:projections}

We now discuss the projections needed to extract $\alpha_3, \ldots, \alpha_7$  in Eq.~\eqref{alphas} 
in the isospin limit $m_u = m_d \equiv m$, in which the matrices $M_2$ and $M_3$ defined in Eq.~\eqref{m2} and Eq.~\eqref{m3} become singular.  

In the case $a=3$, the operators $O_6^{(5)}$ and $O_{10}^{(5)}$ (and the structures multiplying $\alpha_4$ and $\alpha_7$)  vanish. 
To isolate $\alpha_3$, it is sufficient to impose  
\begin{equation}
{\rm Tr}  \left[ \Gamma^{(2)}_C \, \gamma_5 \slashed q  \,  \mathcal M t^3 \right]_S = 0.
\end{equation}
In the isospin limit, for $a=3$,  $\alpha_5$ and $\alpha_6$ are both proportional to $t_3$, and cannot be disentangled with 
a flavor projection. However, the different dependence on $m^2$ and $m^2_s$ can be exploited, by imposing 
\begin{subequations}
\bea
{\rm Tr}  \left[ \left( \frac{\partial^2}{\partial m^2}  - 2 \frac{\partial^2}{\partial m_s^2}  \right) \Gamma^{(2)}_C \, \gamma_5     \,    t^3 \right]_S &=& 0
\\
{\rm Tr}  \left[ \frac{\partial^2}{\partial m_s^2}   \Gamma^{(2)}_C \, \gamma_5     \,    t^3 \right]_S    &=&0 ~. 
\eea
\end{subequations}
The first (second) trace above isolates $\alpha_5$ ($\alpha_6$).

For $a = 0$ and $8$, $M_2$ is not singular even in the isospin limit,
and the mixing of the CEDM with the divergence of the axial current is
found by imposing Eq.~\eqref{mixwithax}.  The structures multiplying
$\alpha_5$, $\alpha_6$ and $\alpha_7$ can be projected on the two
flavor matrices $t^0$ and $t^8$, so that the flavor projections in
$M_3$ are not independent, and the matrix is singular. Also in this
case, one can take advantage of the different dependence on $m^2_s$,
$m^2$ and $m_s m$.  Defining $t_l = (t_8 + \sqrt{2} t_0 )$ and $t_s =
(t_8 - \frac{1}{\sqrt{2}} t_0 )$, $\alpha_5$, $\alpha_6$ and
$\alpha_7$ can be disentangled by the following projections:
\begin{subequations}
 \bea
{\rm Tr}  \left[ \left( \frac{\partial^2}{\partial m^2}  - 2 \frac{\partial^2}{\partial m_s^2} - 4 \frac{\partial^2}{\partial m_s \partial m}\right)  \Gamma^{(2)}_C  \, \gamma_5     t_{l,s}   \right] &=& 0 
\label{isoproj}\\
{\rm Tr}  \left[\frac{\partial^2}{\partial m_s^2} \Gamma^{(2)}_C  \, \gamma_5    \,  t_l  \right]  &=& 0
\\
  {\rm Tr}  \left[ \frac{\partial^2}{\partial m_s \partial m} \Gamma^{(2)}_C  \, \gamma_5   \,  t_s  \right]    &=&0 ~,
\eea
\end{subequations}
where in Eq.~\eqref{isoproj} 
$t_l$ ($t_s$) is to be used for $a=0$ ($a=8$).

\section{Matching coefficients}
\label{app:Matching}

In this appendix we give explicit results for the matching
coefficients from RI-$\tilde{\rm S}$MOM to the $\overline{\rm
  MS}$-${\rm HV}$ scheme (where $\psi$ is defined in
Eq.~\eqref{eq:psi}, $K$ in Eq.~\eqref{eq:K}, $C_A$, $C_F$ and $T_F$ in
Eq.~\eqref{eq:colorfactors}, $\xi$ is the gauge parameter and $n_F$
are the number of flavors).
\begin{subequations} 
\bea
c_{11}  &=&   \frac{C_A  (23 + 9 \xi) - 32 C_F}{12}     \    \psi
+ \frac{C_A - 2 C_F}{2} (1 - \xi)    \ K
-  \frac{C_A - 2 C_F}{2} (1 + \xi) \ \log 2 
\nn \\
&+&  \frac{10}{9} n_F T_F  +  C_F \left( \frac{31}{3}  -\frac{1}{2}  \xi \right)   + \frac{C_A}{72} \,  (-646 - 36  \xi +9  \, \xi^2)        
\\
c_{12}  &  =&    \left( 4 C_F - \frac{C_A}{6} ( 3 + \xi) \right) \ \psi   + (C_A - 2 C_F)  (1 - \xi)  \ K 
+ (C_A - 2 C_F)  (1 + \xi)  \ \log 2   
\nn\\
&+&  C_F  (2 - \xi) + \frac{C_A}{4}  ( - 5 + 2 \xi)
\\
c_{13}  &  =&    \left( - \frac{8}{3}  C_F + \frac{C_A}{12}   ( 3 + \xi) \right) \ \psi   -  \frac{1}{2} (C_A - 2 C_F)  (1 - \xi)  \ K  
-   \frac{1}{2} (C_A - 2 C_F)  (1 + \xi)  \ \log 2
\nn \\
&+&  C_F \left(\frac{25}{3} + \frac{1}{2} \xi \right)  + \frac{C_A}{8}  ( 5 - 2 \xi)
\\
c_{14} &=& 0 
\\
c_{15} &=& - 4
\\
c_{16} &=& 3 \, c_{17} 
\\
c_{17} &=& \frac{C_A \, \xi}{6}   \  \psi   + (C_A - 2 C_F) (1 - \xi) \ K  - \frac{1}{2}  (C_A - 2 C_F) (1 - \xi)  \ \log 2  
\nn \\
&+& C_F ( 8 - \xi)  + C_A \left( -  \frac{13}{4} + \frac{1}{2} \xi \right)  
\\
c_{18} &=&   \frac{8 C_F - C_A (1 + \xi) }{2}  \ \psi     -  (C_A - 2 C_F) (1 - \xi)  \ K  + 2  \, (C_A - 2 C_F) \ \log 2
\nn \\
&+&  C_F \left(\frac{10}{3} + \xi \right)    + C_A \left( \frac{21}{4} - \frac{1}{2} \xi \right) 
\\
c_{19} &=& 0
\\
c_{1,10} &=& 0
\\
c_{22} &=&  c_{88} = c_{99} = c_{10,10} = 
 \frac{C_F}{2} ( 3 + \xi ) \ \psi  -  C_F  (12 + \xi) 
\\
c_{33} &=&  C_F \left[  (1 - \xi)  \left( \frac{4}{3} - \frac{1}{2} \psi \right) + \xi \right]
\\
c_{55} &=&  2 \, C_A \left(1 + \frac{1}{3} \xi \right) \ \psi  + \frac{C_A}{36}  ( - 403   - 18 \xi   + 9 \xi^2)  + \frac{20}{9}  n_F T_F
\\
c_{56} &=&  - 4 C_F \left(  4 -  \psi \right) 
\\
c_{66} &=&   c_{77} =   - 4 \, C_F  ~.
\eea
\end{subequations}

The coefficients in the NDR scheme are given by (we report only the cases where 
$c_{ij}^{\rm NDR} \neq c_{ij}^{\rm HV}$): 
\begin{subequations}
\bea
c_{11}^{\rm NDR} &=& c_{11}^{\rm HV}  + 2 C_A - \frac{4}{3} C_F 
\\
c_{13}^{\rm NDR} &=& c_{13}^{\rm HV} - \frac{4}{3} C_F 
\\
c_{16}^{\rm NDR} &=& c_{16}^{\rm HV} + \frac{C_A - 4 C_F}{6} 
\\
c_{17}^{\rm NDR} &=& c_{17}^{\rm HV} + \frac{C_A - 4 C_F}{2} 
\\
c_{18}^{\rm NDR} &=& c_{18}^{\rm HV} - C_A  - \frac{28}{3} C_F 
\\
c_{66}^{\rm NDR} &=& c_{77}^{\rm NDR}  = 0 
\\
c_{22}^{\rm NDR} &=& c_{88}^{\rm NDR} = c_{99}^{\rm NDR} = c_{10,10}^{\rm NDR} = 
 \frac{C_F}{2} ( 3 + \xi ) \ \psi  -  C_F  (4 + \xi)~.
\eea
\end{subequations}

%%%%%%%%      BIBTEX  
%\bibliographystyle{h-physrev-tanmoy}
\bibliographystyle{h-physrev}
\addcontentsline{toc}{section}{References}\nobreak
\bibliography{edms}

\end{document}